\documentclass[superscriptaddress,preprintnumbers,amsmath,amssymb,prd,nofootinbib,preprint]{revtex4-1}
\pdfoutput=1
\usepackage{graphicx}
\usepackage{epstopdf}
\usepackage{dcolumn}
\usepackage{bm}
\usepackage{hyperref}
\usepackage{color}
\usepackage{amsmath}
\usepackage{cancel}
\usepackage{xpatch}

\begin{document}


\def\a{\alpha}
\def\b{\beta}
\def\c{\varepsilon}
\def\d{\delta}
\def\e{\epsilon}
\def\f{\phi}
\def\g{\gamma}
\def\h{\theta}
\def\k{\kappa}
\def\l{\lambda}
\def\m{\mu}
\def\n{\nu}
\def\p{\psi}
\def\q{\partial}
\def\r{\rho}
\def\s{\sigma}
\def\t{\tau}
\def\u{\upsilon}
\def\v{\varphi}
\def\w{\omega}
\def\x{\xi}
\def\y{\eta}
\def\z{\zeta}
\def\D{\Delta}
\def\G{\Gamma}
\def\H{\Theta}
\def\L{\Lambda}
\def\F{\Phi}
\def\P{\Psi}
\def\S{\Sigma}

\def\o{\over}
\def\beq{\begin{align}}
\def\eeq{\end{align}}
\newcommand{\gsim}{ \mathop{}_{\textstyle \sim}^{\textstyle >} }
\newcommand{\lsim}{ \mathop{}_{\textstyle \sim}^{\textstyle <} }
\newcommand{\vev}[1]{ \left\langle {#1} \right\rangle }
\newcommand{\bra}[1]{ \langle {#1} | }
\newcommand{\ket}[1]{ | {#1} \rangle }
\newcommand{\EV}{ {\rm eV} }
\newcommand{\KEV}{ {\rm keV} }
\newcommand{\MEV}{ {\rm MeV} }
\newcommand{\GEV}{ {\rm GeV} }
\newcommand{\TEV}{ {\rm TeV} }
\newcommand{\1}{\mbox{1}\hspace{-0.25em}\mbox{l}}
\newcommand{\headline}[1]{\noindent{\bf #1}}
\def\diag{\mathop{\rm diag}\nolimits}
\def\Spin{\mathop{\rm Spin}}
\def\SO{\mathop{\rm SO}}
\def\O{\mathop{\rm O}}
\def\SU{\mathop{\rm SU}}
\def\U{\mathop{\rm U}}
\def\Sp{\mathop{\rm Sp}}
\def\SL{\mathop{\rm SL}}
\def\tr{\mathop{\rm tr}}
\def\mpl{M_{\rm Pl}}

\def\IJMP{Int.~J.~Mod.~Phys. }
\def\MPL{Mod.~Phys.~Lett. }
\def\NP{Nucl.~Phys. }
\def\PL{Phys.~Lett. }
\def\PR{Phys.~Rev. }
\def\PRL{Phys.~Rev.~Lett. }
\def\PTP{Prog.~Theor.~Phys. }
\def\ZP{Z.~Phys. }

\def\dd{\mathrm{d}}
\def\ff{\mathrm{f}}
\def\BH{{\rm BH}}
\def\inf{{\rm inf}}
\def\ev{{\rm evap}}
\def\eq{{\rm eq}}
\def\SM{{\rm sm}}
\def\Mpl{M_{\rm Pl}}
\def\GeV{{\rm GeV}}
\def\Myr{\rm Myr}
\def\mw{{\mbox{$\resizebox{.09in}{.08in}{\includegraphics[clip]{Figures/spiral}}$}}}

\newcommand{\Red}[1]{\textcolor{red}{#1}}
\newcommand{\TL}[1]{\textcolor{blue}{\bf TL: #1}}
\newcommand{\updm}{{\Delta'}}
\newcommand{\upex}{{X}}

\title{
CHAMP Cosmic Rays
}
\author{David Dunsky}
\affiliation{Department of Physics, University of California, Berkeley, California 94720, USA}
\affiliation{Theoretical Physics Group, Lawrence Berkeley National Laboratory, Berkeley, California 94720, USA}
\author{Lawrence J. Hall}
\affiliation{Department of Physics, University of California, Berkeley, California 94720, USA}
\affiliation{Theoretical Physics Group, Lawrence Berkeley National Laboratory, Berkeley, California 94720, USA}
\author{Keisuke Harigaya}
\affiliation{School of Natural Sciences, Institute for Advanced Study, Princeton, New Jersey, 08540}
\affiliation{Department of Physics, University of California, Berkeley, California 94720, USA}
\affiliation{Theoretical Physics Group, Lawrence Berkeley National Laboratory, Berkeley, California 94720, USA}

\begin{abstract}

We study interactions of cosmological relics, $X$, of mass $m$ and electric charge $qe$ in the galaxy, including thermalization with the interstellar medium, diffusion through inhomogeneous magnetic fields and Fermi acceleration by supernova shock waves.  We find that for $m \lsim 10^{10} q \; \GeV$,  there is a large flux of accelerated $X$ in the disk today, with a momentum distribution $\propto 1/p^{2.5}$ extending to $(\beta p)_{\rm{max}} \sim 5 \times10^4 q \; \GeV$.  Even though acceleration in supernova shocks is efficient, ejecting $X$ from the galaxy, $X$ are continually replenished by diffusion into the disk from the halo or confinement region.
For $m \gsim 10^{10} q \; \GeV$, $X$ cannot be accelerated above the escape velocity within the lifetime of the shock.  
The accelerated $X$ form a component of cosmic rays that can easily reach underground detectors, as well as deposit energies above thresholds, enhancing signals in various experiments. We find that nuclear/electron recoil experiments place very stringent bounds on $X$ at low $q$; for example, $X$ as dark matter is excluded for $q > 10^{-9}$ and $m < 10^5 \; \GeV$.  For larger $q$ or $m$, stringent bounds on the fraction of dark matter that can be $X$ are set by Cherenkov and ionization detectors.  Nevertheless, very small $q$ is highly motivated by the kinetic mixing portal, and we identify regions of $(m,q)$ that can be probed by future experiments.

\end{abstract}

\date{\today}

\maketitle

\tableofcontents

\section{Introduction}

Theories Beyond the Standard Model may contain exotic stable CHArged Massive Particles, or CHAMPs, of mass $m$ and electric charge $qe$.  They may arise in a variety of ways: for example, from exotic color-neutral matter added to the Standard Model, or even from exotic heavy colored states that bind with the known quarks~\cite{Nomura:2005qg,Harigaya:2016pnu,DeLuca:2018mzn,dunsky2019higgs}.  Another important possibility is that the gauge group is extended to a hidden sector where stable particles couple to a hidden photon that is kinetically mixed with our photon~\cite{Holdom:1985ag}. 

In general, CHAMPs, $X$, are produced in the early universe. The genesis mechanism, and hence the relic abundance, are extremely model-dependent so that, in addition to $(m,q)$, we take the abundance of $X$ normalized by that of dark matter (DM), $f_X \equiv \Omega_X / \Omega_{\rm DM}$, to be a free parameter.  We study cases where CHAMPs account for the entire DM, and where they form a sub-dominant component.

CHAMP DM has been considered for over 30 years.  The kinetic mixing portal~\cite{Goldberg:1986nk} allows DM from a dark sector to become visible by acquiring a small electric charge, $q$.  This charge may be suppressed by a loop factor, involving a heavy connector particle of mass $M$ that carries both charges, suggesting values of $q$ of order $(10^{-2} - 10^{-3})$.  However, in unified theories where hypercharge and hidden $U(1)$s are embedded in non-Abelian factors down to scales $V$ and $V'$ much less than $M$, the charge $q$ receives a power suppression by $V V'/M^2$, and hence may naturally be very small.  Power suppression of $q$ can also arise from an approximate symmetry. One example is a hidden U(1) embedded in a hidden SU(2) that is spontaneously broken by a triplet.  A charge conjugation symmetry from SU(2) forbids kinetic mixing until higher-dimensional operators are added.  In summary,  it is well-motivated to examine a wide range of the $(m,q)$ plane.

A variety of constraints and signals of CHAMP DM with order unity charges were considered in~\cite{DeRujula:1989fe, Dimopoulos:1989hk}, and for CHAMPs with $q \ll 1$ in~\cite{Dobroliubov:1989mr}. An important cosmological bound on CHAMP DM, arising from the era of recombination from constraints on the CMB acoustic peaks and from damping of the density perturbations, requires $m  > 10^{12} q^2 \; \GeV$ for $m > 1$ MeV ~\cite{Dubovsky:2003yn, Burrage:2009yz, McDermott:2010pa,Dolgov:2013una}.  However, this bound disappears if $X$ contributes less than 1\% of the DM.   Chuzhoy and Kolb~\cite{Chuzhoy:2008zy} proposed that, over a certain range of $(m,q)$, supernova (SN) shocks expel CHAMPs from the Milky Way, removing previous constraints on charged dark matter based on bounds from terrestrial observations for these $(m,q)$. 

In this paper, we investigate the evolution of the density and spectrum of CHAMPs in the galactic disk.  We study rates for the three key processes: thermalization of $X$ with the InterStellar Medium (ISM), Fermi acceleration of $X$ by SN shocks, and diffusion of $X$ through magnetic irregularities.   In general, diffusion allows $X$ to both enter and exit the disk.  We find that the $(m,q)$ plane can be divided into three regions
\begin{enumerate}
\item [(I)] $m> 10^{10} \, q$ GeV \hspace{0.25in} The density and spectrum of $X$ in the disk are determined by virialization of the halo. Thermalization, Fermi acceleration and diffusion are negligible so that dark matter signals can be computed by ignoring them.
\item [(II)] $10^5 \, q^2 \, \mbox{GeV} <m <10^{10} \, q$ GeV \hspace{0.25in}X that are initially in the disk are efficiently ejected by SN shocks; however, there is a continual replenishment of $X$ by diffusion from the halo and confinement region to the disk.  The number density and spectrum of $X$ in the disk today follows from a balance between accretion and ejection.
\item [(III)] $ m< 10^5 \, q^2$ GeV \hspace{0.25in}  $X$ collapse with baryons into the disk as it forms.  Thermalization of $X$ with the ISM reduces the efficiency of ejection, leading to large densities of $X$ in the disk today.
\end{enumerate}
 
In both (II) and (III) there are large fluxes of Fermi accelerated $X$s, that we call CHAMP Cosmic Rays. These give signals in a variety of direct detection experiments deep underground as well as on the surface of the Earth, via nuclear recoil, electron recoil, ionization losses and Cherenkov radiation.

The parameter space where halo CHAMPs collapse into galactic disks is studied in Sec.~\ref{sec:gravCollapse}; determining the boundary between (II) and (III).  After a halo virializes, CHAMPs that thermalize with the infalling baryons within a free-fall time are dragged along into the disk. 
CHAMPs that collapse cannot be halo dark matter.  However, they may be a component of dark matter, and the accelerated CHAMP flux is enhanced by a factor of about $10^2$, commensurate with the greater number of CHAMPs  exposed to SN shocks in the disk.

In Sec.~\ref{sec:rates}, we examine three fundamental rates that determine the fate of CHAMPs in the galactic disk: their thermalization rate in the ISM, their encounter rate with SN shocks, and their escape rate from the disk. The three rates generally depend on the CHAMP speed as well as $(m,q)$. The accelerated $X$ spectrum is determined by a competition between thermalization with the ISM plasma, SN shocks that accelerate $\upex$ to shock speeds or beyond, and the interstellar magnetic fields which confine $\upex$ to the disk. 

We investigate the efficiency of the acceleration of CHAMPs and calculate the differential momentum spectrum $f = dn/dp$ of a distribution of shocked CHAMPs during their journey through the ISM in Sec.~\ref{sec:batch}, taking into account thermalization, subsequent shock encounters, and escape losses, as depicted graphically by Figs. \ref{fig:rateHierarchy1Plot} and \ref{fig:rateHierarchy2Plot}. Initially, a single SN shock transforms a batch of thermal CHAMPs into a $p^{-3}$ distribution such that the CHAMP speed is the encountered shock speed and the amplitude is the relative probability of encountering a shock of that speed. Most CHAMPs thermalize quickly, but a small number do not and either escape or encounter additional SN shocks. In the Milky Way the latter dominates and produces a relativistic $p^{-2}$ Fermi spectrum cutoff at $p \sim 5 \times 10^4 ~  q/\beta ~\GeV$, determined by the required acceleration time exceeding the lifetime of the shock. CHAMPs that obtain enough energy escape from the disk, with a high efficiency for $m/q^2 \gsim 10^4$ GeV.

Although CHAMPs are ejected from the disk by Fermi acceleration, they are replenished by diffusion accretion from outside the disk. A balance between ejection and accretion leads to a steady state distribution of CHAMPs in the disk. These accelerated CHAMPs can hit the Earth before escaping from the disk, and we estimate the present flux of such cosmic ray CHAMPs in Sec.~\ref{sec:diffusion}

From the accelerated CHAMP spectrum impinging on the Earth, we calculate signal rates in nuclear recoil, electron recoil, ionization and Cherenkov detectors in Sec.~\ref{sec:DD}. Since the accelerated CHAMPs are moving faster than typically assumed dark matter speeds ($\sim 220 ~{\rm km/s}$),  there are new key features of these signals:  (1) CHAMPs can reach underground detectors easily, even if their charges are large. (2) CHAMPs below $1 ~\GeV$ can impart nuclear recoils above the $\sim ~{\rm keV}$ threshold for direct detection experiments such as XENON1T and CDMS. (3) Similarly, CHAMPs below $10$ MeV impact electron recoils in direct detection experiments such as XENON10. (4) They produce ionization losses in detectors such as MAJORANA, MACRO and other monopole search experiments. (5) Relativistic CHAMPs, or electrons produced by recoils, emit Cherenkov light  when traveling through water, leading to bounds from deep underground detectors such as Super Kamiokande and IceCube. 
These signals of the accelerated cosmic ray CHAMP flux lead to powerful constraints, and point to regions of parameter space where discoveries can be made at future experiments. 

Conclusions are drawn in Sec.~\ref{sec:con}.  Appendices consider CHAMP self interactions from hidden photon exchange, the CHAMP spectrum resulting from repeated shocks, and the diffuse extragalactic CHAMP flux from ejection from galaxies throughout the universe. 
 
\section{Collapse of CHAMPs into the Galactic Disk}
\label{sec:gravCollapse}

Halos that virialize at redshift $z_{\rm{vir}}$ have densities 
\begin{align}
	\rho_{\rm{vir}} = 18 \pi^2 \rho_0 (1 + z_{\rm{vir}})^3  \label{eq:virDens}
\end{align}
where $\rho_0 (1 + z_{\rm{vir}})^3$ is the background density~\cite{1999coph.book.....P}, and temperatures
\begin{align}
	T_{\rm{vir}} = \frac{\mu m_p v_{\rm{vir}}^2}{2} &=  10^7 ~ {\rm K}  ~\left(\frac{M}{10^{12} M_{\odot}} h_0\right)^{2/3}\left(\frac{1 + z_{\rm{vir}}}{10}\right)  \label{eq:virTemp}
\end{align}
where $M$ is the halo mass and $\mu = 0.6$ is the mean molecular weight of the baryonic gas.

The cosmological localities of CHAMPs and baryons can diverge during subsequent cooling of these halos. As noted by the authors in~\cite{DeRujula:1989fe}, $\upex$ is dragged into the galactic disk with the baryons if the thermalization time of $\upex$ with the baryonic plasma is less than the dynamical or collapse time of the halo 
\begin{align}
	t_{\rm{coll}} = \sqrt{3 \pi / 32 G \rho_{\rm{vir}}} &=	\frac{1}{6 \sqrt{2}} \frac{1}{H_0 \sqrt{\Omega_M}} (1 + z_{\rm{vir}})^{-3/2} \label{eq:collapse}	
\end{align}
which is independent of halo mass.

Baryons with a cooling time shorter than $t_{\rm{coll}}$ are able to collapse into the disk on the time scale $t_{\rm{coll}}$.  Pre-reionization ($z \gtrsim 6$), halos with virial temperatures above $10^4 ~ \rm K$ radiatively cool via bremsstrahlung and atomic line emission to $10^4 ~ \rm K$, and then collapse isothermally at this temperature within a time $t_{\rm{coll}}$~\cite{1977MNRAS.179..541R}. Note the gas in unable to cool further since the tail of the Boltzmann distribution becomes insufficient to collisionally excite atoms~\cite{loeb2010}. 

However, post-reionization ($z \lesssim 6$), UV light from the first stars and galaxies permeate the intergalactic medium (IGM), heating up and ionizing halo atoms, making atomic line emission less effective and preventing plasma temperatures from dropping below $\sim 10^{4-4.6} ~\rm K$, depending on the degree of self-shielding which is set by the plasma density~\cite{2010gfe..book.....M,1997ApJ...477....8W} 
. Consequently, post-reionization, only halos with virial temperatures above $10^5 ~ \rm K$ cool and collapse~\cite{Loeb:2006za}.

Since $\upex$ thermalizes with the plasma through electromagnetic collisions via Rutherford scattering, the key difference between the  pre- and post-reionization eras is the ionization fraction in virialized halos. Equating recombination and collisional ionization rates at $10^4 ~ \rm K$ implies the ionization fraction of the plasma is about $10^{-3}$ pre-reionization, while it is near unity above $10^4 ~ \rm K$ post-reionization~\cite{2011piim.book.....D,Loeb:2006za}. 

Specifically, the thermalization time between $\upex$ with mass $m$ and charge $qe$, and a background plasma of temperature $T$ is given by~\cite{DeRujula:1989fe,Spitzer:1956hha}
\begin{align}
	t_{\rm{therm}}	&= \frac{3}{8 \sqrt{2 \pi}} \frac{m \, m_{e,p}}{q^2 \alpha^2 \, n \ln \Lambda} \left(\frac{T_\upex}{m} + \frac{T}{m_{e,p}} \right)^{3/2} \label{eq:thermalization}		
\end{align}
where $T_X = m v^2/3$ is the effective temperature of $\upex$, $n = n_e = n_p$ is the density of protons or electrons with mass $m_{e,p}$, and $\Lambda = 3T/\alpha k_D$ is the IR cutoff where electromagnetic shielding becomes effective beyond an inverse of the Debye momentum of the plasma, $k_D = \sqrt{4 \pi n \alpha / T}$.

Initially, $T_\upex/m = T_{\rm{vir}}/m_p$ since the virial speeds of $\upex$ and protons are identical, being set by gravity. Because the proton and electron plasma quickly cools to a temperature $T_{\rm{min}} \approx 10^4 ~ \rm K$ for halos that virialize before reionization and $T_{\rm{min}} \approx 10^{4-4.6} ~ \rm K$  for halos that virialize after, the second term in parenthesis of \eqref{eq:thermalization} quickly becomes $T_{\rm{min}}/m_{e,p}$. Likewise, the ion number density is given by
\begin{align}
	 n \equiv n_B x_{\rm{ion}} = \frac{\Omega_B}{\Omega_M} \frac{\rho_{\rm{vir}}}{m_p} x_{\rm{ion}}  = 18 \pi^2 \left(\frac{3}{8 \pi G m_p} \Omega_B H_0^2 \right)(1 + z_{\rm{vir}})^3 ~ x_{\rm{ion}} \label{eq:ionDensity}
\end{align}
where $x_{\rm{ion}}$ is the ionization fraction of the plasma and is $10^{-3} ~ (1)$ before (after) reionization.

\begin{figure}
    \centering
        \includegraphics[width=0.45\textwidth]{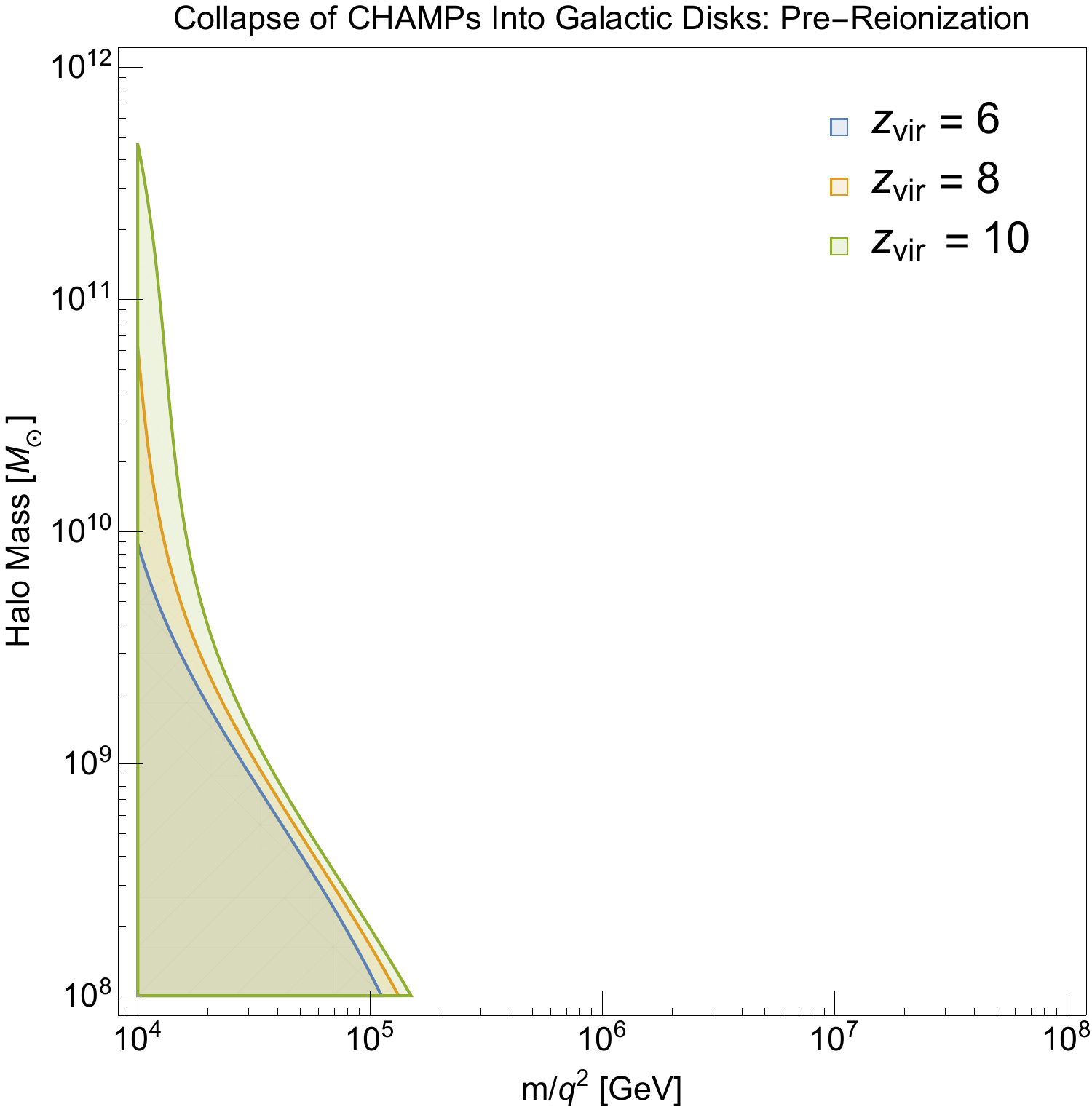}
        \includegraphics[width=0.45\textwidth]{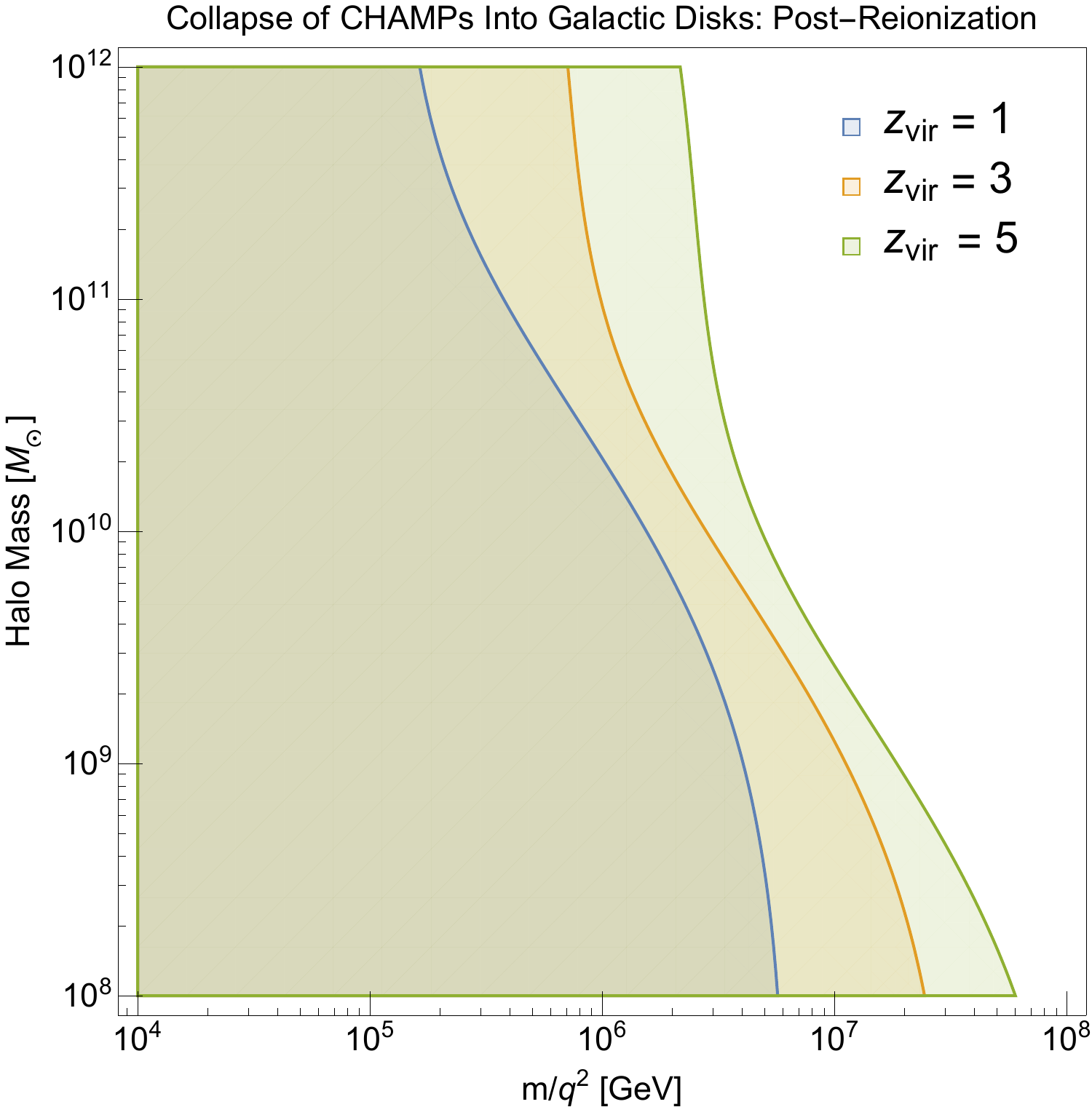}
    \caption{Shaded regions indicate the parameter space where CHAMPs fall into disks with baryons at a halo collapse redshift $z_{\rm{vir}}$, determined by setting $t_{\rm{therm}}(z_{\rm{vir}}) < t_{\rm{coll}}(z_{\rm{vir}})$. Pre-reionization (left), the ion fraction is low and thermalization between $\upex$ and the plasma is difficult. Post-reionization (right) the ion fraction is high and thermalization between $\upex$ and the plasma is enhanced. At high redshifts, the halos are denser, and the thermalization time shorter. The change in concavity for halos $\gtrsim 10^{11} M_\odot$ signifies where thermalization with electrons dominate.}
\label{fig:collapse}
\end{figure}

Inserting the plasma number density $\eqref{eq:ionDensity}$ into \eqref{eq:thermalization}, and demanding $t_{\rm{therm}}<t_{\rm{coll}}$, yields the parameter space where $\upex$ collapses into the galactic disk with the baryons, as a function of halo mass and $m/q^2$, as shown by the shaded regions of Fig.~\ref{fig:collapse}. 

In general, the number of $\upex$ in the disk of a galaxy with halo mass $M$ is approximately
\begin{align}
	N_{\upex} = \frac{M}{m} f_\upex f_D
	\label{eq:hpNumInDiskCollapse}
\end{align}
where $f_\upex \equiv \Omega_\upex/\Omega_{\rm{DM}}$ and $f_D$ the fraction of CHAMPs that actually end up in the disk, exposed to SN. Taking the disk formation efficiency to be similar for CHAMPs and baryons, observations set $f_D \approx 1/4$~\cite{Mo:1997vb} if $\upex$ collapse. The rest of $X$ reside outside the disk.

On the other hand, if $\upex$ do not collapse, the number of CHAMPs in the disk is suppressed, and $f_D$ is approximately 
\begin{align}
	f_D  = \frac{\int_{\rm{disk}} \rho_{\rm{DM}}(x) ~ d^3 x}{\int_{halo} \rho_{\rm{DM}}(x) ~ d^3 x} \; \approx \;  \left(\frac{x_H x_R}{1+ x_R c_N} \frac{c_N^2}{\log({c_N+1}) - 1} \right) \approx \; {\rm few}~10^{-3}
	\label{eq:hpNumInDiskNotCollapse}
\end{align}
where we have evaluated the dark matter mass fraction in the disk using an NFW profile~\cite{Navarro:1995iw}. Here, $x_R$ ($x_H$) is the ratio of disk radius (disk height) to the halo virial radius, and $c_N \approx 12.5(M/10^{12} M_\odot)^{-1/10}(1+z_{\rm{vir}})^{-1}$ the NFW halo concentration parameter~\cite{Loeb:2006za}. Typical values of $x_R$, $x_H/x_R$, and $c_N$~\cite{Mo:1997vb, Loeb:2006za} imply \eqref{eq:hpNumInDiskNotCollapse} is a few $10^{-3}$ in our galaxy.

The result we need for the rest of the paper is that for the Milky Way ($z_{\rm{vir}} \sim 1$, M $\approx 10^{12} M_{\odot}$, $T_{\rm{min}}\simeq 10^{4.6}~{\rm K}$), $\upex$ collapse into the disk for $m/q^2 \lesssim 10^5 ~ \GeV$, and the number density of $X$ inside the disk is about 100 times larger than the naive scaling of the local dark matter density by $f_X = \Omega_X/\Omega_{\rm{DM}}$.
 In this region of parameter space, $X$ cannot be the halo dark matter.  From Fig.~\ref{fig:collapse}, the excluded range of $(m,q)$ for $f_X  =1$ is somewhat larger, since other galaxies also have halo dark matter.   Furthermore, we will discover that the resulting high density of $X$ in the disk leads to a large SN-accelerated CHAMP flux, giving a strong bound on $f_X$. 

One caveat is that when $m < m_p$, thermalization \textit{increases} the speed of $X$ relative to the proton thermal speed. To collapse fully, the orbital radius of $X$ must decrease by a factor $R_0/R_f = v_{\rm{vir}}^2/(3kT_{\rm{min}}/m) \approx 10$, and hence $X$ with $m \lesssim 100 ~\rm MeV$ do not completely collapse for $T_{\rm{min}} \lesssim 10^{4.6} ~\rm K$. However, $X$ with such small masses that do thermalize are already excluded by direct searches (see Sec.~\ref{sec:DD}).

Finally, note that the thermalization time is always shorter than \eqref{eq:thermalization} during the collapse process because the plasma density increases while still maintaining a small, but non-neglible ion fraction. Self-shielding from the background UV light starts becoming effective when $\rm{H_I}$ column densities exceed $N_{HI} \approx 10^{-18} ~{\rm cm^{-2}}$~\cite{2014MNRAS.444..503N}, but transition to neutrality ($x_{\rm{ion}} \lesssim 0.1$) requires column densities two orders of magnitude greater~\cite{2011ApJ...737L..37A}. Since the column density of a collapsing cloud corresponds to a characteristic number density by $N_{HI} \approx n c_s / \sqrt{G \rho}$~\cite{2001ApJ...559..507S,2014MNRAS.444..503N}, densities of order $0.1 -1 ~\rm{cm^{-3}}$ are required for ionization fractions to drop below $0.1$~\cite{2016ASSL..423.....M}. Consequently, for lower mass halos which typically collapse at higher redshifts and densities, the collapse of $X$ into a disk may be partial. However, for the Milky Way, whose self-shielding density coincides with its post-collapse density, it is possible that some CHAMPs with $m/q^2 > 10^5 ~\GeV$ may also thermalize and fall into the disk during the collapse. As we will see in Sec.~\ref{sec:batch}, however, for $m/q^2 > 10^5$ GeV the ejection of $X$ is so efficient that most of $X$ which were initially inside the disk are ejected, and hence the accelerated CHAMPs which hit the Earth in the present universe are dominated by CHAMPs which were initially outside the disk and diffused into the disk later. Therefore for $m/q^2 > 10^5$ GeV, it does not matter whether $X$ collapses into the disk.

\section{Three Key Rates in the Galactic Disk}
\label{sec:rates}
In this section we introduce three key rates, 1) the thermalization rate, 2) the supernova shock rate, and 3) the escape rate from galactic disks. The interplay of these rates determine the probability for $X$ to escape from galactic disks as well as the number density and spectrum of those that remain, as discussed in Sec.~\ref{sec:batch}.

\subsection{Thermalization Rate in the Interstellar Medium}
The $\upex$ that do fall into the disk or happen to reside there are greatly influenced by the environment of the ISM. In our Milky Way, the ISM consists of hot, warm, and cool phases in pressure equilibrium ($nT \approx$ constant), and a self-gravitating molecular phase~\cite{2011piim.book.....D}. The cool phase is composed of small, atomic clouds and the warm and hot phases constitute the intercloud medium and essentially the entire ISM by volume. Typical properties of these phases for our Milky Way are shown in Table \ref{table:ISM}. 
\begin{table}[h!]
\begin{center}
 \setlength{\tabcolsep}{10pt}
 \begin{tabular}{l c c c c} 
 \hline
 ISM Phase & $n_{tot}~ \rm (cm^{-3})$ &	$n_{e}~ \rm (cm^{-3})$ & $T ~ \rm (K) $ & Fractional Volume  $f$\\ [0.5ex] 
 \hline
 Hot Ionized 	& $3\times 10^{-3}$		& $3 \times 10^{-3}$  	& $5 \times 10^{5}$  		&	$0.5$ \\ 
 Warm Ionized 	& $0.3$ 					& $0.2$ 					& $8 \times 10^{3}$  		&	$0.15$ \\ 
 Warm Neutral 	& $0.5$ 					& $\lesssim 0.05$ 				& $8 \times 10^{3}$ 		&	$0.3$ \\ 
 Cold Neutral 	& $50$ 					& $<0.1$ 				& $80$ 						&	$0.04$ \\ 
 Molecular 		& $>300$ 				& $<0.1$ 				& $10$ 						&	$0.01$ \\ 
 \hline
\end{tabular}
\caption{Components of the interstellar medium, taken from~\cite{2005fost.book.....S,McKee:1977dz}.}
\label{table:ISM}
\end{center}
\end{table}

An $\upex$ moving through the ISM at a speed $v$ 
\footnote{Initially, $v$ is set by the thermal speed if $\upex$ is dragged into the disk, or the virial speed if not; later, $v$ is determined by SN shocks.}
thermalizes at an expected rate
\begin{align}
	\Gamma_{\rm{therm}} &= \sum_{{\rm{phase}}~ i} \frac{f_i}{t_{\rm{therm}, i}} \approx \frac{f_{\rm{WIM}}}{t_{\rm{therm}, WIM}} \nonumber \\
	&\approx \left(4 \times 10^7 ~ {\rm yr} \right)^{-1} \left(\frac{m/q^2}{10^6 ~ {\rm GeV}}\right)^{-1} \left(\frac{v}{10^3 ~ {\rm km/s}}\right)^{-3} \left(\frac{n_e}{0.2 ~ {\rm cm^3}}\right)\left(\frac{f_{\rm{WIM}}}{0.15}\right) 
    \label{eq:WIMThermalization}
\end{align}
where $t_{\rm{therm}, i}$ is the thermalization time \eqref{eq:thermalization} of $\upex$ in ISM phase $i$, and $f_i$ its corresponding volumetric filling factor.
The largest ambient electron density implies the shortest thermalization time, and hence the warm ionized medium (WIM) dominates the thermalization rate, as can be seen from Table \ref{table:ISM}. Thus, $\upex$ is most likely to be found in the warm medium and indeed, that is where about half the baryonic mass of the ISM lies~\cite{2011piim.book.....D}. 
Eq.~(\ref{eq:WIMThermalization}) assumes that $v$ is larger than the thermal speed of protons, $v_p \simeq 10$ km/s and electrons, $v_e\simeq 600$ km/s. If $v_p < v < v_e$, $v$ in Eq.~(\ref{eq:WIMThermalization}) should be replaced by $v_e$.

A natural way to obtain small $q$ is to introduce a hidden $U(1)$ under which $X$ is charged, and assume a small kinetic mixing between the hidden $U(1)$ gauge field and the electromagnetic field. Then the interaction between $X$s by the hidden $U(1)$ also contributes to thermalization. As is shown in Appendix~\ref{sec:hidden U1}, this interaction does not change the estimation of the efficiency of the evacuation if $m> O(10)$ GeV or $X$ is produced before the onset of the Big-Bang Nucleosynthesis.

\subsection{Supernova Shock Rate}
\label{subsec:SNEncounter}
CHAMPs are accelerated by SN shocks in the same way a ball is accelerated by reflecting off a moving wall; the moving wall in this case is the moving magnetic field near the shock. When moving slower than the shock, the CHAMP is accelerated to the shock speed. When moving faster than the shock, the CHAMP may repeatedly reflect off the shock, resulting in an exponential momentum gain due to the change in momentum $\Delta p \approx p \times (v_s/v)$ with each reflection. This latter process is known as first-order Fermi acceleration, and the rate at which CHAMPs are accelerated is thus intimately tied to the rate of encountering strong shocks in the ISM. 

\begin{figure}[ht!]
\centering
\includegraphics[width=0.9\textwidth]{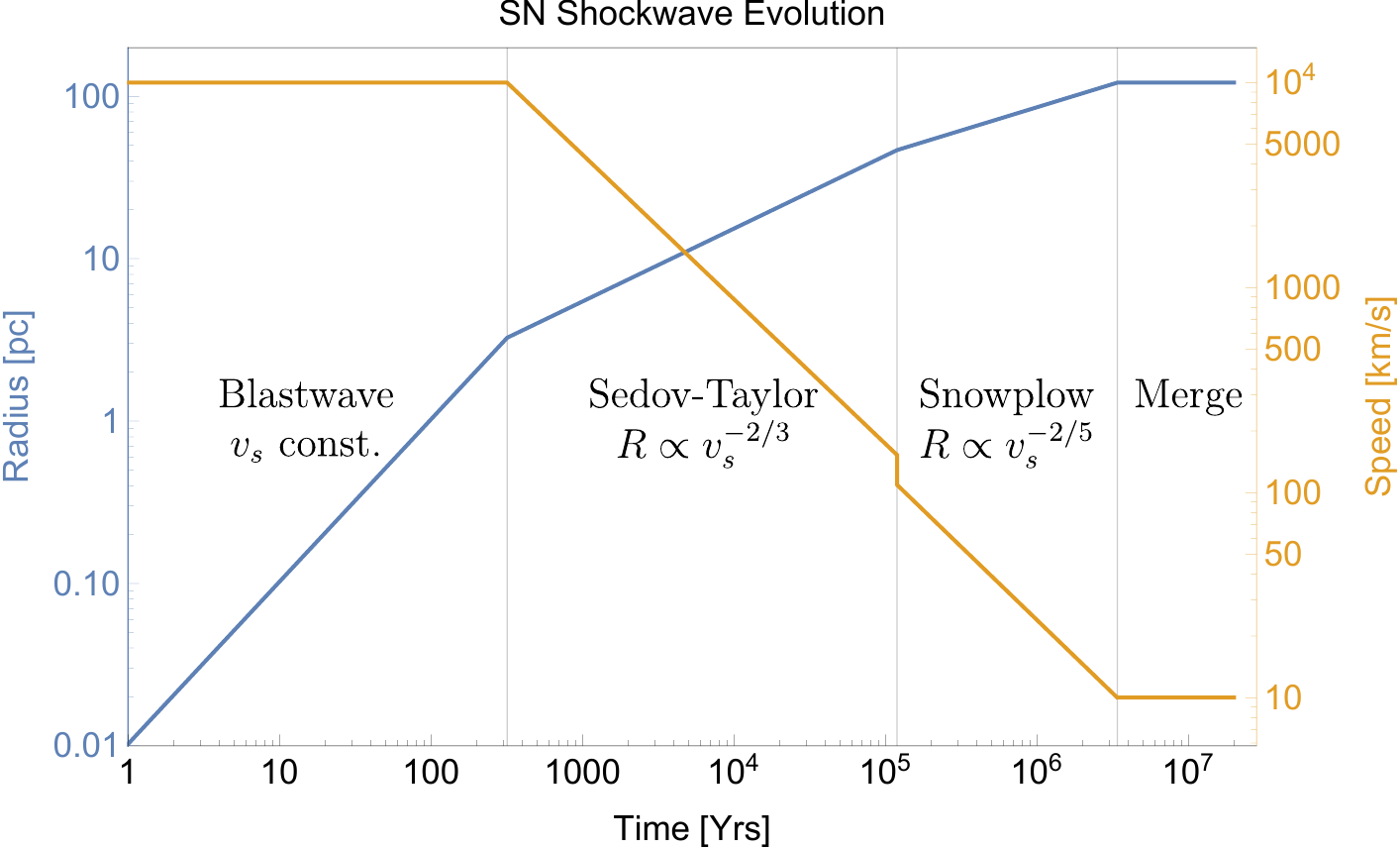} %
\caption{Evolution of a SN remnant in homogenous medium with average density of the intercloud medium $n \approx 0.2 ~{\rm cm^{-3}}$.}
\label{fig:snEvolutionPlot}
\end{figure}

The expected rate of encountering a SN shock of speed $v_s$ is
\begin{align}
	\Gamma_{\rm Enc}(v_s)	&=	\frac{V_{\rm{SN}}}{V_{\rm{disk}}}\Gamma_{\rm{SN}}
    \label{eq:snEncounterRateGeneral}
\end{align}
where $V_{\rm{disk}}$ is the volume of disk, $V_{\rm{SN}} \approx 4\pi R(v_s)^3 / 3$ is the volume of a SN remnant with shock speed $v_s$, and $\Gamma_{\rm{SN}}$ the rate of SN in the galaxy. Note that the shock radius is a decreasing function of shock speed; that is, a CHAMP is much more likely to encounter a slower shock. The SN remnant size and shock speed depend on the medium to which it expands, and the theoretical evolution for a shock expanding into a homogenous ambient medium of density $0.2 ~\rm{ cm^{-3}}$ (the average intercloud density) is shown in Fig.~\ref{fig:snEvolutionPlot}
\footnote{While the average density of the ISM is about $1 ~ {\rm cm^{-3}}$, the shock takes the path of least resistance, propagating primarily through the warm/hot intercloud medium and around the dense atomic clouds~\cite{2015MNRAS.450..504M,2015ApJ...803....7S}. As a result, the mass swept up by the shock is primarily the intercloud mass.}

A SN shock begins life expanding at a constant speed near $10^4 ~{\rm km/s}$. Energy conservation then demands that the shock speed decreases proportional to the square root of mass swept up by the shock. For expansion into a homogenous medium with number density $n$, the radius-velocity relation during this `Sedov-Taylor' phase is~\cite{2011piim.book.....D}
\begin{align}
	R(v_s) &= 39 ~{\rm pc} \left(\frac{v_s}{200 {~\rm km/s}}\right)^{-2/3} \left(\frac{n}{0.2 {~ \rm cm^{-3}}} \right)^{-1/3}\left(\frac{E}{10^{51} ~{\rm erg}} \right)^{1/3} \label{eq:snRadius1}
\end{align}
where $E$ is the SN energy output.

As the shell expands, radiative losses from the shock heated gas become comparable to the energy of the SN and the shell is propelled forward only by the pressure of the hot gas inside. The radius-velocity relation during this `snowplow' phase is
\begin{align}
	R(v_s)	&=	48 ~{\rm pc} \left(\frac{v_s}{100 {~\rm km/s}}\right)^{-2/5} \left(\frac{n}{0.2 {~ \rm cm^{-3}}} \right)^{-.37}\left(\frac{E}{10^{51}~{\rm erg}} \right)^{.32}  \label{eq:snRadius2}
\end{align}
The shock continues to expand until reaching a maximum size before merging with the ISM, at which point its speed equals the thermal sound speed, around $10 ~\rm{km/s}$.

Now, as seen from \eqref{eq:snEncounterRateGeneral}, the largest shock size sets the shock encounter rate. However, the largest shocks are unable to Fermi accelerate $X$. This is because the gas around the shock front must be fully ionized to maintain the strong turbulence necessary for efficient acceleration of $X$, as is shown below. This condition begins to fail early in the snowplow phase, when neutrals begin forming near the shock~\cite{2006MNRAS.371.1975Y}. Thus we take the shock at the end of the Sedov phase to be the largest shock capable of Fermi-accelerating $X$. This corresponds to a minimum shock speed slightly below $200 {~\rm {km/s}}$, a maximum radius \eqref{eq:snRadius1} of $40 ~{\rm pc}$, and hence an expected SN shock rate of
\begin{align}
\Gamma_{\rm{SH}}    &=  \left(2.5 \times 10^7 {~ \rm yr} \right)^{-1} \left(\frac{R_{\rm{max}}}{40 ~{\rm pc}} \right)^3 \left(\frac{R_{\rm{disk}}}{15 {~ \rm kpc}}\right)^{-2} \left(\frac{H_{\rm{disk}}}{300 {~ \rm pc}}\right)^{-1} \left(\frac{\Gamma_{\rm{SN}}}{.03 {~ \rm {yr}^{-1}}} \right).
\label{eq:snEncounterRate}
\end{align}

The strong turbulence is required so that the magnetic fields upstream and downstream of the shock are sufficiently tangled, making the mean free path of $X$ in the shock region as small as its gyroradius (Bohm diffusion), the minimum possible mean free path~\cite{2013APh....43...56B,2006MNRAS.371.1975Y}. The necessity for such a small mean free path near the shock can be understood by calculating the maximum possible rigidity, (the ratio of momentum to charge, $p/q$), a SN shock can impart to a CHAMP.
The maximum rigidity is set by spatial and temporal constraints. Spatially, the shock cannot accelerate a CHAMP anymore once its mean free path grows larger than the size of the shock region. For a SN of radius $R$ in the Sedov phase, hydrodynamic simulations show the thickness of the shock region is $\approx 0.05R$~\cite{2011piim.book.....D}. A SN of radius $R_{\rm{max}} = 40 ~{\rm pc}$, then cannot accelerate $X$ beyond 
\begin{align}
	\left(\frac{p}{q}\right)_{\rm{max}} \approx 3 \times 10^7 ~ \GeV \left(\frac{B}{15 ~{\rm \mu G}} \right) \left(\frac{R_{\rm{max}}}{40 ~{\rm pc}}\right)
	\label{eq:maxRigiditySpace}
\end{align}
where we have taken the shock magnetic field about three times the ambient ISM field due to shock compression~\cite{2009MNRAS.397.1410T}.
Temporally, the shock cannot accelerate a CHAMP for longer than the age of the remnant. The acceleration timecale to Fermi-accelerate a particle to rigidity $p/q$ and speed $v$ is approximately $t_{\rm acc} \approx 8 D_{\rm{s}}/{v_{\rm{s}}^2}$ where $D_{s} = \frac{1}{3}\lambda_{s}v$ the diffusion constant near the shock, and $\lambda_s = r_{\rm{gyro}}$ the mean free path \cite{2013APh....43...56B}. Equating the acceleration time with the age of the remnant, $\tau_{\rm{SN}} = (2/5)R/ v_s$ implies a SN of radius $R_{\rm{max}} = 40 ~{\rm pc}$ cannot accelerate $X$ beyond 
\begin{align}
	\left(\frac{p}{q}\right)_{\rm{max}} \approx \frac{5.5 \times 10^4 ~ \GeV}{\beta} \left(\frac{B}{15 ~{\rm \mu G}} \right) \left(\frac{R_{\rm{max}}}{40 ~{\rm pc}}\right)	\left(\frac{v_s}{200 ~{\rm km/s}}\right)
	\label{eq:maxRigidityTime}
\end{align}
The maximum rigidities of \eqref{eq:maxRigiditySpace} and \eqref{eq:maxRigidityTime} imply particles are unaffected by SN shocks in the limit $q \to 0$. This condition must be true since SN shocks transfer momentum to encountered particles solely through electromagnetic scatterings.

Note the factor of $\beta \equiv v/c$ in the denominator of \eqref{eq:maxRigidityTime} compared to \eqref{eq:maxRigiditySpace}. This is because first-order Fermi acceleration is more efficient at slower speeds since the momentum change upon reflection is greater for smaller $v$. For CHAMPs with $\beta> \beta_{\rm{esc}} \simeq 2\times 10^{-3}$, the temporal constraint \eqref{eq:maxRigidityTime} dominates.%
\footnote{Ordinary cosmic rays are believed to be injected when the shock is young and the magnetic field is nearly a milligauss, which gives a maximum rigidity near $3 \times 10^6 ~\GeV$, exactly where the proton "knee" is observed in the cosmic ray spectrum. Further evidence for the validity of \eqref{eq:maxRigidityTime} is the iron knee, which drops at a momentum 26 times higher. CHAMPs which encounter young SN remnants can be accelerated above the rigidity \eqref{eq:maxRigidityTime}, but we do not consider such a process.}

\begin{figure}[tb]
\centering
\includegraphics[width=0.7\textwidth]{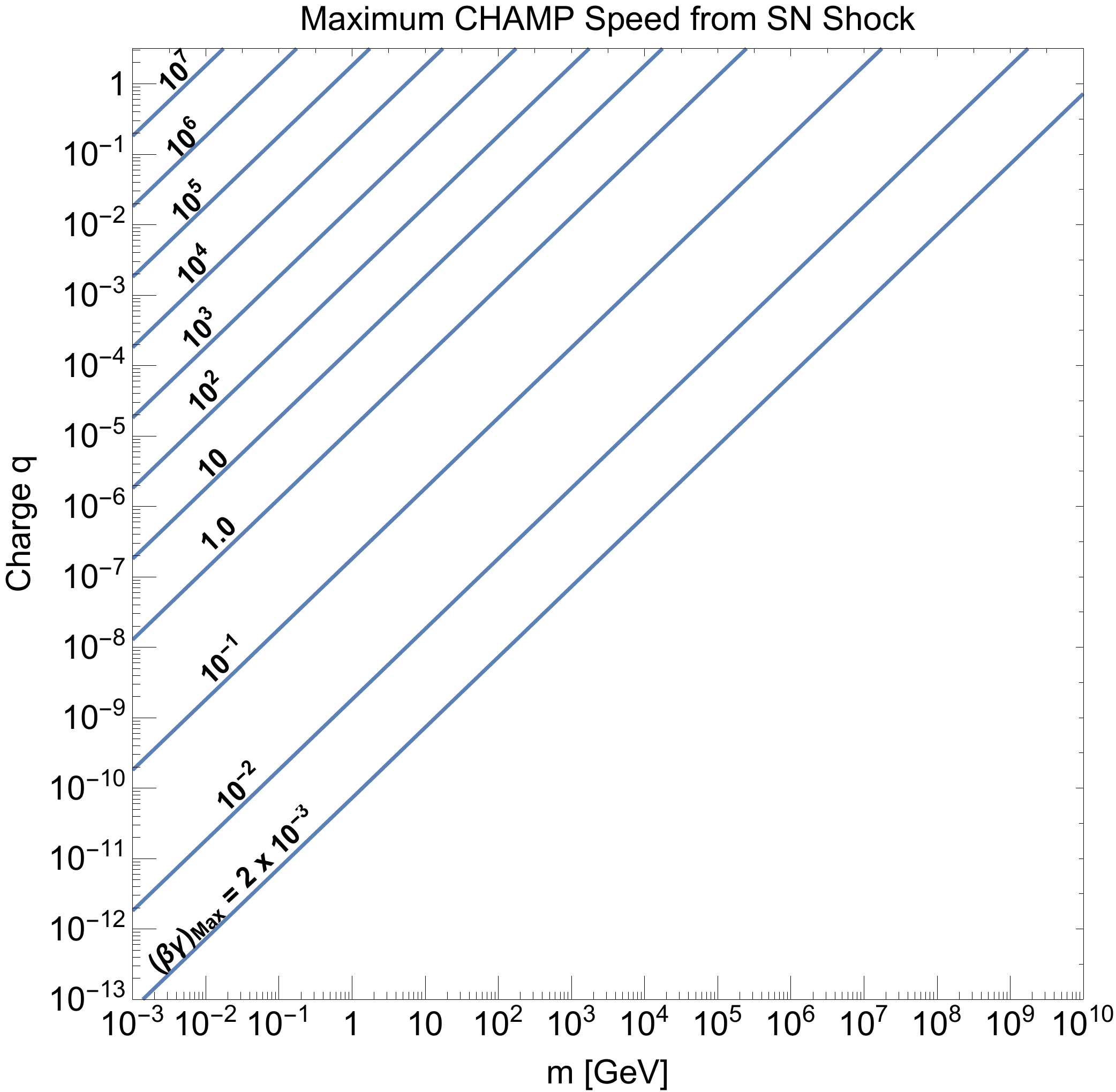} %
\caption{Contours of the maximum $\gamma \beta$ a CHAMP aquires from Fermi acceleration by a SN shock.}
\label{fig:maxRigidity}
\end{figure}

Contours showing the largest possible $\gamma \beta$ for a given CHAMP mass and charge is shown in Fig.~\ref{fig:maxRigidity}.
CHAMPs with $m/q \gtrsim 10^{10}$ GeV cannot be ejected from the Milky Way as $\beta_{\rm{max}} < \beta_{\rm{esc}} \simeq 2\times 10^{-3}$, and hence remain throughout the halo and disk with the virialized velocity distribution.

\subsection{Escape Rate from the Disk}

CHAMPs diffuse through the ISM by resonantly scattering off magnetic irregularities on the scale $k = 2\pi/r_{\rm{gyro}}$, where $r_{\rm{gyro}} = \gamma mv/qB$. This scattering leads to a mean free path $\lambda \propto R^a$, where $R \equiv r_{\rm{gyro}} B = \gamma mv/q$ is the magnetic rigidity, and $a$ is set by the magnetic field power spectrum~\cite{Longair:1981jc,Jones:2000qd}. The observed steady-state cosmic ray secondary to primary spallation ratios at various rigidities implies $a \sim 0.5$ and leads to a mean free path
\begin{equation}
    \lambda \, \simeq \, 10\; \mbox{pc} \left( \frac{v}{10^3 \, \mbox{km/s}} \right)^{1/2}   \left( \frac{m/q}{10^6 \, \mbox{GeV}} \right)^{1/2} \gamma^{1/2}
        \label{eq:diffusionmfp}
\end{equation}
for most cosmic ray propagation models~\cite{Strong:2007nh,Jones:2000qd} 
\footnote{The mean free path becomes rigidity independent at rigidities below $\sim \GeV$. However, the smallest CHAMP rigidities we consider, $m v_1 /q$ (see Eq. \ref{eq:x1}) are alway greater than a $\GeV$, except in parameter space already excluded by collider searches and $N_{\rm{eff}}$.}

For CHAMPs with speeds above the gravitational escape speed, $v_{\rm{esc}}$, the typical  rate to diffuse out of the disk is $\Gamma_{\rm{esc}} = 2D/H_{\rm{disk}}^2$, where the diffusion constant $D = \lambda v/3$.  As with cosmic rays, diffusion has the effect of increasing the time it takes for CHAMPs to escape the disk.  
The resulting escape rate from the disk is
\footnote{The cosmic ray lifetime in the entire galaxy, not just the disk, is determined from the relative abundance of cosmic ray radioactive isotopes to their children, and is about $10$ times longer than the lifetime in the disk \eqref{eq:escapeRate}~\cite{2005ppfa.book.....K,Longair:1981jc}. This is because diffusion continues above the disk into a $\sim 3 ~ \rm kpc$ high hot gas region, so called the confinement region. However, the lifetime in the stellar disk, where $\upex$ will encounter SN, is bounded by the grammage of matter traversed as observed from spallation products, and agrees well with \eqref{eq:escapeRate}~\cite{2005ppfa.book.....K}.}  
\begin{align}
    \Gamma_{\rm{esc}} \approx \left(2 \times 10^7 ~ {\rm yr} \right)^{-1}  \left(\frac{v}{10^3 ~{\rm km/s}}\right)^{3/2} \left(\frac{m/q^2}{10^6 ~{\rm GeV}} \right)^{1/2} \left(\frac{H_{\rm{disk}}}{300 {~ \rm pc}}\right)^{-2} q^{1/2} \gamma^{1/2}  \theta \left(v - v_{\rm{esc}}\right) \label{eq:escapeRate}
\end{align} 

This escape rate breaks down when $\lambda > H_{\rm{disk}}$ and hence is valid only for particles with rigidities $\gamma mv/q \lesssim 10^7 ~\rm{GeV}$. However, SN can only accelerate CHAMPs marginally beyond this rigidity anyway \eqref{eq:maxRigiditySpace}, so the ISM mean free path essentially always remains below $H_{\rm{disk}}$.

\section{Acceleration and Ejection from the Galaxy}
\label{sec:batch}

\subsection{The Accelerated Spectrum}

To understand the interplay between the thermalization rate in the ISM \eqref{eq:WIMThermalization}, the SN shock rate \eqref{eq:snEncounterRate}, and the escape rate from the disk \eqref{eq:escapeRate}, we define the parameter 
\begin{align}
    x \equiv \frac{v}{10^3 ~\rm{km/s}} \left(\frac{m/q^2}{10^6 ~{\rm GeV}}\right)^{1/3},
    \label{eq:x}
\end{align}
an independent-variable nearly mutual to each rate, and three values $\{x_1, ~ x_2, ~\overline{x}\}$ such that $\Gamma_{\rm{SH}}(x_1) = \Gamma_{\rm{therm}}(x_1)$, $\Gamma_{\rm{SH}}(x_2) = \Gamma_{\rm{esc}}(x_2)$, and $\Gamma_{\rm{therm}}(\overline{x}) = \Gamma_{\rm{esc}}(\overline{x}) \equiv \overline{\Gamma}$. These critical points are given by
\begin{align}
    x_1             &=  0.9 \times \left(\frac{H_{\rm{disk}}}{300 ~{\rm pc}}\right)^{1/3}  \left(\frac{R_{\rm{disk}}}{15 ~{\rm kpc}}\right)^{2/3}  \left(\frac{R_{\rm{max}}}{40 ~{\rm pc}}\right)^{-1} \left(\frac{\Gamma_{\rm{SN}}}{.03 ~{\rm yr^{-1}}}\right)^{-1/3} \label{eq:x1} \\
    \overline{x}    &=  0.9 \times \left(\frac{H_{\rm{disk}}}{300 ~{\rm pc}}\right)^{4/9} q^{-1/9} ~ \gamma(\overline{v})^{-1/9}  \label{eq:xBar} \\
    x_2             &=  0.9 \times \left(\frac{H_{\rm{disk}}}{300 ~{\rm pc}}\right)^{2/3} \left(\frac{R_{\rm{disk}}}{15 ~{\rm kpc}}\right)^{-4/3}  \left(\frac{R_{\rm{max}}}{40 ~{\rm pc}}\right)^{2} \left(\frac{\Gamma_{\rm{SN}}}{.03 ~{\rm yr^{-1}}}\right)^{2/3} q^{-1/3} ~ \gamma(v_2)^{-1/3} \label{eq:x2}
\end{align}
We have normalized $H_{\rm{disk}}$, $R_{\rm{disk}}$, $R_{\rm{max}}$ and $\Gamma_{\rm{SN}}$ to values for the Milky Way.  Accidentally this leads to comparable prefactors when $q=1$.  However, for $q<1$ the hierarchy of speeds for the Milky Way is $v_2 > \bar{v} > v_1$ for any value of $m/q^2$.

\begin{figure}[ht!]
\centering
\includegraphics[width=0.9\textwidth]{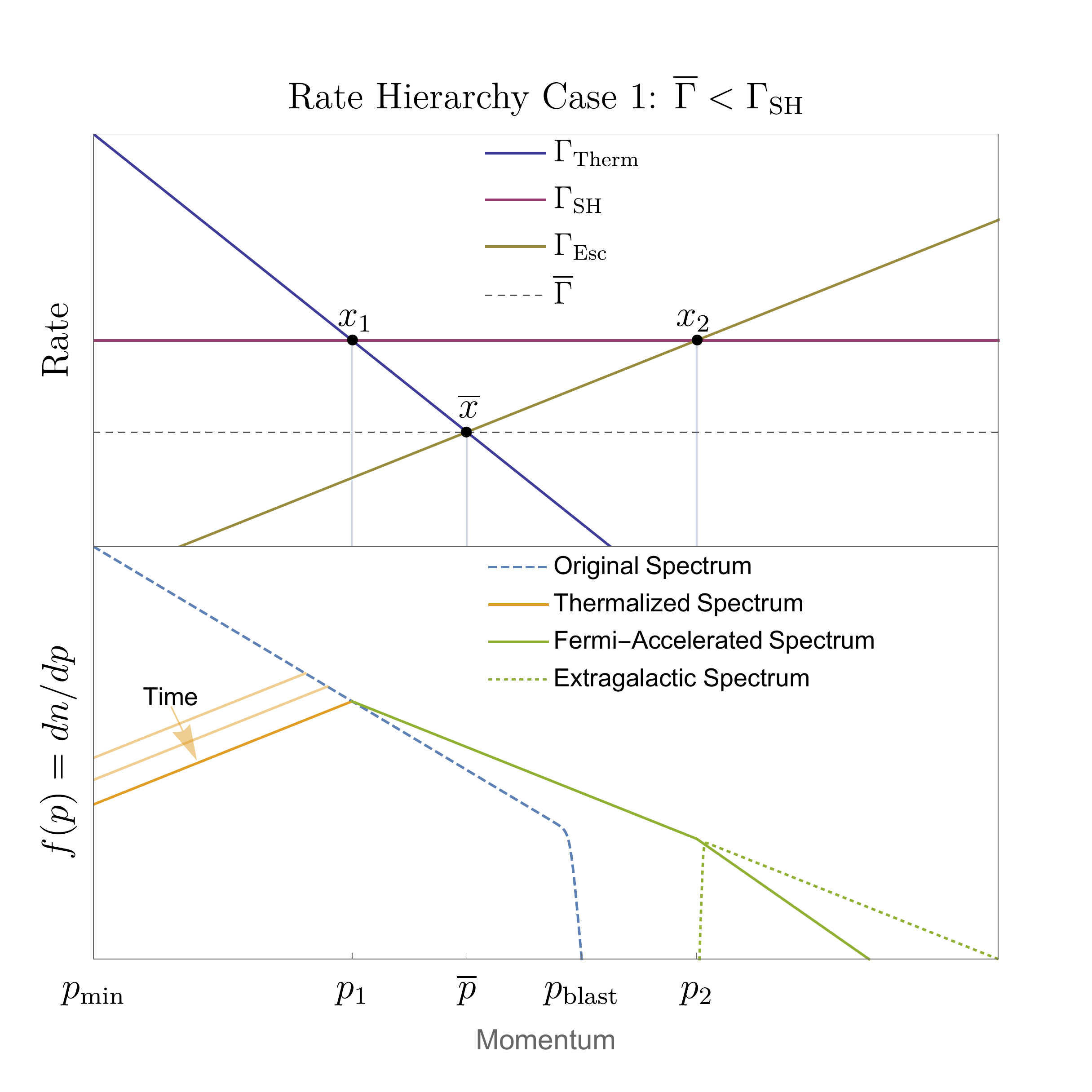} %
\caption{Comparison of the three key rates and the spectrum of accelerated CHAMPs for case 1.}
\label{fig:rateHierarchy1Plot}
\end{figure}

Galaxies with different disk and ISM properties will have different $v_{1,2}$ and $\bar{v}$; however, there are only two possible orderings of these speeds corresponding to the two cases $\overline{\Gamma} < \Gamma_{\rm{SH}}$ and $\overline{\Gamma} > \Gamma_{\rm{SH}}$, as shown in top panels of Figs. \ref{fig:rateHierarchy1Plot}, \ref{fig:rateHierarchy2Plot}. Equivalently, the two cases correspond to whether a CHAMP that surpasses the thermalization bottleneck is more likely to encounter another SN shock on the way out of the disk (and be Fermi-accelerated to relativistic speeds) or to escape the disk without meeting any further shocks (and remain non-relativistic upon escape). Our galaxy belongs to the first case for all $q \lesssim 1$ which we investigate in the following.

Boom. A SN goes off and its shock expands in the ISM. We first consider the case that $X$ are thermal with speeds much less than $v_1$. Since their speeds are also below the shock speed, when hit by a first shock they are accelerated only to the shock speed.
Since the probability of a shock encounter goes as the shock radius cubed, we see from \eqref{eq:snRadius1} and \eqref{eq:snRadius2} that a SN remnant produces a CHAMP number density spectrum $dn/d\ln v \propto v^{-2}$ in the Sedov-Taylor phase (i.e. $v \gtrsim 200 ~{\rm km/s}$) and $\propto v^{-6/5}$ in the snowplow phase. The latter thermalize so quickly that they are irrelevant to the following discussion. 

We will now discuss the spectrum of this batch in terms of the differential momentum spectrum $f = dn/dp$, since it is the momentum $p$ which is the fundamental quantity that describes the spectrum from non-relativistic to relativistic regimes. Because $\{x_1, ~ x_2,~ \overline{x}\}$ all occur at non-relativistic speeds, there is no loss in generality between $x_i$, and its associated momentum $p_i$.

Thus, the relevant differential spectrum $f(p) = dn/dp$ of this batch is initially proportional to $p^{-3}$, up to $p_{\rm blast} = m \times 10^{4} ~{\rm km/s}$,%
\footnote{For sufficiently large $m/q$, $X$ cannot be accelerated to $p_{\rm blast}$ for reasons discussed in Sec.~\ref{subsec:SNEncounter}. However, the galactic spectrum remains the same as long as $p_{\rm max} > p_1$, which we find always true. There also exists a momentum $p_{\rm break}$ such that $X$ with $p > p_{\rm break}$ diffusively catch up the same shock that initially accelerated $X$. When this occurs, $X$ is Fermi-accelerated and the spectrum becomes $p^{-2}$ above $p_{\rm break}$. This again does not change the galactic spectrum since we also confirm that $p_{\rm break} > p_1$.}
as shown by the dashed blue line in Fig \ref{fig:rateHierarchy1Plot}. However, as time progresses, energy losses from thermalization even alter this spectrum, chipping away at the slower moving CHAMPs which thermalize first. The evolution of this differential spectrum due to thermalization obeys $\partial f/\partial t = (1/2)\partial(p \Gamma_{\rm{therm}}(p) f)/ \partial p$~\cite{Longair:1981jc}, whose solution implies $f \propto p^2$ for $t \Gamma_{\rm{therm}}(p) \gtrsim 1$ and unchanged for $t \Gamma_{\rm{therm}}(p) \lesssim 1$. Therefore, when this batch encounters a second SN shock with speed $v_s \approx 200 {~\rm km/s}$ a time $t \approx 1/\Gamma_{\rm{SH}}$ later, its spectrum is peaked at $p = p_1$, dropping as $p^{2}$ for $p < p_1$ and $p^{-3}$ for $p > p_1$, as shown by the orange and dashed blue lines in Fig.~\ref{fig:rateHierarchy1Plot}.

The CHAMPs in the batch moving faster than the approaching shock can convectively and diffusively travel back and forth across the shock front. The expected momentum gain for each cycle as well as the probability of completing $n$ cycles can be calculated, which together yield the post-shock distribution~\cite{drury1983introduction}. The above physics is encoded in a transformation of the original spectrum, $f_{\rm{pre}}(p)$, to a final spectrum, $f_{\rm{post}}(p)$, by~\cite{bell1978acceleration2}
\begin{align}
    f_{\rm{post}}(p) &=   (\mu - 1)p^{-\mu}\int_{p_{\rm{min}}}^p dk ~ k^{\mu -1} f_{\rm{pre}}(k).
    \label{eq:fermiSpec}
\end{align}
Here, $p_{\rm{min}} \approx m \times (200 ~{\rm km/s}) \ll p_1$ and $\mu = 2$ is the theoretically predicted power dependence from Rankine-Hugoniot plasma boundary conditions. Performing the convolution \eqref{eq:fermiSpec} on the $t \approx 1/\Gamma_{\rm{SH}}$ spectrum, we find the effect of the second shock is to leave unchanged the $p^2$ spectrum below $p_1$ but to change the $p^{-3}$ spectrum above $p_1$ to a Fermi-accelerated $p^{-\mu} = p^{-2}$ spectrum, as shown by the green and orange lines in Fig.~\ref{fig:rateHierarchy1Plot}. Qualitatively, this is because the largest number of particles have initial momenta $p_1$. Note the resulting $p^{-2}$ spectrum now includes CHAMPs with relativistic speeds. 

Those CHAMPs with momenta now above $p_2$ will quickly leave the disk and contribute to the extragalactic spectrum, as shown by the dotted green line in Fig.~\ref{fig:rateHierarchy1Plot}. Meanwhile, CHAMPs with momenta between $p_1$ and $p_2$ are more likely to stay in the disk and encounter additional SN shocks before escaping and reaching the momentum $p_2$.%
\footnote{The escape probability for $\upex$ with $t < \Gamma_{\rm{esc}}^{-1}$ is exponentially suppressed. Only for $\upex$ with $t \approx \Gamma_{\rm{esc}}^{-1}$ is the escape proability non-negligible, with value $t \, \Gamma_{\rm{esc}}$. Consequently, the probability galactic CHAMPs with momenta less than $p_2$ escape before encountering repeated shocks is negligible.}
The evolution of the CHAMP spectrum by the repeated encounters is investigated in Appendix~\ref{sec:repeatedShocks}, and it is shown that CHAMPs with momenta below $p_2$ eventually escape from the disk with a time scale $\sim \Gamma_{\rm SH}^{-1}$.

The two-stage acceleration mechanism we consider becomes ineffective once $v_1$ is above $v_{\rm blast} \approx 10^{4} ~{\rm km/s}$, as almost all CHAMPs accelerated by the first shock are thermalized before encountering the next shock and hence cannot be `injected', or Fermi-accelerated, at the second shock. This is the case if $m/q^2 \lesssim 600$ GeV, which is excluded by direct searches for CHAMPs (see Sec.~\ref{sec:DD}). Note that nuclei belong to this parameter region and cannot be accelerated by the two-stage injection process described above. It is currently not understood how a very small fraction of thermal nuclei \textit{are} directly injected from a single, young, shock (the so-called `injection-problem'~\cite{2005ppfa.book.....K,2011piim.book.....D}) to become cosmic rays. It is likely too, that a very small fraction of thermal CHAMPs are also directly injected by a single shock, though large uncertainties exist since the process is not understood for even ordinary cosmic rays. Nevertheless, the authors of~\cite{2017PhLB..768...18H} assume that CHAMPs are Fermi-accelerated in the same manner as nuclei, and obtain the spectrum of CHAMP cosmic rays from that of protons with the same rigidity. Since the efficiency of direct-injection is much less than two-stage injection, the resultant CHAMP cosmic ray abundance is much smaller than ours.

In the above discussion we assumed that $X$ are thermalized and have speeds below $v_1$ before encountering a SN remnant. However, if $m/q^2 > 3\times 10^6$ GeV, the thermalization does not occur and $X$ have velocities of $v_{\rm{vir}}$, which is larger than the shock speed. On encountering the first SN shock, $X$ undergo Fermi acceleration.  Hence, whether or not there is initial thermalization, the accelerated spectrum always has the form  $f(p) \propto p^{-2}$, cutoff at low speeds at 
\begin{align}
v_0 \equiv \begin{cases}
v_1 & m/q^2 < 3\times 10^6~{\rm GeV} \\
v_{\rm{vir}} & m/q^2 > 3\times 10^6~{\rm GeV}.
\end{cases}
\end{align}

\subsection{Efficiency of Expulsion}

\begin{figure}[tb]
\centering
\includegraphics[width=0.7\textwidth]{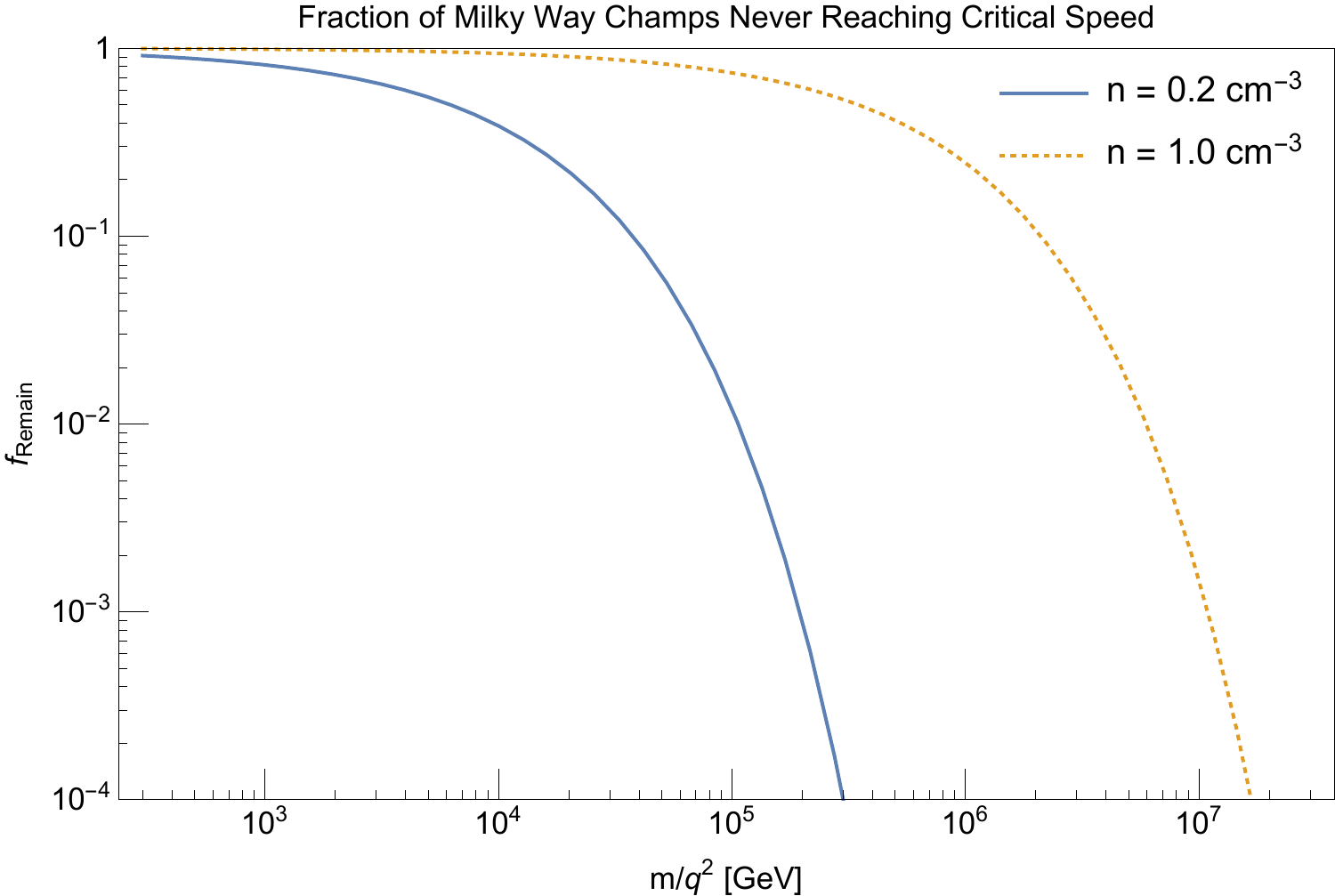} %
\caption{The fraction of CHAMPs which do not encounter a shock with velocity above the critical velocity and hence remain in the disk.}
\label{fig:fracRemainingPlot}
\end{figure}

For now we ignore the diffusion of CHAMPs from outside the disk and compute the fraction of CHAMPs that escape the disk. The fraction is given by the probability to encounter a critical shock to overcome the thermalization bottleneck within the disk. As discussed in Sec~\ref{subsec:SNEncounter}, for $m/q^2 > 3\times 10^6$ GeV it is enough to encounter a shock at the end of the Sedov-Taylor phase because of the inefficient thermalization. For $m/q^2 < 3\times 10^6$ GeV, encounter with a shock with a velocity $v_c > v_1$ is required. Shock speeds capable of reaching $v_c$ occur during the early Sedov-Taylor phase, where the shock radius-velocity relation~\eqref{eq:snRadius1} implies the expected encounter rate for a critical shock~\eqref{eq:snEncounterRateGeneral} is
\begin{align}
    \Gamma_{\rm SH,c} &= \begin{cases}
     \left(10^{8} ~{\rm yr}\right)^{-1} \left(\frac{m/q^2}{3\times 10^6 ~{\rm GeV}}\right)^{2/3} 
     & m/q^2 < 3\times 10^6~{\rm GeV} \\
     \Gamma_{\rm SH} & m/q^2 > 3\times 10^6~{\rm GeV}
    \end{cases}
    \label{eq:critRate}
\end{align}
Since the chance of encountering a critical shock is rare, we expect the number of critical shocks encountered to be a Poisson random variable with an expected rate given by \eqref{eq:critRate}. Consequently, the fraction of CHAMPs that never encounter a critical shock and thus remain in the disk after a time $T$ is
\begin{align}
    f_{\rm{rem}} =  \exp{\left(-\int_{0}^T \Gamma_{\rm SH,c}(t) ~ dt \right)}  
    \label{eq:frem}
\end{align}
Assuming $\Gamma_{\rm SH,c}$ is independent of time, $T \sim 10^{10} ~ {\rm yr}$, and the shock expands into a homogenous medium of density $0.2 ~ {\rm cm^{-3}}$, the fraction of CHAMPs that remain in the disk as a function of $m/q^2$ is shown in blue in Fig.~\ref{fig:fracRemainingPlot}. Note that while an order one fraction of the CHAMP population in the disk may be ejected after one folding-time $\Gamma_{\rm SH,c}^{-1}$, efficient removal from the disk requires many folding-times.

\section{Diffusion into the Disk and the Local CHAMP Flux}
\label{sec:diffusion}
Although ejection from the disk is efficient for $m/q^2 \gtrsim 10^4$ GeV, this does not imply the absence of CHAMPs in the disk.  There is a continuous replenishing of CHAMPs in the disk by diffusive accretion from the halo and the confinement region.  The balance between accretion and ejection leads to a (quasi-) steady state.  Hence, even after $10^{10}$ years, there are CHAMPs that have been recently accelerated by SNe and hit the Earth before escaping from the disk. In this section we estimate the present flux of such accelerated CHAMPs.

\subsection{CHAMPs that do not Collapse into the Disk}
We first consider $m/q^2 > 10^5$ GeV, where CHAMPs do not collapse into the disk when it forms. Most galactic CHAMPs are outside the disk, so that their diffusion into the disk plays an important role.  The disk of the Milky Way, which we take to have a width of $H_d = 300$ pc, is surrounded by the confinement region, which we take to have a width of $H_c = 6$ kpc~\cite{Longair:1981jc,Strong:2007nh}.  This region has a random magnetic field similar to that of the disk, so we take the CHAMPs to diffuse in this region with the same mean free path as in the disk.  Diffusion through the confinement region plays a key role in determining the accelerated CHAMP flux hitting the Earth today.

We solve the following equations for $n(t,z)$, the $X$ number density inside the confinement region with the virial speed, and for $n_A(t)$, the number density of accelerated $X$s in the disk
\begin{align}
\label{eq:diff eq}
\frac{\partial n (t,z)}{\partial t} \;=\; &  D \, \frac{\partial^2 n (t,z)}{\partial z^2} - \Gamma_A \; \theta(z+ \frac{H_{\rm disk}}{2}) \; \theta(\frac{H_{\rm disk}}{2}-z) \, n (t,z),  \\
n(0,z) \;=\; & n(t, \pm H_{c}/2) = n_0, \nonumber\\
\Gamma_A \; = \; & \Gamma_{\rm SH,c} \\
\frac{d n_A(t)}{d t} \;=\; &\Gamma_A \, n(t,0) - \Gamma_{\rm SH} \, n_A(t),~~~
n_A(0)=0,
\label{eq:diff eqA}
\end{align}
where $n_0\simeq 0.3 \,f_X/m$ GeV/cm$^3$ is the initial local $X$ number density, and the diffusion constant $D= \lambda(v_{\rm{vir}}) v_{\rm{vir}}/3$, with $\lambda$ given in (\ref{eq:diffusionmfp}).

We take the escape rate of the accelerated CHAMPs to be the SN shock rate $\Gamma_{\rm SH}$, as the number density of accelerated Xs is dominated by ones with low momenta, $p < p_2$, and these typically escape by encountering SNe and are rapidly accelerated to momentum $p_2$, where the escape rate is equal to the shock rate. Moreover, as shown in Appendix \ref{sec:repeatedShocks}, CHAMPs with momentum $p_0 < p < p_2$ are repeatedly shocked and quickly evacuate the disk in a time $\sim \Gamma_{\rm SH}$ as well.

\begin{figure}[tb]
\centering
\includegraphics[width=0.7\textwidth]{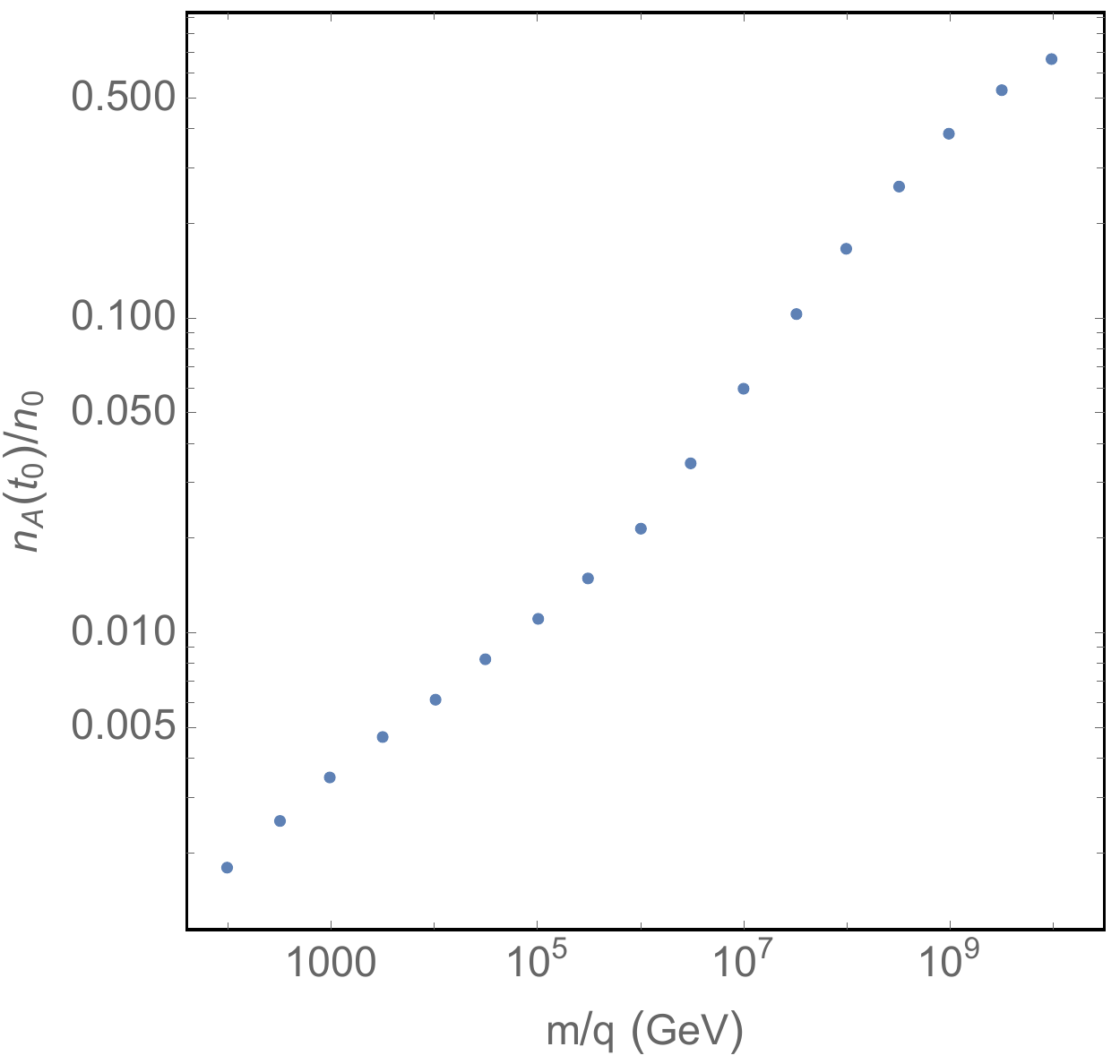} %
\caption{The accelerated $X$ number density, normalized to the original local number density.}
\label{fig:nlocal}
\end{figure}

Numerical results for the number density of the accelerated CHAMPs, $n_A(t_0)$, are shown in Fig.~\ref{fig:nlocal} as a function of $m/q$.  These results, including the slopes and the kink at $m/q = 10^6$ GeV, can be understood from a simple analytic estimate.  The acceleration of $X$ in the disk creates a gradient $dn/dz$ in the confinement region that drives a diffusion current of $X$ into the disk, from above and below
\begin{align}
J \; =\;  n_0 \, 
\begin{cases}
\left( \frac{d(t_0)}{2 t_0} \right) &  \hspace{0.5in} m/q < 10^6~{\rm GeV} \\
  \frac{D}{H_c/2} & \hspace{0.5in} m/q > 10^6~{\rm GeV} 
\label{eq:J}
\end{cases}
\end{align}
where $\lambda$ is the mean free path and $d(t_0) \sim \sqrt{t_0 \lambda v_{\rm{vir}}}$ the diffusion distance in time $t_0$. For $m/q > 10^6$ GeV, $d(t_0) > H_c$ and we reach a steady state where $X$ from the halo diffuse through the confinement region to reach the disk. Even for the largest $m/q$ that lead to shock acceleration, this does not substantially alter the density of $X$ in the halo. For $m/q < 10^6$ GeV, $d(t_0) < H_c$ so that $J$ is time dependent;  $J(t_0)$ arises from transporting $X$ from the confinement zone a distance $d(t_0)$ from the disk. Finally, note $J$ must always be less than $n_0 v_{\rm{vir}}$ which it is, since the mean free path $\lambda < H_{\rm{disk}} < H_{c}$.

In the disk, Eq.~(\ref{eq:diff eq})-(\ref{eq:diff eqA}) then reduce to $\dot{n} \simeq 2J/H_d - \Gamma_A\,n$ and $\dot{n}_A \simeq \Gamma_A n-\Gamma_{\rm SH}n_A$ leading to the (quasi-) steady state solutions $n_A =2J/H_d \Gamma_{\rm SH}$, $n =2J/H_d \Gamma_A$.  Inserting $J$ from (\ref{eq:J}) gives
\begin{align}
\label{eq:nA_nocollapse}
\frac{n_A}{n_0} \; \simeq \; 0.02
\begin{cases}
\left(\frac{m/q}{\rm10^6 \,GeV}\right)^{1/4} &  \hspace{0.5in} m/q < 10^6~{\rm GeV} \\
\left(\frac{m/q}{\rm10^6 \,GeV}\right)^{1/2} & \hspace{0.5in} m/q > 10^6~{\rm GeV}
\end{cases} \\
\frac{n}{n_0} \simeq \frac{n_A}{ n_0} \times 
\begin{cases}
1 & m/q^2 > 3\times 10^6~{\rm GeV}\\
6  \left( \frac{m/q^2}{3\times 10^6~{\rm GeV}} \right)^{-2/3} &  m/q^2 <  3\times 10^6~{\rm GeV}
\end{cases}
\end{align}
where the differing powers of $m/q$ in \eqref{eq:nA_nocollapse} result from the different powers of $\lambda$ in $J$ for the two cases. The continual accretion of CHAMPs onto the disk, followed by their acceleration and expulsion, has led, remarkably, to a large accelerated cosmic ray flux of CHAMPs today on Earth.

We take the CHAMP velocity to be the virial velocity to estimate the diffusion constant $D$. Since $X$ are efficiently ejected from the disk, refilling by diffusion from outside the disk determines the present number density of CHAMPs. The diffusion of unaccelerated CHAMPs inside the disk is not important. Outside the disk, thermalization is ineffective on cosmological time scales for $m/q^2 > 2 \times 10^{6}$ GeV and we may safely take the virial speed for the above estimation. For $m/q^2 < 2 \times 10^{6}$ GeV, the thermalization occurs and their velocity decreases down to the thermal velocity. If the velocity is below the Alfven velocity $\sim 50$ km/s, the scattering by the turbulent magnetic field accelerates CHAMPs up to the Alfven velocity with the rate as large as the gyrofrequency~\cite{1997PhPl....4..856W}, and hence the CHAMP velocity is at the smallest the Alfven velocity. The diffusion constant for the Alfven velocity is about 8 times smaller than that for the virial velocity. For $m/q^2 < 2 \times 10^{6}$ GeV, $m/q\lesssim 10^6$ GeV and $n_A$ is proportional to $D^{1/2}$. The accelerated number density decreases at the most by a factor of three because of the thermalization. We neglect the small suppression.

Finally, it is worthwhile to mention that the steady-state accelerated spectrum is fairly insensitive to whether or not $X$ collapses. This is because $n(t, \pm H_c/2) \sim n_0$ either way, since the baryon disk formation efficiency is only $\sim 25 \%$. Moreover, since acceleration out of the disk is efficient for CHAMPs that do not collapse, $m/q^2 \gtrsim 10^5 ~\GeV$, the same steady-state spectrum is quickly reached regardless of the initial disk density. Similarly, since acceleration out of the disk and diffusion into the disk are less efficient for CHAMPs that do collapse, $m/q^2 \lesssim 10^5 ~\GeV$, the  same steady-state spectrum is reached regardless of the density at the confinement-halo interface. We find that, even if we assume the collapse of CHAMPs (see below), $n_A$ is enhanced by a factor of few for $m/q^2$ just above $10^5 ~\GeV$.


\subsection{CHAMPs that do Collapse into the Disk}

For $m/q^2 < 10^5$ GeV, CHAMPs collapse into the disk. We solve Eqs.~(\ref{eq:diff eq}, \ref{eq:diff eqA}) with the initial and boundary conditions
\begin{align}
n(0,z) = 100 \, n_0 \; \theta(z+ H_{\rm disk}/2) \; \theta(H_{\rm disk}/2-z) + 0.1n_0,~~~
n(t, \pm H_{c}/2) \approx 0.1 n_0,
\end{align}
where $100n_0$ is the initial concentration from collapse at disk formation and $0.1n_0$ the concentration that remain in the halo near the confinement interface \cite{Gaensler:2008ec}. 
For $m/q^2 < 10^5$ GeV, even CHAMPs outside the disk but inside the confinement region are thermalized. We take the velocity of the unaccelerated CHAMPs to be the maximal of the thermal velocity and the Alfven velocity. We find that $n$ and $n_A$ in the present universe are approximated by the following semi-empirical formulae,
\begin{align}
\label{eq:nA_collapse}
\frac{n_A}{n_0} = & 100 \times  {\rm exp}\left( - \Gamma_A t_0 \times\frac{H_d}{H_d + 2 \sqrt{D t_0}} \right) \times \frac{\Gamma_{A}}{\Gamma_{\rm SH}}, \\
\frac{n}{n_0} = & 100 \times  {\rm exp}\left( - \Gamma_A t_0 \times\frac{H_d}{H_d + 2 \sqrt{D t_0}} \right).
\end{align}
This result can be understood as follows. The large charge and the low velocity implies that diffusion is ineffective, so that the number density $n$ inside the disk is basically given by $100 \, n_0 {\rm exp}(-\Gamma_A t_0)$, which is corrected by the second factor in the exponent taking into account the suppression of the ejection by small diffusion out from the disk. The number density of accelerated CHAMPs is then determined by the quasi-steady state solution with $dn_A/dt = 0$.

\subsection{The Local CHAMP Flux and Spectrum}
The accelerated CHAMPs initially have the spectrum $dn_A/dp \propto 1/p^2$. CHAMPs with momentum above $p_2$ have an escape rate larger than $\Gamma_{\rm SH}$ by a factor of $(p/p_2)^{3/2}$ for non-relativistic $p$ and $(m/p_2)^{3/2}(p/m)^{1/2}$ for relativistic $p$. Taking account the larger escape rate, the flux of the accelerated CHAMPs is given by
\begin{align}
\label{eq:final spectrum}
\frac{dn_A}{dp} \, v \; =\;  n_0 \, v_0 \times &
\begin{cases}
\frac{n_A}{n_0}~\mbox{of Eq.}~(\ref{eq:nA_nocollapse})& m/q^2 > 10^5~{\rm GeV} \\
\frac{n_A}{n_0}~\mbox{of Eq.}~(\ref{eq:nA_collapse}) &m/q^2 < 10^5~{\rm GeV}
\end{cases}\nonumber \\
\times & 
\begin{cases}
\frac{1}{p} &  m v_0 < p  < {\rm max}(p_2, m v_{\rm{vir}}) \\
 \frac{{\rm max}(p_2, m v_{\rm{vir}})^{3/2}}{p^{5/2}} &  p > {\rm max}(p_2, m v_{\rm{vir}}).
\end{cases}
\end{align}

\begin{figure}[tb]
\centering
\includegraphics[width=0.7\textwidth]{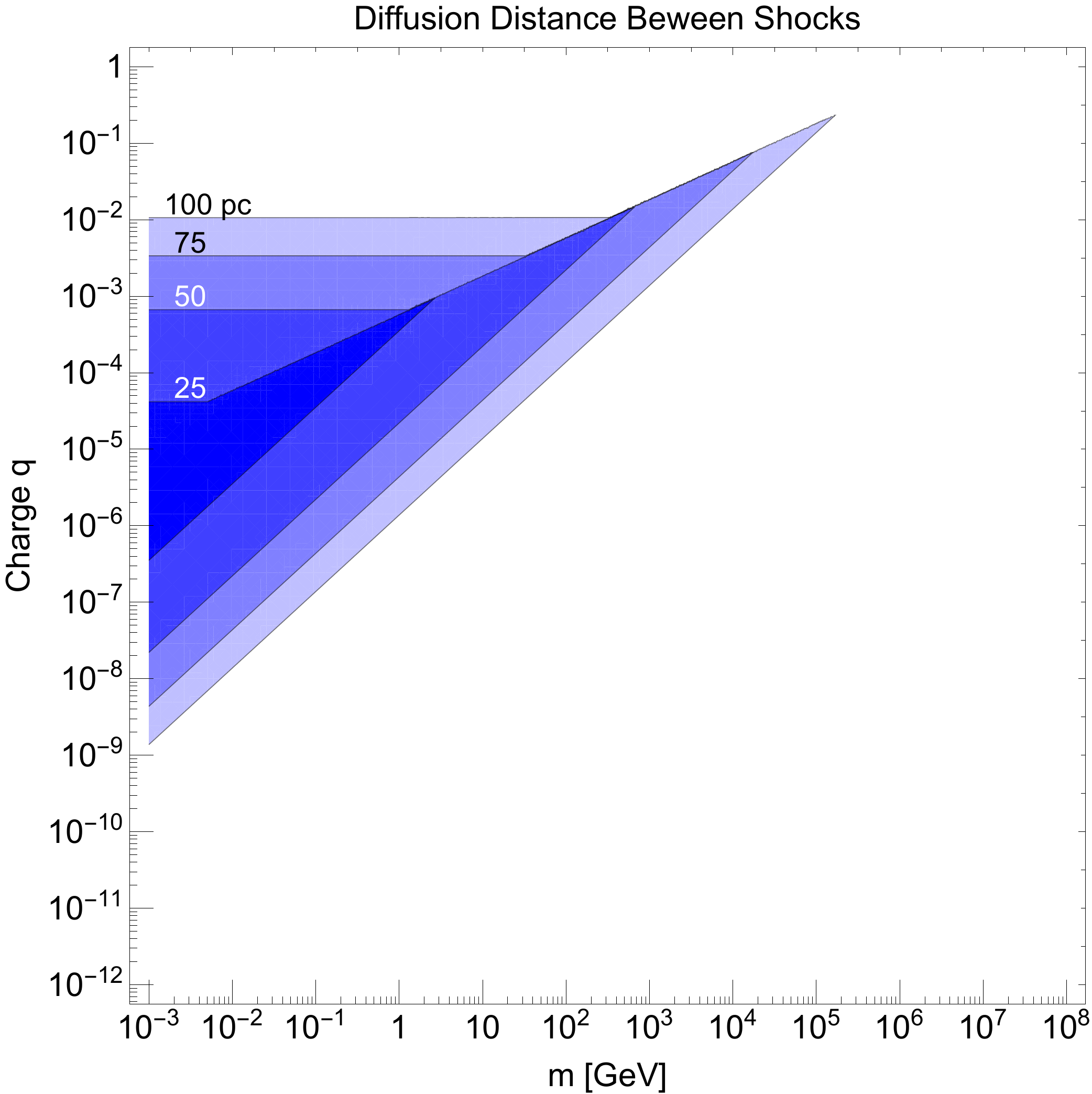} %
\caption{The diffusion distance an $X$ with a momentum $p_0$ travels in a time $\Gamma_{\rm SH}^{-1}$.}
\label{fig:difDist}
\end{figure}

We note that the spectrum in Eq.~(\ref{eq:final spectrum}) may be further modified during the diffusion between the acceleration site and the Earth.
In Fig.~\ref{fig:difDist}, we show the typical distance $X$ with a velocity $v_0$ can travel before encountering another shock, $\sqrt{2D/\Gamma_{\rm SH}}$. In the shaded region, the distance is smaller than the typical distance between the acceleration site and the Earth, $\sim 100$ pc, and $X$ is likely to encounter multiple shocks before hitting the Earth. We define the momentum of $X$ above which the encounter typically does not occur as $\tilde{p}_0$, which is at the most as large as $p_2$. Then the spectrum is the one with $m v_0$ in Eq.~(\ref{eq:final spectrum}) replaced by $\tilde{p}_0$, with a subdominant spectrum in $p \lesssim \tilde{p_0}$. This does not weaken the constraints derived in the next section, since the stopping by the Earth crust and/or the energy threshold of the searches require the momentum of detectable $X$ to be above $p_2$ in the parameter region with inefficient diffusion. Rather, the signal rates may be enhanced by a factor of $\tilde{p}_0/(mv_0)$ (which is at the most $10$). Since we are not able to determine $\tilde{p}_0$ in a reliable manner, we do not consider this possible enhancement in this paper. One should take care of this issue if momenta $p < \tilde{p}_0$ are important for $X$ searches.

CHAMPs that escape from other galaxies and reach our galaxy also contribute to the accelerated CHAMP spectrum. The spectrum of the extragalactic CHAMP background is estimated in Appendix~\ref{sec:diffuseEGBackground} and is found to be subdominant.

\section{Direct Detection of Accelerated CHAMP Cosmic Rays}
\label{sec:DD}
In this section we discuss direct detections of CHAMPs on the Earth. We first investigate two possible barriers for the detection: the solar wind and stopping in the Earth before reaching detectors. Then we compute signal rates in experiments sensitive to nuclear recoil, ionization and Cherenkov radiation.  We assume that $X$ couples to nucleons, electrons and photons dominantly through the charge $q$. If $X$ feels the strong interaction, as is the case with heavy colored states that bind with the known quarks, the constraint is altered.

\subsection{The Solar Wind and Stopping by the Earth}
\label{sec:obstacles}

The direct detection of CHAMPs on Earth can dramatically be affected by the solar wind, an outflow of charged particles and associated magnetic fields from the sun which suppress the flux of interstellar charged particles that reach the earth.

The net flux of CHAMPs a distance $r$ from the Sun is given by a convection-diffusion equation~\cite{Longair:1981jc}
\begin{align}
    J(r) = n(r)v_w(r) - D \frac{\partial n(r)}{\partial r} 
    \label{eq:netFlux}
\end{align}
where $n$ is the number density of CHAMPs, $v_w$ the solar wind speed, and $D = \frac{1}{3}\lambda(R)v$ the rigidity-dependent diffusion constant of charged particles in the interplanetary magnetic field. The net flux \eqref{eq:netFlux} is zero in the steady-state regime and leads to the solution
\begin{align}
    n(r)    &=  n_0(r_0) \exp{\left(-\int_r^{r_0}\frac{v_w(r) dr}{\frac{1}{3}\lambda(R) v}\right)}
    \label{eq:lowEFlux}
\end{align}
Observations of the low-energy cosmic ray flux on Earth find \eqref{eq:lowEFlux} to be well fit by~\cite{ucb.b1836189520110101,Meyer:1969we}
\begin{align}
    n_E = n_\infty \exp \left(-\frac{\eta(t)}{\beta g(R)}\right)
\end{align}
where $\eta(t)$ parameterizes the modulation of the solar wind and the interplanetary magnetic field, $g(R)$ the rigidity dependence of the particle mean free path, and $n_E$ ($n_\infty$) the number density of CHAMPs on Earth (far away in the ISM).

\begin{figure}[tb]
\centering
\includegraphics[width=0.7\textwidth]{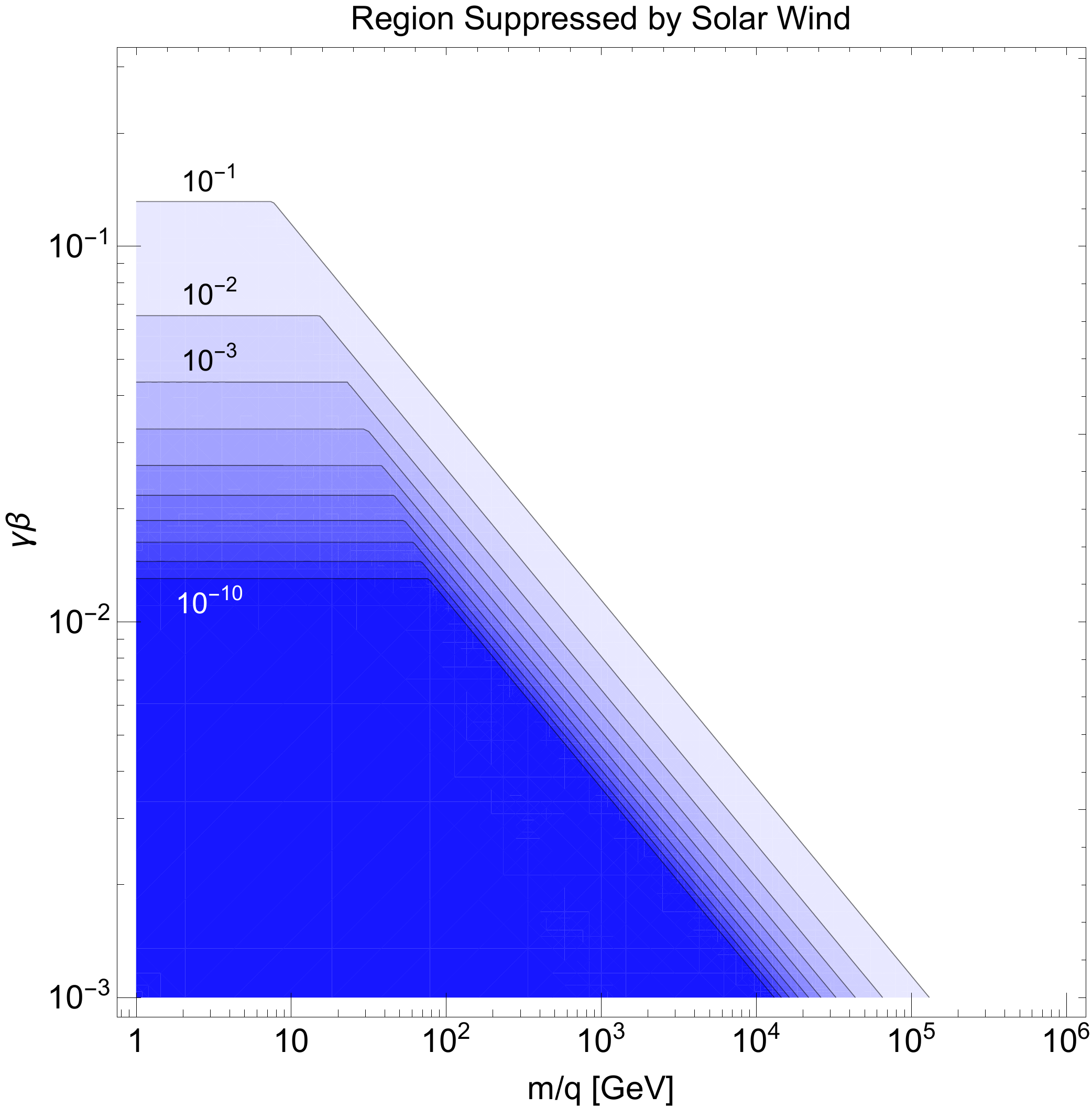} %
\caption{Contours of the fractional flux of $X$ that penetrate the solar wind for a given $m/q$ and $\gamma \beta$.}
\label{fig:solarSuppressionPlot}
\end{figure}

During the $11$-year solar cycle minimum, when the solar wind suppression on the cosmic ray flux is weakest, measurements indicate $\eta(t) \approx 0.3 ~{\rm GeV}$, and~\cite{ucb.b1836189520110101,Meyer:1969we}\footnote{The change in the rigidity dependence of the mean free path can be explained by a change in the interplanetary magnetic field power spectrum. Measurements from the Mariner 4 spacecraft indicate the power spectrum changes its power dependence at wavenumbers near $k_c \approx 2\pi \times  6 \times 10^{-12} ~ {\rm cm ^{-1}}$~\cite{1971RvGSP...9...27J}, corresponding to a scattering gyroradius of $r_{gyro,c} = 2 \pi/k_c \approx  10^{11} ~\rm{cm}$~\cite{ucb.b1836189520110101}. Since the solar wind magnetic field is around $50 ~\rm{\mu G}$, the critical rigidity occurs at $R_c \approx 1.5 ~\rm{GeV}$, in excellent agreement with \eqref{eq:solarMFP}.}
\begin{equation}
g(R) = 
\left\{
\begin{aligned}
&   R    && \quad      \text{ for $R > R_c \approx 1 ~\rm{GeV}$} \\
&   R_c  && \quad     \text{ for $R < R_c$}. \label{eq:solarMFP}
\end{aligned}
\right.
\end{equation}
CHAMPs with $\eta/\beta g > 1$ scatter frequently enough with the magnetic fields carried by the solar wind that they cannot travel `upstream' from the outskirts of the heliosphere to the Earth. The parameter space where the solar wind suppression is significant is shown by the shaded region of Fig.~\ref{fig:solarSuppressionPlot}.

CHAMPs that penetrate the solar wind must also penetrate the Earth's atmosphere/ crust to the depth of the detector. 
CHAMPs with $\beta \gtrsim 0.01$ passing through matter slow down chiefly from electron ionization. The stopping power is well described by the Bethe equation for $\beta > 0.1$,
\begin{align}
	-\left\langle\frac{dE}{dx}\right\rangle	= 0.15 ~{\rm MeV ~ cm^2/g} \left(\frac{q}{\beta}\right)^2 \left(\frac{Z/A}{1/2}\right) \ln \left(\frac{2 m_e \gamma^2 \beta^2}{10Z {~\rm eV}}\right).
\end{align}
For CHAMPs slower than the Fermi-velocity $(\beta \lesssim \alpha = 1/137$), energy losses from collisions with electrons are proportional to the CHAMP velocity~\cite{ZIEGLER19883}. Unlike ions, which are partially ionized in this velocity regime and must be assigned an effective nuclear charge as described by the Lindhard-Scharff equation, the effective CHAMP charge remains $q$ and hence its stopping power through a material is just $q^2$ times the proton stopping power~\cite{ZIEGLER19883}, which is given in the NIST Database~\cite{Berger:399381}.  We use the tabulated stopping power for $\beta < 0.1$ and the Bethe equation for $\beta > 0.1$.

\begin{figure}[tb]
\begin{center}
\includegraphics[width=0.7\textwidth]{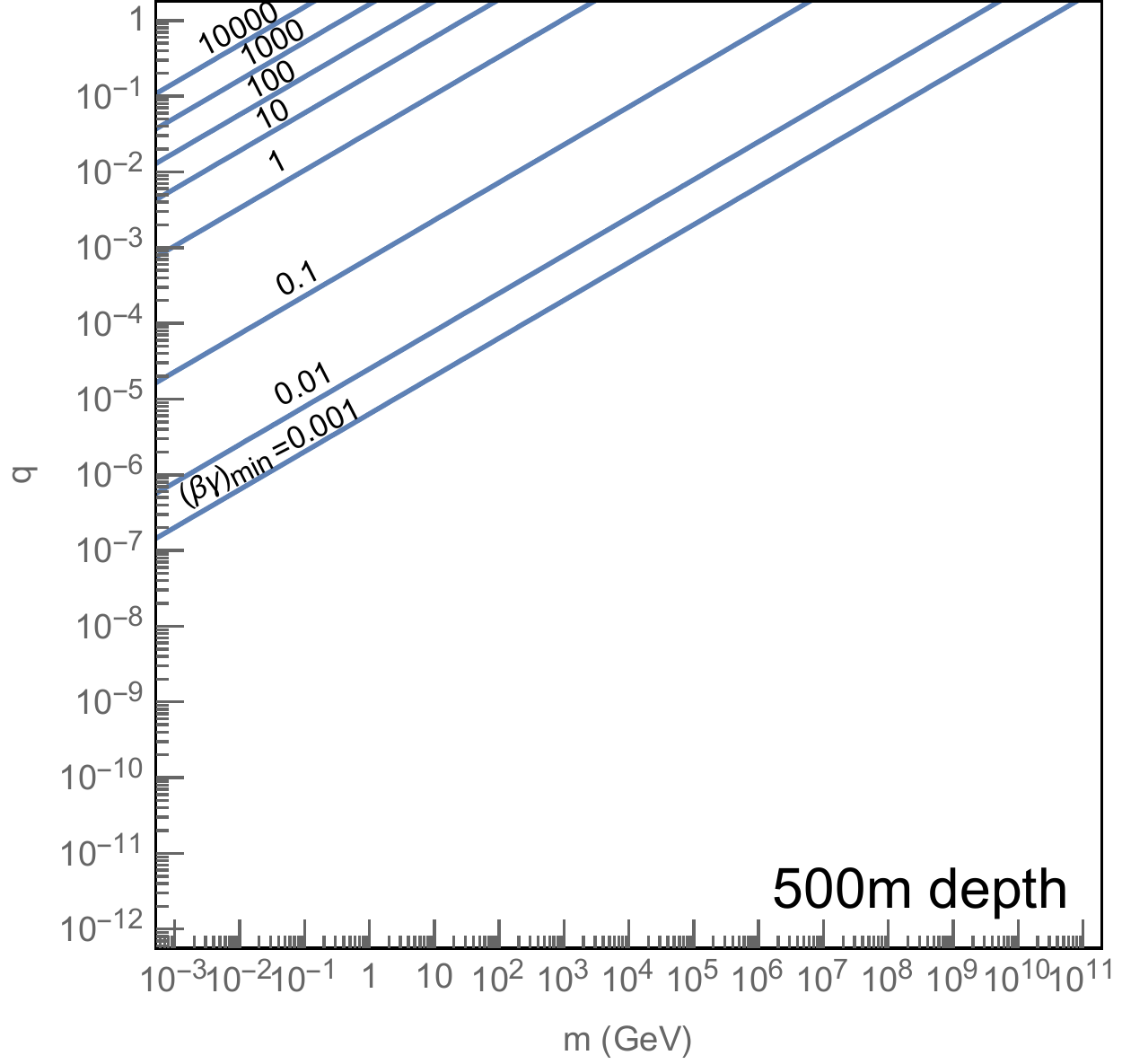}
\caption{Contours of the minimum $\beta \gamma$ for CHAMPs to traverse $500 \rm{m}$ of Earth crust. Contours for penetrating the Earth's atmosphere may be obtained by shifting these contours up above by a factor of 10 in charge $q$.  The solar wind constraint is subdominant, and is thus not displayed. }
\label{fig:minimumSpeed}
\end{center}
\end{figure}

Contours in the $(m,q)$ plane of the minimum $(\beta \gamma)_{\rm{min}}$ to reach underground detectors 500 m below the Earth's surface are shown in Fig.~\ref{fig:minimumSpeed}. 
Contours for penetrating the Earth's atmosphere may be obtained by shifting these contours up above by a factor of 10 in charge $q$. At high $\beta \gamma$ radiative losses dominate over electron ionization, but we find radiative losses are not important for values of $(m,q)$ that are allowed by direct searches and astrophysics.

\subsection{Nuclear Recoil at Deep Underground Detectors}

The scattering cross section between $X$ and a nucleus of mass $m_N$ and charge $Z$ is
\begin{align}
\frac{d \sigma}{ d\Omega} = \frac{\alpha^2 Z^2 q^2}{\mu^2 v^4 (1- \cos\theta)^2}|F(Q)|^2,
\end{align}
where $\mu$ is the reduced mass, $v$ is the speed of $X$, $Q$ is the momentum transfer and $F(Q)$ is the nuclear form factor. The recoil energy of the nucleus is
\begin{align}
E_{R} = \frac{\mu^2}{m_N} v^2 (1- \cos\theta)
\end{align}
and the minimum speed to obtain such a recoil energy is
\begin{align}
v^2_R = \frac{E_{R} m_N}{2\mu^2}.
\end{align}
The integrated cross section above a threshold $E_{R,\rm{th}}$ for fixed speed $v$ is
\begin{align}
\sigma(E_R> E_{R,\rm{th}}) = \frac{2\pi \alpha^2 Z^2 q^2}{m_N E_{R,\rm{th}} v^2} f(E_{R,\rm{th}})\, \Theta (v- v_{R,\rm{th}}),
\end{align}
where $v_{R,\rm{th}}$ is $v_R$ evaluated at the threshold recoil energy and $f(E_{R, \rm th})$ takes into account the suppression of the scattering by the form factor
\begin{align}
f(E_{R,\rm th}) = \left[ \int_{Q_{R,\rm th}}^{Q_{R,\rm max}} d Q |F(Q)|^2 Q^{-3} \right] / \left[  \int_{Q_{R,\rm th}}^{Q_{R,\rm max}} d Q \, Q^{-3} \right], \nonumber \\
Q_{R,\rm th} = \sqrt{2 m_N E_{R,\rm th}},~ Q_{R,\rm max}= 2 m_N v_{\rm rel}.
\end{align}
Assuming the Helm form factor~\cite{Helm:1956zz,Lewin:1995rx}, we find $f(E_{R,\rm th})\simeq 0.3$. The signal rate in a given detector with a number of target nuclei $N_N$ is
\begin{align}
\label{eq:xenon1t_rate_NR}
\Gamma_{\rm{Sig}} =  N_N \int  dv \, \sigma(E_R> E_{R,\rm{th}}) v \, \frac{d n_A}{dv}  \simeq  N_N \left[ \sigma(E_R> E_{R,\rm{th}}) v \frac{d n_A}{d{\rm ln}v} \right]_{v=v_-},
\end{align}
where $v_-$ is the minimum detectable $X$ speed.

\begin{figure}[tb]
\centering
\includegraphics[width=0.7\textwidth]{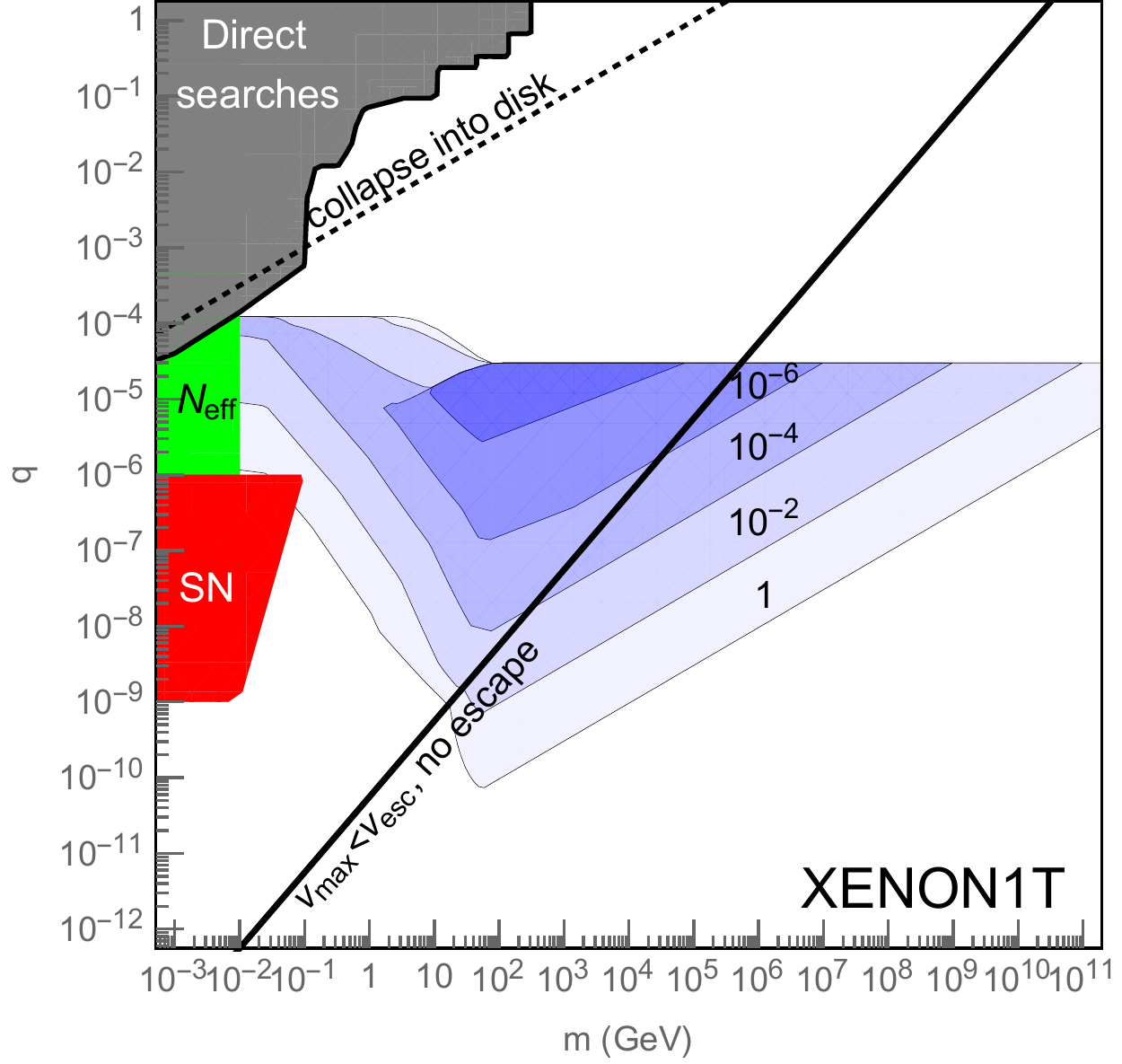} %
\caption{Upper bounds on the fraction of $X$ as dark matter from XENON1T.}
\label{fig:xenon1t}
\end{figure}

Using Eqs.~(\ref{eq:xenon1t_rate_NR}),
we compute the signal event rate at XENON1T~\cite{Aprile:2018dbl} with $E_{\rm{th}} = 10$ keV, and require fewer than 16 expected events for a 1 ton-year exposure, putting an upper bound on the fraction of $X$ as dark matter, as shown in Fig.~\ref{fig:xenon1t}. In the analysis of~\cite{Aprile:2018dbl}, events with extra ionization electrons are vetoed. Thus we conservatively require that the ionization energy loss of $X$ passing through 1m of liquid Xenon is below $10$ eV, so that typically no electron recoils occur. The minimal velocity $v_-$ is determined by this requirement through the dependence of the ionization energy loss on the velocity, the threshold energy, the minimal velocity to reach the detector, and $v_0$.

Below the thick solid line of Fig.~\ref{fig:xenon1t}, the maximum speed $X$ can gain from SNe is below the escape velocity, and the standard constraint is applicable. Above the dashed line $X$ collapse into the disk. Note that a bound exists even if $m<10$ GeV, where XENON1T is insensitive to dark matter with a virial speed due to the threshold. The accelerated CHAMPs have speeds much larger than the virial speed, and easily deposit energies above the threshold. The larger velocity also help CHAMPS to reach the underground detector, strengthening the constraint at larger values of $q$. For $q>$ few $10^{-5}$, electron recoils typically occur while $X$ pass through the detector, and hence $X$ scattering events may be vetoed.

We also show bounds on the parameter space from direct searches~\cite{Davidson:1991si,Prinz:1998ua,Davidson:2000hf,CMS:2012xi,Magill:2018tbb}, SN cooling~\cite{Chang:2018rso} and from the dark radiation abundance in the universe. The constraint from dark radiation is weaker than the one in~\cite{Vogel:2013raa}, as entropy production could occur near the MeV scale for $m \gsim 10$ MeV.

\begin{figure}[tb]
\centering
\includegraphics[width=0.7\textwidth]{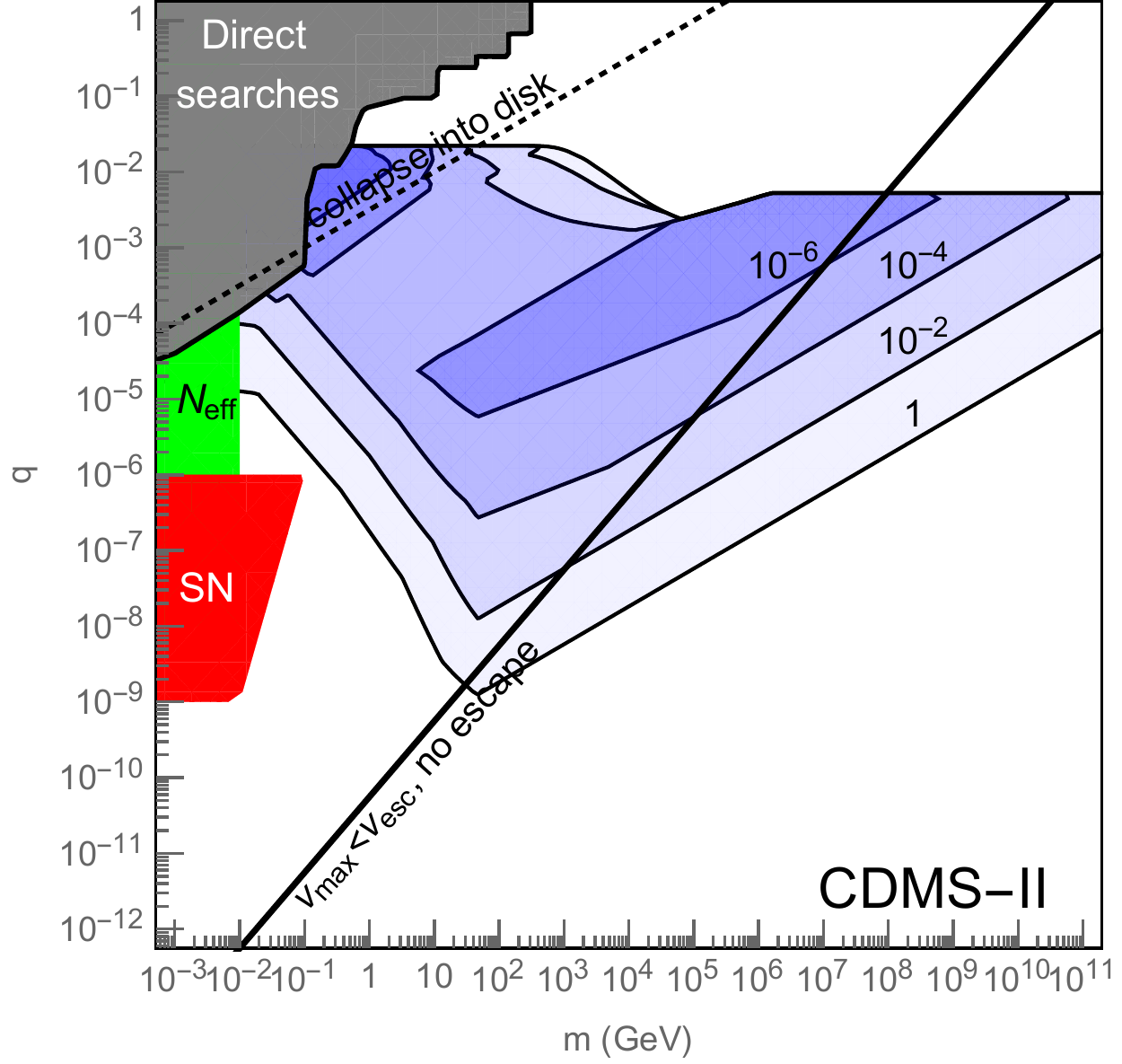} %
\caption{Upper bounds on the fraction of $X$ as dark matter from CDMS-II.}
\label{fig:cdms2}
\end{figure}

We compute the signal rate at CDMS-II~\cite{Ahmed:2009zw} with $E_{\rm{th}} = 10$ keV, and require fewer than 10 expected events for a 600 kg-day exposure. The constraint is shown in Fig.~\ref{fig:cdms2}. Signal regions are defined by a small ionization yield, below 30\% of the recoil energy. Thus we require that the ionization energy loss of $X$ passing through $1$ cm of germanium is below $3$ keV. A muon veto is also imposed, but we find that as long as the energy loss in the germanium is below $3$ keV $X$ signals evade the veto.
Although the constraint is weaker then that from XENON1T, CDMS-II constrains a region with larger values of $q$, up to $10^{-2}$.

\subsection{Electron Recoil at Deep Underground Detectors}

\begin{figure}[tb]
\centering
\includegraphics[width=0.7\textwidth]{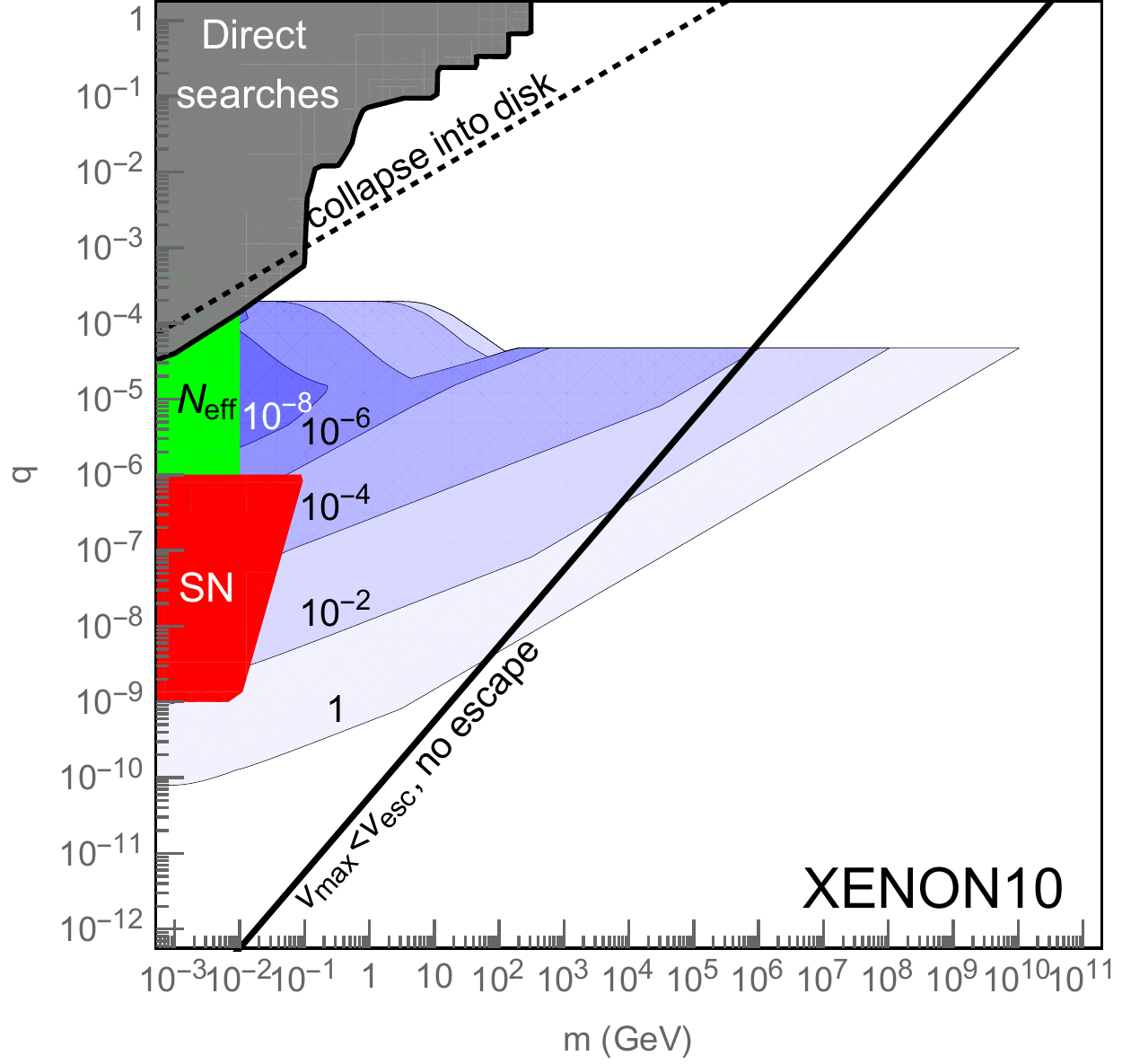} %
\caption{Estimated upper bound on the fraction of $X$ as dark matter from XENON10.}
\label{fig:xenon10}
\end{figure}

Nuclear recoil experiments cannot probe the region with small $m$ and $q$, since the maximum velocity of CHAMPs are still below the threshold. Such a region can be probed by searches for electron recoils at deep underground detectors with low thresholds. The estimation of the precise signal rate requires a computation involving an atomic form factor and  is beyond the scope of the paper. Instead, we obtain a rough estimation of the constraint by scaling the constraint in~\cite{Essig:2017kqs} in the following way.~\cite{Essig:2017kqs} defines a DM-free electron scattering cross section with the matrix element artificially evaluated at the momentum transfer of $m_e \alpha$ as $\bar{\sigma}_e$. For CHAMPs, it is given by
\begin{align}
\bar{\sigma}_e = \frac{16\pi q^2 \mu^2}{\alpha^2 m_e^4},
\end{align}
where $\mu$ is the reduced mass between an electron and a CHAMP. Assuming that the CHAMP is the dominant component of dark matter and has a virial velocity, the upper bound is $\bar{\sigma}_e < 3\times 10^{-34}~{\rm cm}^2 (m/{\rm GeV}) \equiv \bar{\sigma}_{e,limit}$ for $m\gg 10$ MeV. The bound becomes rapidly weaker for $m<10$ MeV as the kinetic energy of CHAMPs becomes smaller than the typical electron binding energy. We adopt the following as a rough estimation of the bound,
\begin{align}
\bar{\sigma}_e < \bar{\sigma}_{e,limit} \times \left[ \frac{n_0 v_{\rm{vir}}}{n_A v}\times \frac{dE/dx(v_{\rm{vir}})}{dE/dx(v)}  \right]_{v=v_-},
\end{align}
where $v_-$ is the minimal detectable speed. It is the maximum of $v_0$, $\beta_{\rm{min}} c$, $v_{\rm{vir}} \sqrt{10~{\rm MeV} / m}$ for sufficient kinetic energy, and the minimal velocity to deposit less than 10 eV by ionization while passing 40 cm of liquid Xenon.  The constraint is shown in Fig.~\ref{fig:xenon10}, which covers the small $m$ and $q$ region.

\subsection{Relativistic Electron Recoil and Subsequent Cherenkov Light}

\begin{figure}[tb]
\centering
\includegraphics[width=0.7\textwidth]{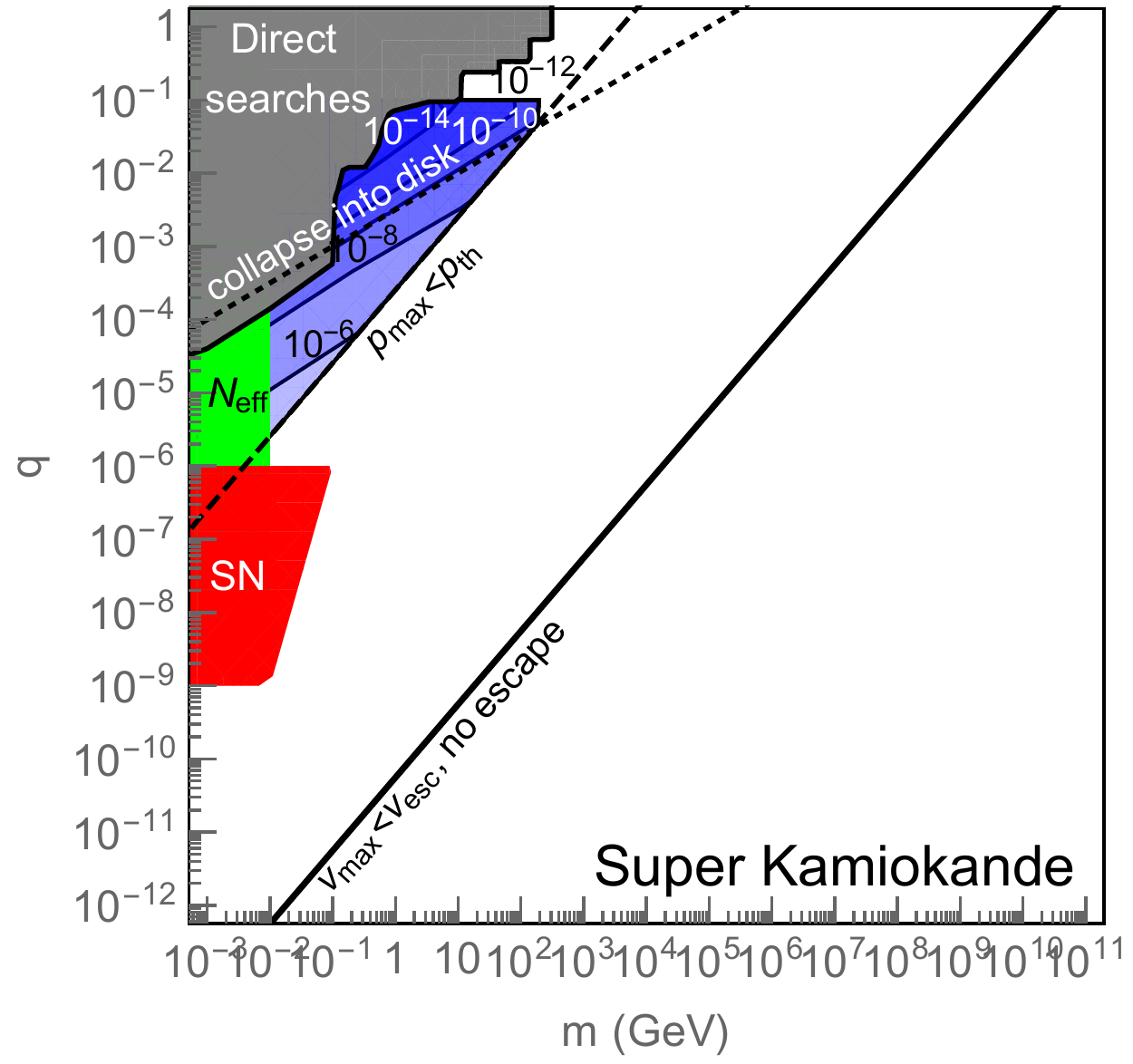} %
\caption{Upper bounds on the fraction of $X$ as dark matter from Super Kamiokande.}
\label{fig:SK}
\end{figure}

Relativistic CHAMPs passing through water may deposit enough energy to accelerate electrons to relativistic speeds. If the speed of these recoiling electrons is $> 0.75c$, they emit detectable Cherenkov light. Such events are detected by Super-Kamiokande for deposition energies above the threshold of 100 MeV~\cite{Kachulis:2017nci}.
The main target of the search is dark matter coming from the center of the galaxy, and constraints are put on events within a cone with a certain opening angle measured from the center of the galaxy. The accelerated CHAMPs come isotropically, and hence we use the bound on the signal rate for the largest cone, giving limits on $f_X$ shown in Fig.~\ref{fig:SK}.
Below the dashed line, the maximum momentum of accelerated CHAMPs is below the threshold, $m \sqrt{100~{\rm MeV}/m_e}$. For $q>0.1$, photomultiplier tubes (PMTs) in the outer detector typically receive more than one photon from the Cherenkov radiation of CHAMPs, giving events that are vetoed in the analysis of~\cite{Kachulis:2017nci}.

\subsection{Cherenkov Light from Relativistic CHAMPs}
Relativistic $X$ with speeds above $\beta_{C} =0.75$ produce Cherenkov light when traveling through water. For $q \ll 1$, the intensity of Cherenkov light is typically too low to observe individual tracks of $\upex$ in the ice. Nevertheless, the $(m,q)$ parameter space can be constrained when the total number of Cherenkov photons emitted from relativistic $\upex$ surpasses the observed $540 ~{\rm Hz}$ background count rate of the IceCube PMTs.%
\footnote{Radioactive decays are dominantly responsible for the remaining dark counts~\cite{2011A&A...535A.109A}.}

\begin{figure}[tb]
\centering
\includegraphics[width=0.7\textwidth]{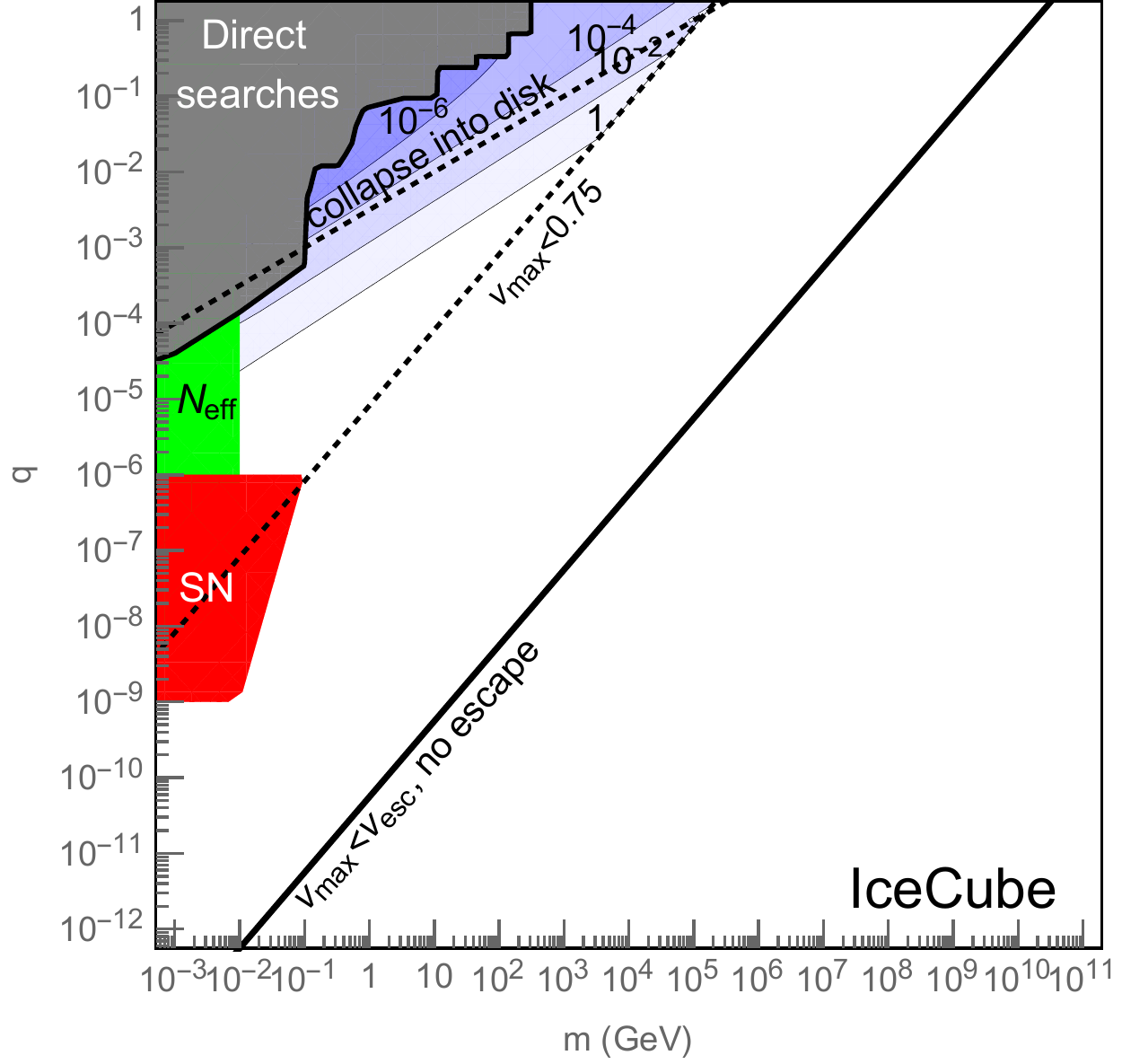} %
\caption{Upper bounds on the fraction of $X$ as dark matter from IceCube.}
\label{fig:icecube}
\end{figure}

The integrated flux of atmospheric muons $2 ~{\rm km}$ below the Antarctic ice is $\Phi_\mu / 4\pi \approx 10^{-7} ~\rm{cm^{-2} ~ s^{-1} ~ sr^{-1}}$~\cite{PhysRevD.98.030001} and contributes $3 \%$~\cite{2011A&A...535A.109A} of the background rate.  The number of Cherenkov photons emitted per unit wavelength and unit pathlength of $X$ is proportional to $q^2$~\cite{Jackson:1998nia}.  Requiring CHAMPs to give a signal below the observed PMT dark count rate constrains the integrated CHAMP flux above $\beta = \beta_C$ 
\begin{align}
 	\Phi  < 30 \; \frac{\Phi_\mu}{q^2} \hspace{0.5in} \mbox{where}  \hspace{0.5in}
	\Phi(\beta > \beta_C) \simeq \Phi(\beta \sim \beta_C) \simeq p \left. \frac{dn_A}{dp}v \right|_{\beta = \beta_C}.
 	\label{eq:chrenkovConstraint}
\end{align} 
The constraint \eqref{eq:chrenkovConstraint} is shown in Fig.~\ref{fig:icecube}. It is generally weaker than constraints from nuclear recoils in XENON1T or from energy deposition in MAJORANA.

\subsection{Ionizing Particle Searches}

\begin{figure}[tb]
\centering
\includegraphics[width=0.7\textwidth]{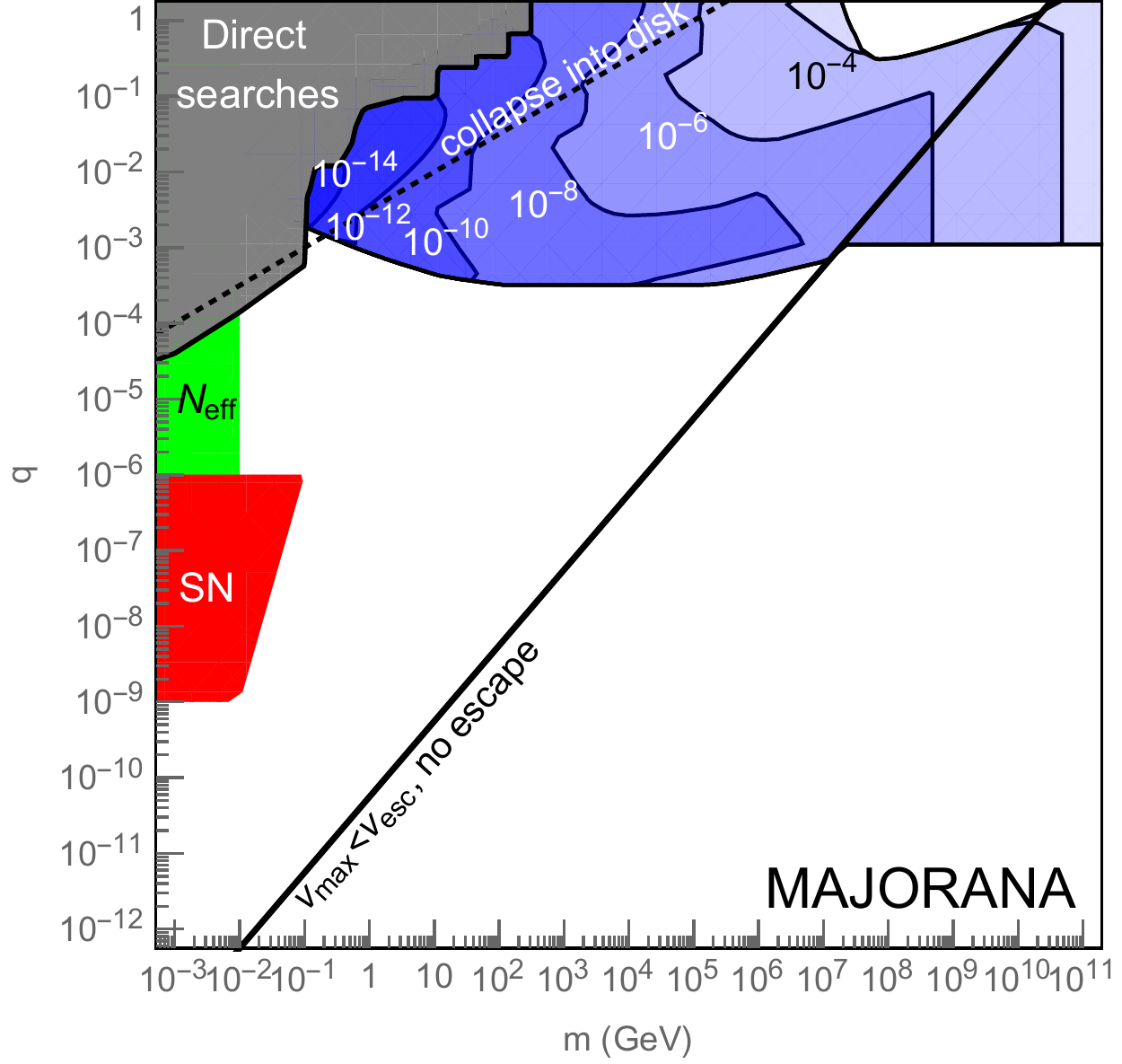} %
\caption{Upper bounds on the fraction of $X$ as dark matter from MAJORANA. Here we require that the typical energy deposit is above the threshold of $1$ keV. A larger parameter region will be constrained once accidentally large energy deposits are taken into account.}
\label{fig:majorana}
\end{figure}

As $q$ grows, CHAMPs yield significant ionization. The
MAJORANA experiment searches for such ionizing particles with a threshold of $1$ keV, and puts an upper bound on the flux, $\Phi_{\rm{MAJORANA}} <10^{-9}~ {\rm cm}^{-2}~{\rm s}^{-1}~{\rm sr}^{-1}$~\cite{Alvis:2018yte}. Taking $\beta = \max(\beta_{\rm{min}},v_1,v_{\rm{vir}},\beta_{\rm{ion}})$, where $\beta_{\rm{ion}}$ is the minimum velocity to exceed the threshold, as the minimum $\upex$ speed that can reach MAJORANA and yield signals, we find the upper bound on the fraction of $X$ as dark matter as shown in Figure~\ref{fig:majorana}.
The bound complements that from nuclear recoil experiments. There is no constraint for $q\lesssim 10^{-3.5}$, since even maximally ionizing $X$, with a velocity $\beta \sim 0.01$, cannot deposit an energy above the threshold. Here we use the NIST Database~\cite{Berger:399381} to calculate the typical energy deposit on germanium. In the region close to the solid line, $v_{\rm max} <0.01$ and $q$ must be larger to deposit enough energy. Similarly, for small $m_X$, the required value of $q$ becomes larger, as $X$ must have a larger velocity to reach MAJORANA, and is less ionizing.

We expect that larger parameter regions are actually constrained.  Even if the typical energy deposit is below a keV, there is a probability for $X$ to deposit an energy above the threshold, as computed in~\cite{Alvis:2018yte} for a minimally ionizing speed. This effect will lead to constraints in broader parameter regions, but is beyond the scope of our paper.

\begin{figure}[tb]
\centering
\includegraphics[width=0.7\textwidth]{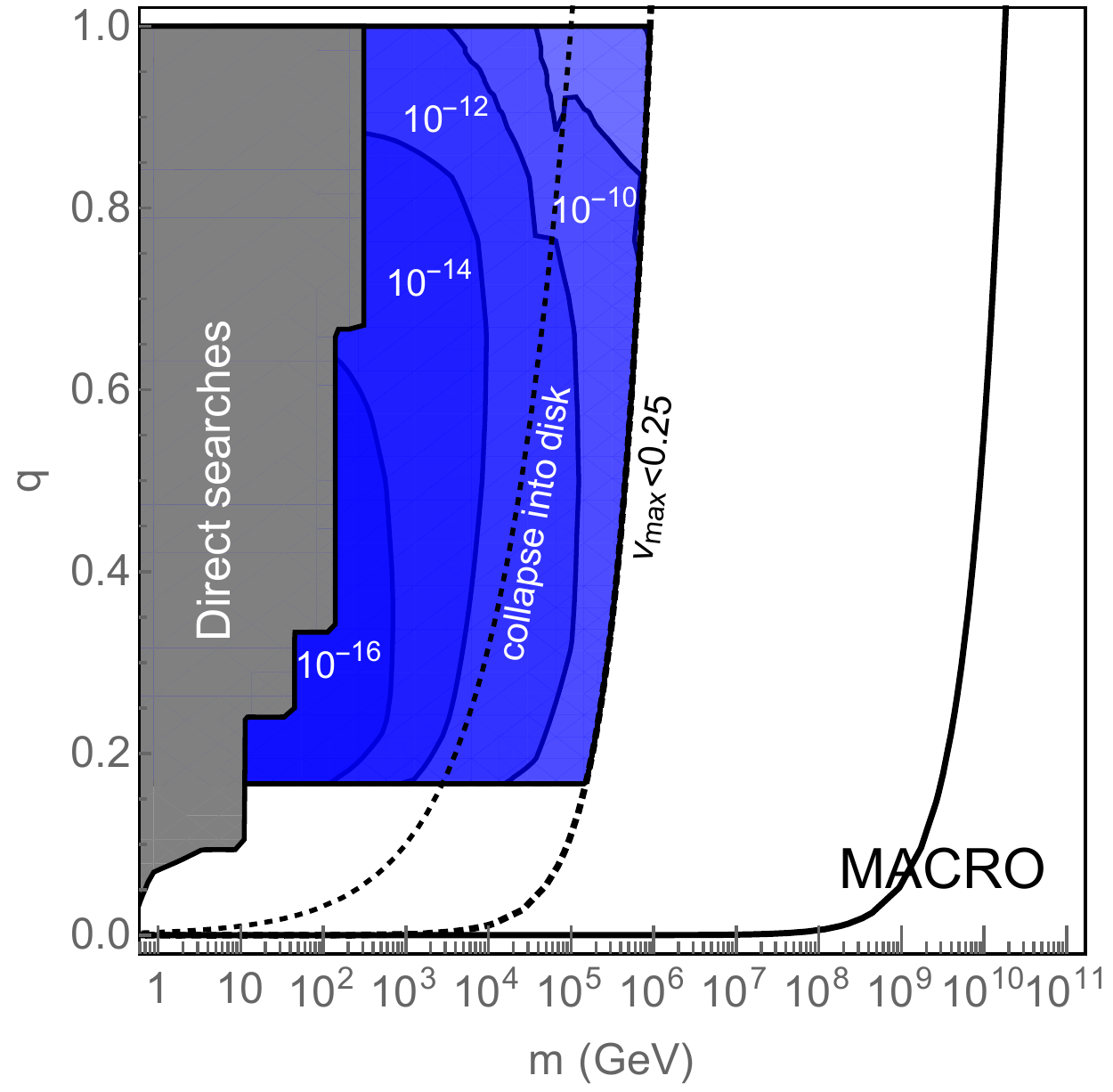} %
\caption{Upper bounds on the fraction of $X$ as dark matter from MACRO.}
\label{fig:macro}
\end{figure}
Relativistic CHAMPs with charges between $0.2 < q \lesssim 1$ produce visible tracks in MACRO's scintillation and streamer detectors that can be distinguished from integer charged cosmic rays through the $q^2$ dependence of` $dE/dx$. The upper bound on the flux of CHAMPs with $\beta > 0.25$ is $\Phi_{\rm{MACRO}} < 6.1 \times 10^{-16} ~ {\rm cm^{-2} ~ s^{-1} ~ sr^{-1}}$~\cite{Ambrosio:2004ub} for $1/4 < q < 1/2$ and weaker outside this range. Taking the lowest $\upex$ speed that MACRO can detect to be $\max(0.25, \beta_{\rm{min}})$, we find the upper bound on the fraction of $X$ as dark matter as shown in Figure~\ref{fig:macro}. For $m \gtrsim 10^6$ GeV, $v_{\rm max} < 0.25$ and hence the constraint is absent. 

\begin{figure}[tb]
\centering
\includegraphics[width=0.45\textwidth]{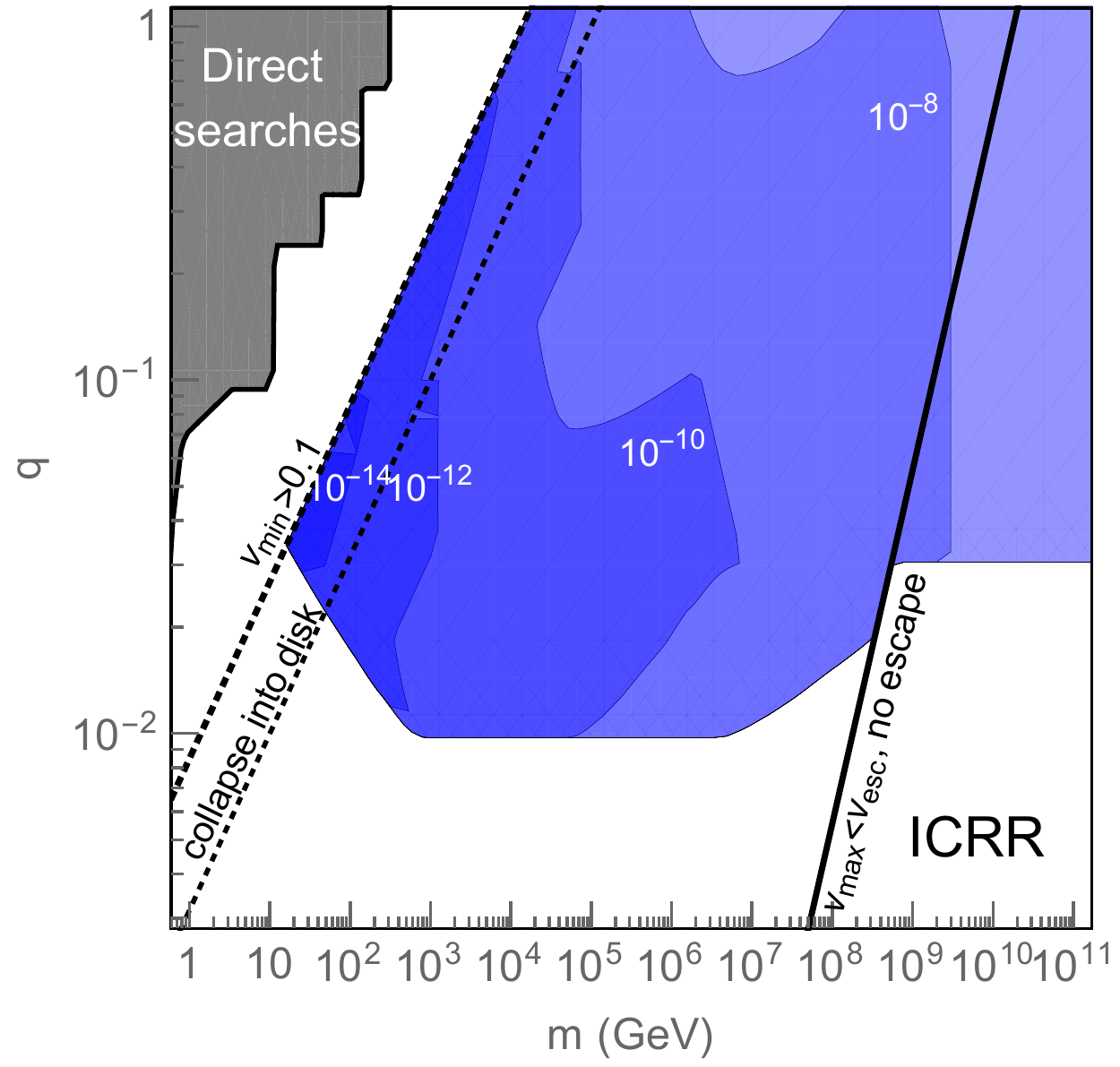} %
\includegraphics[width=0.45\textwidth]{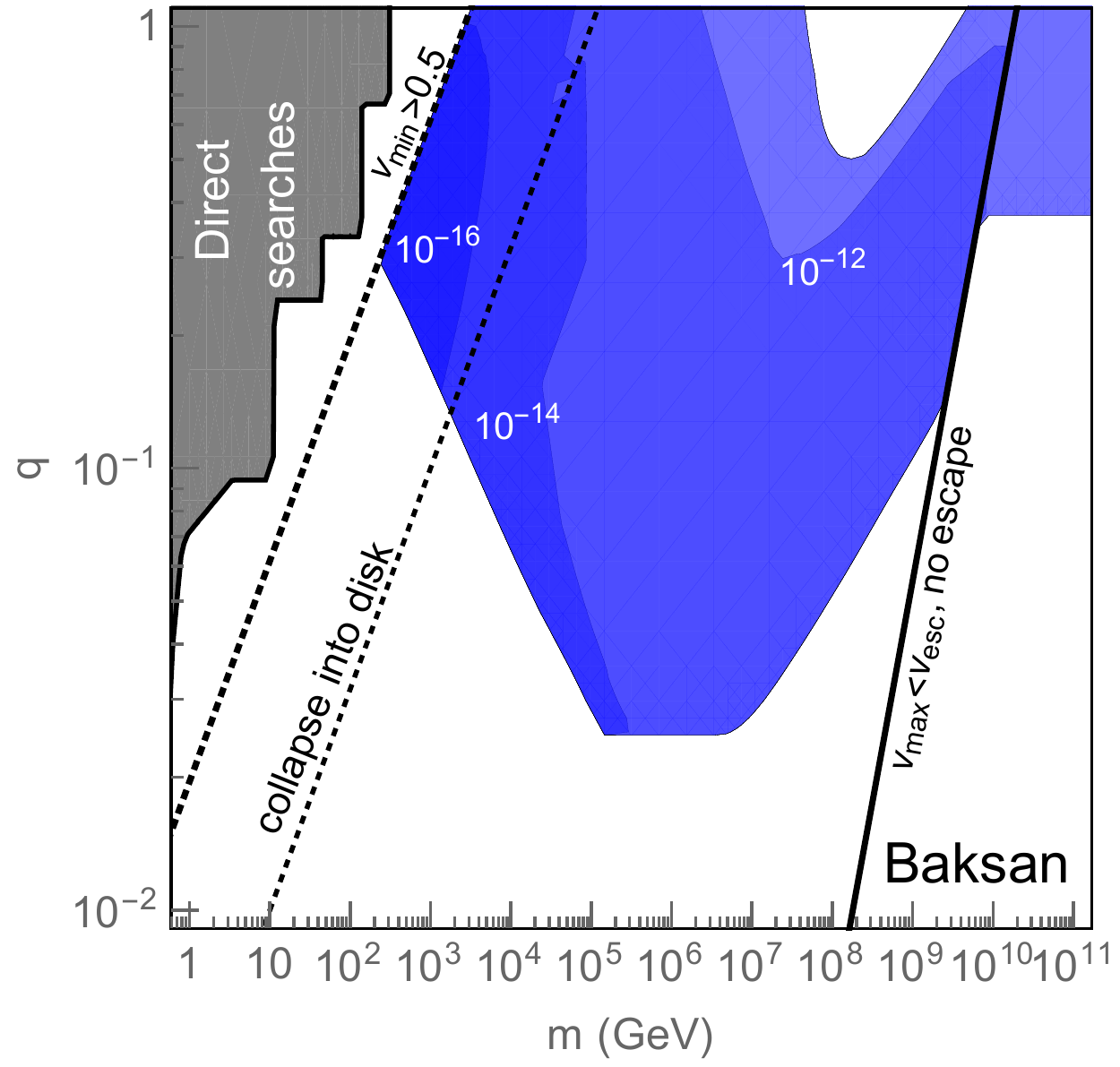} %
\caption{Upper bounds on the fraction of $X$ as dark matter from ICRR and Baksan experiments.}
\label{fig:iccr&baksan}
\end{figure}

While the flux constraints from MACRO are strong, the trigger efficiency of the MACRO hardware is sensitive only to relativistic $X$ with $\beta > 0.25$. Moreover, the MACRO experiment is over one km underground which prevents slower moving CHAMPs from reaching the detector (see Fig.~\ref{fig:minimumSpeed}). An experiment at the Institute for Cosmic Ray Research (ICRR), designed to look for the scintillation light of slow, penetrating, and highly ionizing particles on the surface of the Earth, constrains the flux of CHAMPs with $2.5 \times 10^{-4} \lesssim \beta \lesssim 1.0 \times 10^{-1}$ to $\Phi_{\rm{ICRR}} \lesssim 1.8 \times 10^{-12}  ~ {\rm cm^{-2} ~ s^{-1} ~ sr^{-1}}$~\cite{Kajino:1984ug}. The ICRR experiment is sensitive to ionization deposits greater than $1/20$ the minimum ionization $I_{\rm min}\sim 1.6 ~{\rm MeV/cm}$~\cite{1987PhRvD..36.2641B}. As discussed in Sec \ref{sec:obstacles}, the ionization losses of $X$ with $\beta < 0.1$ can be read from the experimental stopping power of protons, scaled by $q^2$, and imply charges as low as $10^{-2}$ can be detected. From the observed stopping power of protons through plastic scintillators~\cite{Berger:399381}, we find the upper bound on the fraction of $X$ as dark matter as shown in Figure~\ref{fig:iccr&baksan}. The lower edge of the constrained region is determined by ionization losses. To the left of the dotted lines, the velocity of CHAMPs which reach the detector is larger than $0.1c$. To compute $v_{\rm min}$, we use the stopping power of air, shown in the NIST Database~\cite{Berger:399381} and a column depth of $10^3$ g/cm$^2$.

Last, the Baksan experiment~\cite{1983ICRC....5...52A,1985ICRC....8..250A}, an underground scintillation detector searching for slowly-moving ionizing particles, complements MACRO by providing comparable flux constraints to non-relativistic CHAMPs with $\beta < 0.1$. The upper bound on the flux of CHAMPs with $2 \times 10^{-4} \lesssim \beta \lesssim 10^{-1}$ is $\Phi_{\rm{Baksan}} \lesssim 2 \times 10^{-15} ~ {\rm cm^{-2} ~ s^{-1} ~ sr^{-1}}$. The Baksan experiment is sensitive to ionization deposits greater than $\min(1, .02/\beta)\times .25 \, I_{\rm{min}}$~\cite{1983ICRC....5...52A}, and implies $q$ as low as $\sim 1/40$ can be constrained for $\beta > 0.02$. CHAMPs with $\beta < 0.02$ cannot traverse the length of the detector within one integration time and thus must have greater $dE/dx$ (that is, greater $q$) to be detected.  Taking into account the stopping by the Earth and the stopping power of protons through liquid scintillators,%
\footnote{Liquid scintillators  have similar $dE/dx / \rho$ to plastic-based ones since both are organic compounds.}
we find the upper bound on the fraction of $X$ as dark matter as shown in Figure~\ref{fig:iccr&baksan}.
Note while Baksan and ICRR both probe similar velocity ranges, Baksan has stronger flux constraints but is not as sensitive to small $q$ as ICRR, nor can it detect the slowest CHAMPs which stop in the Earth before reaching it. 

\subsection{Constraints on DM or Thermally Produced CHAMPs}
\begin{figure}[tb]
\centering
\includegraphics[width=0.7\textwidth]{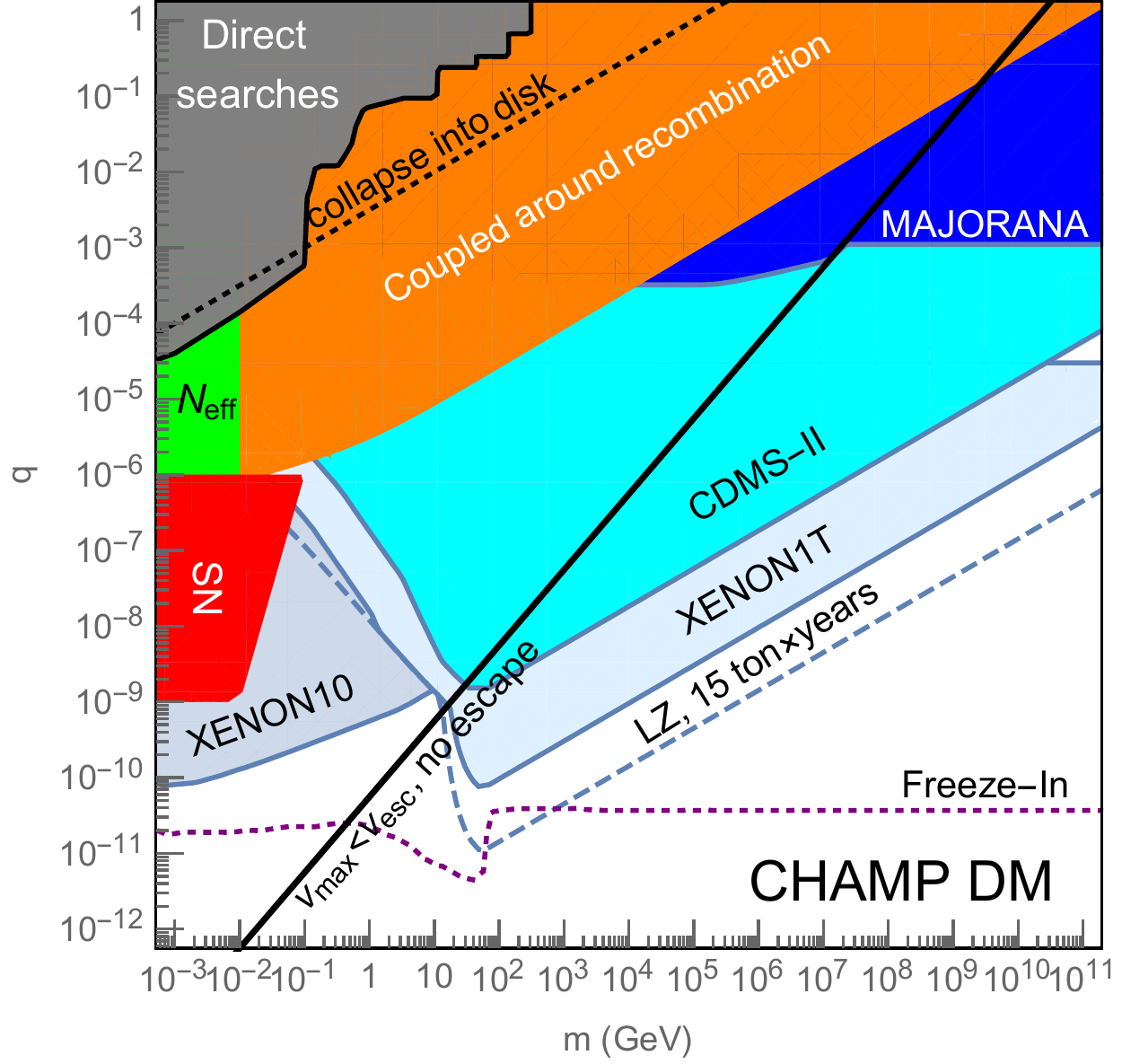} %
\caption{Constraints on $(m,q)$ assuming all of dark matter is $X$.}
\label{fig:constraint_dm}
\end{figure}

Fig.~\ref{fig:constraint_dm} summarizes the constraint on $(m,q)$ assuming all of dark matter is $X$. In the orange-shaded region indicated as ``Coupled around recombination", $X$ couples to baryons around the era of recombination, and the fluctuations of the cosmic microwave background are altered~\cite{Dubovsky:2003yn,Dolgov:2013una}.
We also show the prospected sensitivity of the LZ experiment~\cite{Mount:2017qzi} assuming a 15 ton-years exposure with a threshold energy of $10$ keV.
The purple dotted line shows the prediction for the charge $q$ from the Freeze-In production of $X$ dark matter~\cite{Chu:2011be}; see below for a rough estimation. Nuclear recoil experiments have just begun to reach the sensitivity to probe Freeze-In production.

We take a closer look at the thermally produced $X$. We assume that initially only the Standard Model sector is thermalized. The Freeze-In abundance of $X$ from pair production by s-channel photon exchange is
\begin{align}
\frac{\rho_{X_{FI}}}{s} \simeq 0.01\frac{4 \pi \alpha^2 q^2}{\sqrt{g_s(m)}} \mpl,
\end{align}
where $\rho_X$ is the energy density of $X$, $s$ is the entropy density, and $g_s$ is the effective number of degrees of freedom.  It is almost independent of $m$ but grows with $q^2$.  This gives $f_X = 1$ for $q \sim 10^{-11}$. We use the precise estimation of~\cite{Chu:2011be} in the following. For large enough $q$ this abundance becomes sufficient for pair-annihilation of $X$ to occurs, so that the final yield is then given by Freeze-Out
\begin{align}
\frac{\rho_{X_{FO}}}{s} \simeq \frac{m/T_{\rm FO}}{ g_s^{1/2}(T_{\rm FO}) \mpl \, \sigma v },
\end{align}
where $\sigma v$ is the annihilation cross section, and $T_{\rm FO}$ is the temperature at Freeze-Out. If Freeze-Out occurs then it determines the final abundance, otherwise it is determined by Freeze-In.

\begin{figure}[tb]
\begin{center}
\includegraphics[width=0.7\textwidth]{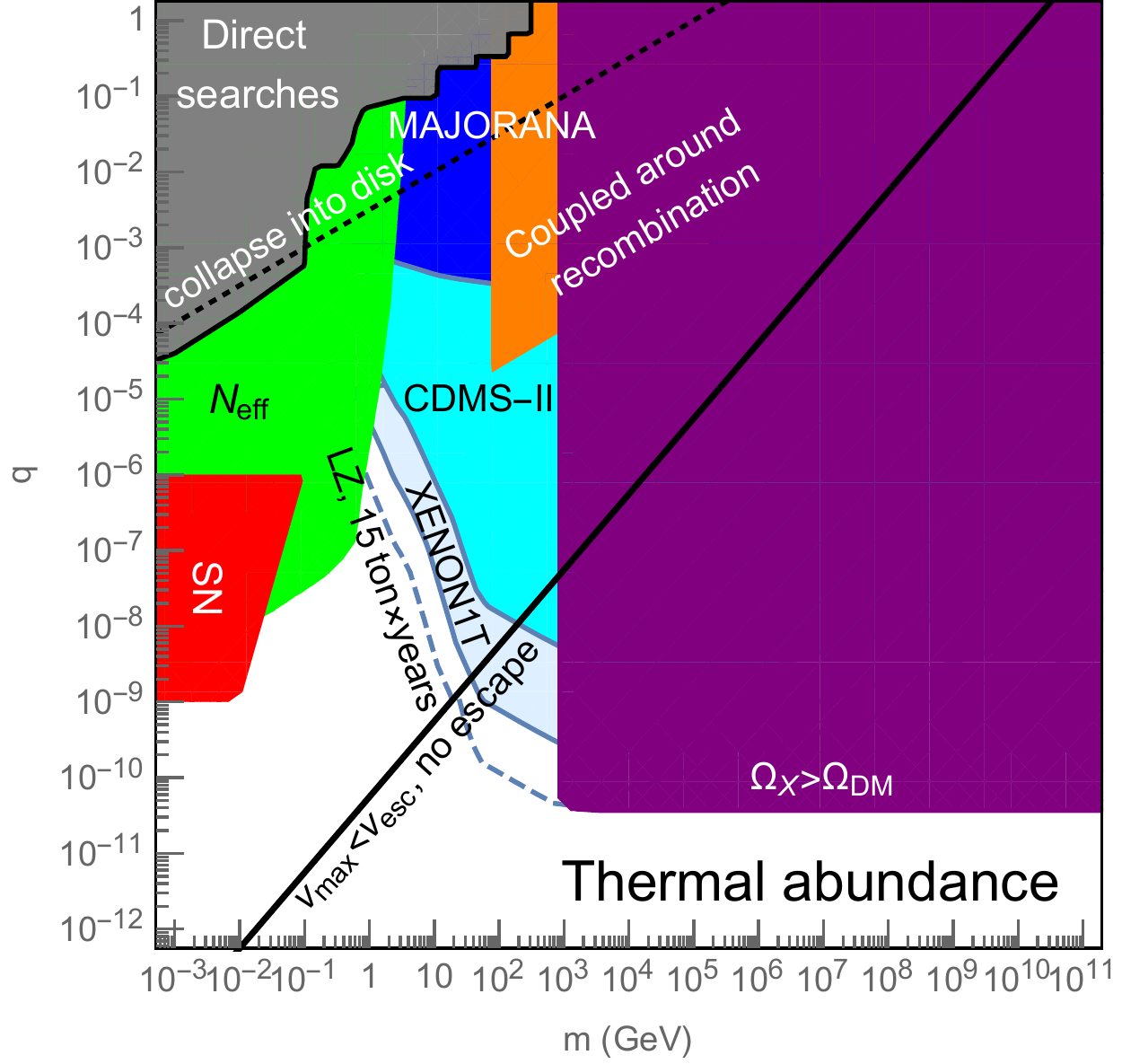}
\caption{Constraints on $(m,q)$ assuming that $X$ is produced thermally in the early universe. Here we assume that the gauge coupling constant of the dark photon is the same as the electro-magnetic coupling.}
\label{fig:thermal}
\end{center}
\end{figure}

If $X$ is taken to be the only addition to the Standard Model then almost the entire region of interest having $q$ larger than $10^{-11}$ is excluded because $f_X >1$. Hence we add a massless dark photon so that $X$ can pair-annihilate into dark photons.   This simple scheme for dark matter has been studied for general values of the $U(1)'$ gauge coupling in some depth~\cite{Chu:2011be}.  Here, for simplicity, we fix the gauge coupling of $X$ to the dark photon to be the same as between an electron and photon.  Fig.~\ref{fig:thermal} shows the constraints on $(m,q)$ resulting from this thermal abundance of $X$.   The abundance exceeds the dark matter abundance in the purple-shaded region at the right of the figure.  At the edge of this region $f_X = 1$, with production from Freeze-In along the bottom edge and from Freeze-Out along the left edge. Moving to the left, $f_X$ drops as $m^2$.  The constraint from fluctuations of the cosmic microwave background, the orange-shaded region, is applicable if $X$ comprises more than 1\% of dark matter~\cite{Dolgov:2013una}.  The constraint from dark radiation is taken from~\cite{Vogel:2013raa}, with the latest bound $N_{\rm eff} \lesssim 3.5$~\cite{Aghanim:2018eyx}.

Last, we comment on the possible effect of CHAMPs on the structure of the halo for $f_X\sim 1$. For $m/q \lesssim 10^{12}$ GeV, the mean free path and the gyro-radius of $X$ with $v= v_{\rm{vir}}$ is smaller than the height of the confinement region. This means that the dynamics of $X$ are not only governed by the gravitational force but are also affected by the magnetic field in the confinement region, which may change the distribution of CHAMP dark matter in the inner part of the halo, and possibly lead to further limits/signals.  The ejection of CHAMP by SNe may further affect the halo structure. We do not pursue this possibility further in this paper.

\section{Summary and discussion}
\label{sec:con}

Cosmological relics, whether comprising all of dark matter or just a component, are almost always considered to be electrically neutral.  However, charged relics may have escaped detection either because their electric charge $qe$ is very small, or because their mass, $m$, is very large.  While CHAMPs may arise as simple additions to the Standard Model, kinetic mixing provides a window to dark sectors that contain a $U(1)$ gauge group, greatly enhancing the importance of CHAMP searches, and strongly motivating searches over a wide range of the $(m,q)$ plane.

For $m> 10^{10} q$ GeV, $X$ form a virialized halo that is not disturbed on cosmological time scales by interactions with the interstellar medium or by magnetic fields. However, for smaller values of $m/q$, Fermi acceleration by shock waves of supernova remnants, diffusion through magnetic inhomogeneities, and thermalization via Rutherford scattering with ionized interstellar matter play crucial roles in determining the number density and spectrum of $X$ hitting the solar system today.  We have discovered that, over a wide region of $(m,q)$, a steady state is established balancing efficient ejection of $X$ from the galaxy by SN shocks with continuous diffusion of $X$ into the disk from the halo.   The resulting accelerated cosmic ray flux at the solar system today is shown in Eq.~(\ref{eq:final spectrum}). It has a $1/p^{5/2}$ spectrum to a maximum momentum determined by the size and the lifetime of the shock-wave accelerator, as is shown in Fig.~\ref{fig:maxRigidity}.  The corresponding local number density of the accelerated $X$ is very large, $(10^{-3} - 1)$ of the halo density for $m/q = (10^2 - 10^{10})$ GeV.  Hence, in this region of  
$(m,q)$, limits from direct detection experiments are very powerful, whereas previously, neglecting diffusion in from the halo, they were believed to be absent. 

For $m<10^5 q^2$ GeV, $X$ collapse into the disk as it forms. Clearly $X$ cannot be halo dark matter; however, they may still provide a window to the dark sector.
The inefficient diffusion in from the halo because of the thermalization or the small rigidity of $X$ suggests that constraints on a component of $X$ might be weak.
This is incorrect: $X$ are strongly coupled with the ISM giving a thermalization bottleneck, inhibiting Fermi acceleration and ejection from the galaxy.  Although most $X$ are ejected by today, there remains a local flux of accelerated $X$, and for $m>$ MeV the charge $q$ is sufficiently large that direct detection limits are extremely powerful.

Over the entire $(m,q)$ plane with $m <  10^{10} q$ GeV, the accelerated CHAMPs have speeds larger than typically assumed for dark matter, opening up new signals and regions of parameter space to be probed by experiments.  We have derived constraints from XENON1T, CDMS II, XENON10, Super Kamiokande, IceCube, MAJORANA, MACRO, ICRR as well as Baksan.  Over a large part of the $(m,q)$ plane, the most powerful constraints on $f_X$, the fraction of dark matter that can be $X$, arise from direct detection limits from nuclear/electron recoil.  Indeed, for $q< 10^{-6}$, the only limits come from nuclear/electron recoil.   Limits on $f_X$ from the XENON1T, CDMS II and XENON10 experiments are shown in Figures \ref{fig:xenon1t}, \ref{fig:cdms2} and \ref{fig:xenon10}.  At larger $q$ the most powerful bounds on $f_X$ arise from signals from Cherenkov light and ionization; frequently these bounds are extremely powerful, although they often apply to only a small region of the $(m,q)$ plane, as shown in Figures \ref{fig:SK}-\ref{fig:iccr&baksan}. 

We briefly comment on the EDGES detection of an enhanced absorption feature in the 21-cm line at $z\sim 17$ \cite{bowman2018absorption}. Such an anomaly can be explained if a fraction $f_X \sim 10^{-3}$ of DM are CHAMPs with mass $10-80 ~ {\rm MeV}$ and charge $10^{-6} - 10^{-4}$ \cite{Berlin:2018sjs}. However, such a scenario is ruled out by XENON10 and Super-K experiments by $3-5$ orders of magnitude according to Figs. \ref{fig:xenon10} and \ref{fig:SK}.

Constraints on CHAMPs comprising all of dark matter, no matter what the production mechanism, are severe, as shown in Figure \ref{fig:constraint_dm}. $q>10^{-9}$ is excluded for any $m < 10^5 \; \GeV$. It will be exciting to see how much of the Freeze-In region can be reached by future experiments.

There are two clear signal regions for thermally produced CHAMPs that contribute all of dark matter in theories with a dark photon.  In Figure \ref{fig:thermal}, these are along the edges of the purple region that is excluded by overproduction of dark matter. The first arises from Freeze-In production of the CHAMPs from the Standard Model sector, and has $q \sim 4 \times 10^{-11}$ and $m > 1$ TeV.  The second arises from Freeze-Out annihilation to dark photons and has $m \sim 1$ TeV, and $q$ in the fairly narrow range of $4 \times 10^{-11}-10^{-10} $, with larger $q$ being excluded by XENON1T.  Future nuclear recoil experiments will continue to probe the Freeze-Out region and may eventually reach the Freeze-In region.  For $m < 1$ TeV, Freeze-out gives $f_X \propto m^2$, and future nuclear recoil experiments will probe significant regions of the $(m,q)$ plane where $X$ is a sub-dominant component of dark matter.

\let\oldaddcontentsline\addcontentsline
\renewcommand{\addcontentsline}[3]{}

\section*{Acknowledgement}
The authors thank Simon Knapen, Daniel  McKinsey, Eliot Quataert, and Nick Rodd for useful discussion.
This work was supported in part by the Director, Office of Science, Office of High Energy and Nuclear Physics, of the US Department of Energy under Contract DE-AC02-05CH11231 and DE-SC0009988 (KH), as well as  by the National Science Foundation under grants PHY-1316783 and PHY-1521446. 

\let\addcontentsline\oldaddcontentsline

\appendix

\section{Interaction between CHAMPs}
\label{sec:hidden U1}

In computing the ejection of $X$ from the disk, we have ignored any $XX$ scattering between CHAMPs.  In particular, after $X$ particles  are accelerated by a SN shock, they could be slowed down by scattering from ambient $X$ in the disk.  In this appendix we consider such scattering to arise from massless hidden photon exchange, and derive the condition such that the scattering does not change our estimation of the accelerated CHAMPs.
We show that our previous results are not affected if $X$ is produced before Big-Bang Nucleosynthesis, or $m>10$ GeV.

For a hidden charge $Q e$, the thermalization rate of $X$ via hidden photon exchange is
\begin{align}
\Gamma_{\rm th,X} \simeq \frac{8 \sqrt{2 \pi}}{3} \frac{\rho_XQ^4 \alpha^2 }{m^3 v^3}.
\end{align}
The velocity of $X$ above which the thermalization rate is smaller than the encounter rate with SNe is
\begin{align}
v_1' = 3000 \; {\rm km}/s  \;  (f_X\frac{n}{n_0})^{1/3} Q^{4/3} \left(\frac{{\rm GeV}}{m}\right),
\end{align}
which is analogous to $v_1$ derived in the main text based on the $X$-baryons scattering. In order for our previous results not to be affected, $v_1'$ should be smaller than $v_0$;
\begin{align}
\label{eq:condition_minv}
Q^4 f_X \frac{n}{n_0} < 3 \times 10^{-4} \left( \frac{m}{\rm GeV} \right)^{3} \times
\begin{cases}
20 \left( \frac{m/q^2}{3\times 10^6~{\rm GeV}} \right)^{-1}  & m/q^2 < 3\times 10^6~{\rm GeV} \\
1 & m/q^2 > 3\times 10^6~{\rm GeV}
\end{cases}
\end{align}
If this condition is violated, $v_0$ in Eq.~(\ref{eq:final spectrum}) is replaced with $v_1'$.

The  $XX$ scattering can also affect the estimation of $n_A/n_0$. The encounter rate with SNe with a shock velocity above $v_1'$ is
\begin{align}
\Gamma_{\rm{SH},c}' = (7 \times 10^{9}~{\rm years})^{-1} \left(\frac{m}{\rm GeV}\right)^2 \left( Q^4 f_X \frac{n}{n_0} \right)^{-2/3}.
\end{align}
For $m/q^2 < 10^5$ GeV, $n_A/n_0$ is not affected as long as this encounter rate is larger than $\Gamma_{\rm{SH},c}$, which requires that
\begin{align}
\label{eq:condition_encounter_collapse}
Q^4 f_X \frac{n}{n_0} < 0.1  \left( \frac{m}{\rm GeV} \right)^{3} \left( \frac{m/q^2}{10^5~{\rm GeV}} \right)^{-1}.
\end{align}
If this condition is violated, $\Gamma_A$ in Eq.~(\ref{eq:nA_collapse}) is replaced with $\Gamma_{\rm{SH},c}'$.
For $m/q^2 > 10^5$ GeV, $n_A/n_0$ is insensitive to $\Gamma_A$ and hence to $\Gamma_{\rm{SH},c}' $. This is however not the case if $n/n_0$ becomes close to 1 because of inefficient acceleration. Requiring that $n/n_0 <1$, we obtain
\begin{align}
\label{eq:condition_encounter}
Q^4 f_X \frac{n}{n_0} <  0.1\left( \frac{m}{\rm GeV} \right)^{3} \times
\begin{cases}
\left( \frac{m/q}{10^6~{\rm GeV}} \right)^{-3/8}& m/q < 10^6~{\rm GeV}\\
  \left( \frac{m/q}{10^6~{\rm GeV}} \right)^{-3/4}  & m/q > 10^6~{\rm GeV}.
\end{cases}
\end{align}
If this condition is violated, $n/n_0 \simeq 1$, and $n_A/n_0 \simeq \Gamma_{\rm{SH},c}'/\Gamma_{\rm{SH}}$.

In the parameter space of our interest, $m/q < 10^{10}$ GeV, all of the above conditions are weaker than the restriction $Q<1$ if $m \gtrsim 10$ GeV.

Note that the fraction $f_X$ is bounded from above for a given $Q$ and the production temperature. We assume that $X$ is produced at the temperature of $T_{\rm pro}$. The number density of $X$ is bounded by the annihilation of $X$ into hidden photons,
\begin{align}
n_X \frac{\pi Q^4\alpha^2}{m_X^2} \lsim \frac{T_{\rm pro}^2}{\mpl}
\end{align}
The fraction $f_X$ and the charge $Q$ is bounded as
\begin{align}
\label{eq:condition_ab}
Q^4 f_X < 3 \times 10^{-4} \left(\frac{m}{\rm GeV}\right)^{3} \frac{4~{\rm MeV}}{T_{\rm pro}}.
\end{align}
Let us assume that $T_{\rm pro}$ is above 4 MeV, as required if the $X$ production mechanism also produces Standard Model particles with an energy density comparable to that of $X$. For CHAMPs satisfying the condition (\ref{eq:condition_ab}), all off the conditions above are satisfied.

\section{Repeated Shock Encounters}
\label{sec:repeatedShocks}

Here we consider the effect of repeated shocks on the galactic CHAMP spectrum, which occurs for CHAMPs residing in case 1 galaxies.
We show that $X$ with a momentum below $p_2$ is ejected from the disk within a time $\sim \Gamma_{\rm SH}^{-1}$. Consequently, using  $\Gamma_{\rm SH}$ as the escape rate in \eqref{eq:diff eqA} is a good approximation.

First note the spectrum of a batch of CHAMPs after one shock can be written as a transformation on the original spectrum ~\cite{1978MNRAS.182..147B}
\begin{align}
    f_1(p) &=   (\mu - 1)p^{-\mu}\int_0^p dk ~ k^{\mu -1} f_0(k)
    \label{eq:vlasovSpectrum}
\end{align}
where $f_0$ is the original spectrum and $\mu = 2$ the theoretical value from Rankine-Hugeniot plasma boundary conditions at the shock discontinuity. Without loss of generality, let $f_0(q) = n_0 \delta(k - p_0)$ so that the spectrum after the one shock is
\begin{align}
    f_1(p) &= n_0 (\mu - 1)p_0^{\mu - 1}p^{-\mu}~ \theta(p - p_0)
\end{align}
which is the standard Fermi spectrum. Now, the CHAMPs in this spectrum with momentum above $p_2$ are more likely to escape the disk before encountering another shock, while those with momentum below $p_2$ more likely to encounter another shock before escaping. Thus, the spectrum after one shock approximately bifurcates into a galactic and extragalactic spectrum:
\[
f_1(p) \rightarrow  
\left\{
\begin{aligned}
&   f_1(p)_{\rm{In}}    &=& \quad   n_0 (\mu - 1)p_0^{\mu - 1}p^{-\mu}~ \theta(p_2 - p)\\
&   f_1(p)_{\rm{Out}}   &=& \quad   n_0 (\mu - 1)p_0^{\mu - 1}p^{-\mu}~ \theta(p - p_2)
\end{aligned}
\right.
\]
After the next shock, the galactic spectrum is
\begin{align}
    f_2(p)  &=  (\mu - 1)p^{-\mu}\int_{p_0}^p dk ~ k^{\mu -1} f_1(k)_{\rm{In}} \\
            &=  n_0(\mu - 1)^2 p_0^{\mu - 1} p^{-\mu} \ln \left(\frac{\min\{p,p_2\}}{p_0}\right)
\end{align}
Which again bifurcates into a galactic (where $\min\{p,p_2\} = p$) and extragalactic (where $\min\{p,p_2\} = p_2$) spectrum and so on. After $n$ shocks, the galactic spectrum is 
\begin{align}
    f_n(p)_{\rm{In}}    &=  n_0(\mu - 1)^n p_0^{\mu - 1} p^{-\mu} \ln \left(\frac{p}{p_0}\right)^{n-1} \frac{1}{(n-1)!} ~\theta(p_2 - p)
    \label{eq:galacticSpectrum}
\end{align}
which quickly approaches $0$ for $n > \ln (p/p_0)$, which is $O(1)$ in the parameter space of the interest. As a result CHAMPs with an initial momentum $p < p_2$ escape from the disk with the time scale $\sim \Gamma_{\rm SH}^{-1}$. Meanwhile, the extragalactic spectrum is
\begin{align}
    f(p)_{\rm Out}  &=  \sum_{i \geq 1} f_i(p)_{\rm Out} \\
                    &=  n_0 ~ p_0^{\mu - 1} p^{-\mu} ~\theta(p - p_2) \sum_{i \geq 1} (\mu - 1)^i \ln \left(\frac{p_2}{p_0}\right)^{i-1} \frac{1}{(i-1)!} \\
                    & \xrightarrow{i \gg 1} n_0 ~ p_0^{\mu - 1} p^{-\mu} ~\theta(p - p_2) (\mu - 1) \exp\left[(\mu - 1)\ln (p_2/p_0 )\right] \\
                    &=  n_0(\mu - 1) p_{2}^{\mu - 1} p^{-\mu} ~\theta(p - p_2)
                    \label{eq:extragalacticSpectrum}
\end{align}
which approaches the Fermi accelerated spectrum of single shock with an input spectrum $n_0 \delta(k - p_2)$. Note the extragalactic spectrum is independent of initial conditions, given that $\upex$ initially reaches a speed greater than or equal to $v_0$ for case 1.

\section{The Diffuse Extragalactic CHAMP Background}
\label{sec:diffuseEGBackground}
In this appendix we investigate the ejection of CHAMPs in galaxies and estimate the extragalactic CHAMPs spectrum.

In case one galaxies, with rate orderings of Fig.~\ref{fig:rateHierarchy1Plot}, CHAMPs with momentum $p_1 < p < p_2$ repeatedly encounter shock and are accelerated. Once the momenta are above $p_2$, they escape from the disks before encountering another shock, ultimately producing an extragalactic spectrum 
\begin{align}
\label{eq:EGSpectrum1}
f_1(p) = \; N_X p_2 \;\; \frac{1}{ p^2} \quad     \text{ for $p > p_2$},
\end{align}
where $N_X$ is the total number of ejected CHAMPs from that galaxy.
\begin{figure}[ht!]
\centering
\includegraphics[width=0.9\textwidth]{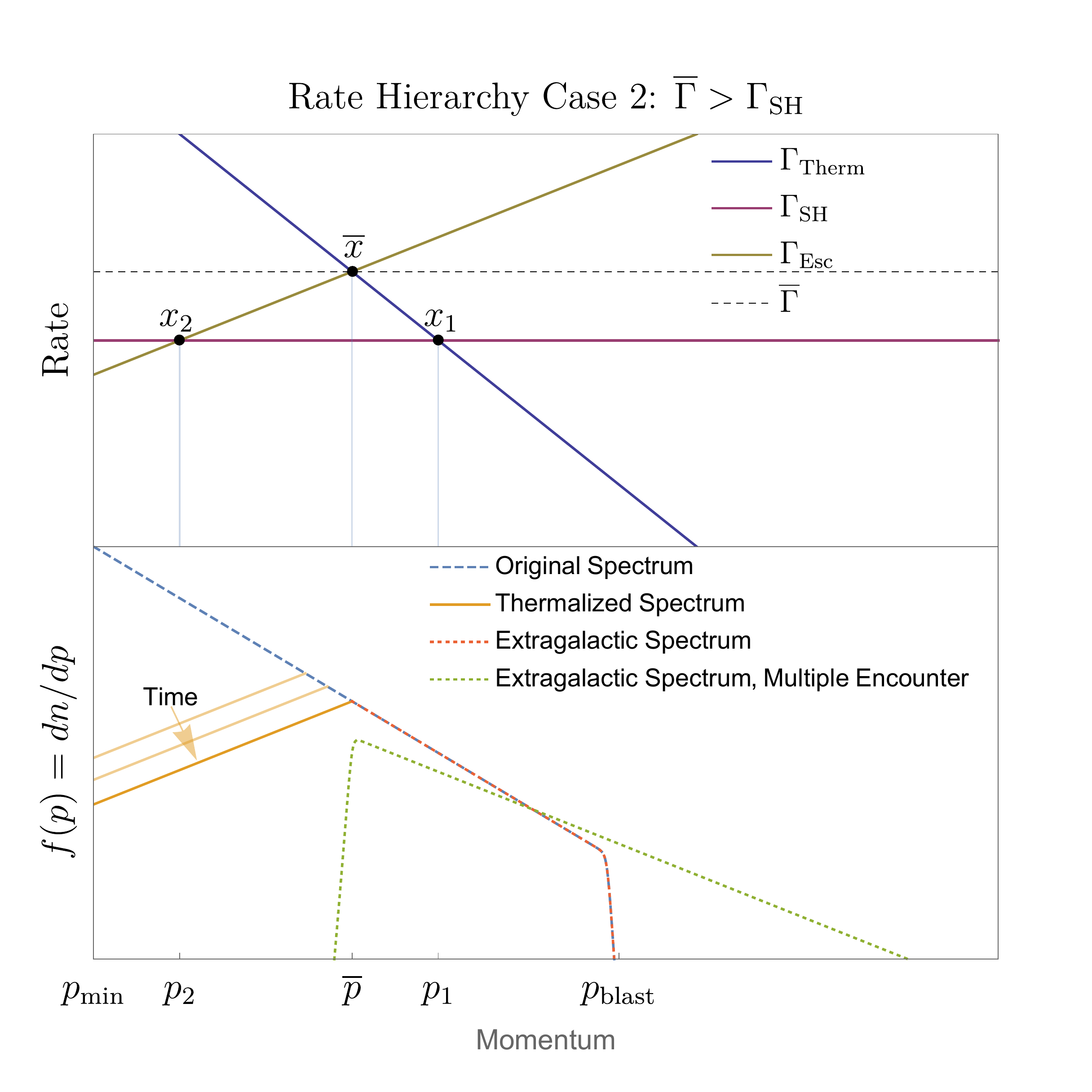} %
\caption{Comparison of the three key rates and the spectrum of accelerated CHAMPs for case 2.}
\label{fig:rateHierarchy2Plot}
\end{figure}

In case two galaxies, with rate orderings of Fig.~\ref{fig:rateHierarchy2Plot}, CHAMPs with momentum $p_2 < p < p_1$ generally enconter one shock before escape. Just as before, when a SN explodes, its Sedov-Taylor shock produces a batch of CHAMPs with a $p^{-3}$ differential spectrum as shown by the dashed blue line of Fig.~\ref{fig:rateHierarchy2Plot}. As time progresses, the numerous slower CHAMPs at the far left of the spectrum again thermalize first, and are converted to a $p^2$ spectrum which moves to the right with time. Concurrently, the scarce number of faster CHAMPs at the far right of the spectrum escape first, and are converted to a non-relativistic extragalactic spectrum which moves to the left with time. The last, and greatest number of CHAMPs to escape have momentum $p = \overline{p}$, the momentum when the escape and thermalization rates are equal. When this occurs at time $t \approx 1/\Gamma_{\rm{esc}}(\bar{p})$, the galactic and extragalactic spectrums are peaked at $p = \bar{p}$, with the galactic spectrum dropping as $p^2$ for $p < \bar{p}$ and the extragalactic spectrum dropping as $p^{-3}$ for $p > \bar{p}$, as shown by the orange and dotted red lines in Fig.~\ref{fig:rateHierarchy2Plot}. Finally, note that a fraction, $(p_2/p)^{3/2}$ of the escaping CHAMPs will encounter a second SN shock and be Fermi-accelerated before escaping. The probability is peaked at $p = \overline{p}$, which leads to a subdominant $p^{-2}$ extragalactic spectrum, as shown by the dotted green line in Fig.~\ref{fig:rateHierarchy2Plot}. Consequently, the final extragalactic spectrum produced from galaxies in case $2$ is given by the sum of the following two contributions,
\begin{equation}
f_2(p) = 
\left\{
\begin{aligned}
&   \frac{1}{2}N_X \overline{p}^2 \;\; \frac{1}{ p^3}                              && \quad     \text{ for $\overline{p} < p < p_{blast}$} \\
&   \frac{1}{2}N_X \overline{p} \left(\frac{p_2}{\overline{p}}\right)^{3/2}\;\; \frac{1}{ p^2}         && \quad     \text{ for $\overline{p} < p < p_{\rm{max}}$}.
\label{eq:EGSpectrum2}
\end{aligned}
\right.
\end{equation}

The key difference between galaxies of case 1 \eqref{eq:EGSpectrum1} and galaxies of case 2 \eqref{eq:EGSpectrum2} is the resulting relativistic extragalactic CHAMP spectrum. In case 1, CHAMPs faster than the thermalization bottleneck generally encounter an additional SN shock before escaping, resulting in a dominant $p^{-2}$ relativistic extragalactic spectrum. However, in case 2, CHAMPs faster than the thermalization bottleneck generally escape before encountering an additional SN shock, resulting in a dominant $p^{-3}$ non-relativistic extragalactic spectrum and a subdominant $p^{-2}$ relativistic spectrum.

As can be seen from \eqref{eq:x1}, \eqref{eq:xBar}, and \eqref{eq:x2}, the critical points $\{x_1, x_2, \overline{x}\}$ depend on various ratios of the disk volume, disk height, SN rate, and SN volume. Consequently, for a fixed  $(m,q)$, the momenta $p_1$, $p_2$, and $\overline{p}$ may vary between galaxies, implying each galaxy produces a potentially different $p^{-2}$ extragalactic spectrum. However, galaxies, unlike snowflakes, are not completely unique. Scaling relations allow us to estimate how the three fundamental rates depend on a given halo mass.

First, let us consider the disk radius. In current theories of disk formation, the baryonic mass in the halo which cools and falls cannot collapse to a central point due to angular momentum conservation. Instead, the baryons pancake into a disk with a radius proportional to the halo virial radius~\cite{2013fgu..book.....L,Mo:1997vb}. Thus, 
\begin{align}
   R_{\rm{disk}} \propto R_{\rm{vir}} \propto M^{1/3} ~ (1+z_{\rm{vir}})^{-1}
\end{align}
which is supported observationally~\cite{2013fgu..book.....L}.

Similarly, energy conservation dictates that the disk height $H_{\rm{disk}}$ is proportional to $\sigma^2 / \Sigma_{\rm{disk}}$, where $\sigma$ is the velocity dispersion of gas in the disk and $\Sigma_{\rm{disk}} = M_{\rm{disk}}/(2\pi R_{\rm{disk}}^2)$ the mass surface density. If the centrifugal force cannot balance the gravitational force, the disk is subject to fragmentation and collapse. This instability occurs when the Toomre parameter $Q = \sqrt{2}\sigma \Omega/ \pi G \Sigma_{\rm{disk}} < 1$~\cite{1964ApJ...139.1217T,2012MNRAS.421.3488H}, where $\Omega = v_c/R_{\rm{disk}}$ is the angular frequency. Amazingly, via self-regulation, star formation and SN feedback maintain $Q = 1$ over the disk~\cite{2012MNRAS.421.3488H}.%
\footnote{Disk regions where $Q > 1$ do not initially collapse but eventually cool and then collapse; hence $Q$ drops below $1$. Regions where $Q < 1$ collapse and form stars which inject energy so that $\sigma$ increases and collapse is halted; hence $Q$ rises above $1$.}
Combining these two conditions imply 
\begin{align}
    H_{\rm{disk}} \propto \frac{M_{\rm{disk}}}{v_c^2} \propto \frac{M_{\rm{disk}}}{M^{2/3}(1+z_{\rm{vir}})}     
\end{align}
where we assume the asymptotic circular speed of the disk $v_c$ is proportional to the virial speed.

To determine how the SN rate $\Gamma_{\rm{SN}}$ depends on galactic parameters, we first note that $\Gamma_{\rm{SN}}$ is dominated by core-collapse SN so that on cosmological timescales, there is little delay between the star formation rate and the SN rate and hence they are proportional. The global star formation rate per area is empirically observed to be proportional to $\Sigma_{\rm{gas}}^{3/2}$~\cite{2011piim.book.....D}, where $\Sigma_{\rm{gas}} = m_p n_{\rm{gas}} H_{\rm{disk}}$ is the gas surface density of the disk,%
\footnote{This relation is known as the Schmidt-Konneticut Law and is observed over a wide-range of galactic environments. One theoretical motivation for the $3/2$ power is the star formation rate should be proportional $m_{\rm{gas}}/t_{\rm{dyn}} \propto m_{\rm{gas}}\sqrt{G \rho} \propto m_{\rm{gas}}^{3/2}$, where $m_{\rm{gas}}$ is the mass of the gas in the disk.}
which implies
\begin{align}
    \Gamma_{\rm{SN}} \propto R_{\rm{disk}}^2 H_{\rm{disk}}^{3/2} n_{\rm{gas}}^{3/2}   
\end{align}
Last, since the SN radius goes as $n_{\rm{gas}}^{-1/3}$, the max SN volume goes as 
\begin{align}
    V_{\rm{SN}} \propto n_{\rm{gas}}^{-1}   
\end{align}
Putting this altogether, we find
\begin{alignat}{2}
    \Gamma_{\rm{SH}} \enskip &=  \enskip \frac{V_{\rm{SN}}\Gamma_{\rm{SN}}}{V_{\rm{disk}}} && \enskip \propto\enskip  \frac{M_{\rm{disk}}^{1/2}}{M^{1/3}} ~ (1+z_{\rm{vir}})^{-1/2} ~ n_{\rm{gas}}^{1/2}  \\
    \Gamma_{\rm{esc}} \enskip &= \enskip \frac{2D}{H_{\rm{disk}}^2} && \enskip \propto \enskip \frac{M^{4/3}}{M_{\rm{disk}}^2} ~(1+z_{\rm{vir}})^2 ~ D(R) \\
    \Gamma_{\rm{therm}}\enskip  &\approx \enskip \frac{f_{\rm{WIM}}}{t_{\rm{therm, WIM}}} && \enskip \propto \enskip f_{\rm{WIM}} ~ n_{\rm{WIM}} \label{therm1}
\end{alignat}
To simplify these rates further, we note: 
\begin{enumerate}
    \item  Simulations indicate that the number densities, temperatures, and filling factors of the ISM in other star-forming galaxies are very similar to our own (Tab. \ref{table:ISM}), being self-regulated by SN shocks~\cite{2012MNRAS.421.3488H}. Thus $f_{\rm{WIM}}$, $n_{\rm{WIM}}$ and $n_{\rm{gas}}$ are independent of galaxy and $\Gamma_{\rm{therm}}$ \eqref{therm1} is constant.
    \item Simulations show the turbulent $\sim {\rm \mu G}$ magnetic field in our disk is similar to those in other galaxies and forms very early in the formation of the disk. Thus we assume $D(R)$ is universal in other galaxies.
    \item Observations and simulations show that $M_{\rm{disk}} \propto M$ for halo masses $M \gtrsim 10^{11} M_\odot$, but falls much more steeply ($\geq \frac{3}{2}$ power) for lower mass halos~\cite{2015MNRAS.454.1105S}.%
    \footnote{Since the size and escape speed of these galaxies are small, one idea to explain their lack of gas (and hence stars and low luminosity) is the SN remnants from their first stars blew out of the disk and into the halo, expelling or disrupting the disk and halo gas and severely hampering subsequent star formation~\cite{dekel1986origin,2013fgu..book.....L}.}
    Therefore, in this lower mass halo regime, the SN encounter rate mildly drops as $M^{5/12}$ while the escape rate sharply rises as $M^{-5/3}$, implying CHAMPs in these disks are typically far in the case 2 regime and easily escape non-relativistically, just as the baryons do. Moreover, even  CHAMPs with a very small charge that remain in the case 1 regime for small mass galaxies still have an extragalactic spectrum dominated by high mass galaxies since $p_2$ is greater for heavier mass galaxies (and $p_1$ is nearly constant). Thus, we will only consider the extragalactic flux of CHAMPs from disks with halo masses $M \gtrsim 10^{11}M_\odot$, which at worst, underestimates the diffuse CHAMP flux.  Lighter galaxies can dominate $p < p_{2}(M=10^{11} M_\odot)$.
    \item Likewise, we do not consider halo masses $M \gtrsim 10^{12} M_\odot$ since such massive halos have difficulty cooling and cannot form disks within $10^{10} ~\rm{yr}$~\cite{1977MNRAS.179..541R}. 
    \item Since the relevant halo mass range to consider is $10^{11}M_\odot \lesssim M \lesssim 10^{12} M_\odot$, which typically virialize near $z_{\rm{vir}} = 2$ (see Fig.~\ref{fig:collapse}), we have $z_{\rm{vir}}$ a constant.
\end{enumerate}
Putting this all together, the three fundamental rates scale as
\begin{align}
    \Gamma_{\rm{SH}} &\propto M^{1/6} \approx \text{const} \\
    \Gamma_{\rm{esc}} &\propto M^{-2/3} \\
    \Gamma_{\rm{therm}} &\propto \text{const}
\end{align}
so that 
\begin{align}
    v_1 &=  900 ~{\rm km/s} \left(\frac{m/q^2}{10^6 ~\GeV}\right)^{-1/3} \left(\frac{M}{M_{MW}}\right)^{-1/18} \theta(10^4 ~{\rm km/s} - v_1) \label{eq:v1Final} \\
    \overline{v} &=  900 ~{\rm km/s} \left(\frac{m/q^2}{10^6 ~\GeV}\right)^{-1/3}  \left(\frac{M}{M_{MW}}\right)^{4/27} q^{-1/9} ~ \gamma(\overline{v})^{-1/9} \label{eq:vBarFinal} \\
    v_2 &=  \max\left\{{900 ~{\rm km/s} \left(\frac{m/q^2}{10^6 ~\GeV}\right)^{-1/3} \left(\frac{M}{M_{MW}}\right)^{5/9}  q^{-1/3} ~ \gamma(v_2)^{-1/3}, ~ v_{\rm{esc}}}\right \} \label{eq:v2Final}
\end{align}

After escaping from a galaxy, the CHAMPs diffuse into the intergalactic medium (IGM). If the CHAMPs traverse intergalactic distances ($R_{\rm{sep}} \sim 1 ~{\rm Mpc}$) within the lifetime of the universe, the extragalactic fluxes from different galaxies overlap and produce a steady-state, diffuse background of CHAMPs. That is, the CHAMPs ejected from galactic disks get smeared over the entire universe so that disk densities are essentially diluted by a factor $(V_{\rm{disk}}/R_{\rm{sep}}^3) \sim 10^{-8}$. More rigorously, consider the transport equation for the total extragalactic spectrum expelled from a disk
\begin{align}
    \frac{\partial f}{\partial t} = \nabla \cdot (D\nabla f)  
    \label{eq:IGMDifEqn}
\end{align}
where $D$ is the IGM diffusion constant, which may depend on particle rigidity and position. Taking the initial spectrum to be a point source in space and time (on cosmological scales),%
\footnote{The condition of the spectrum initially being localized in space and time relative to cosmological scales just simplifies the solution of \eqref{eq:IGMDifEqn} to be the Green's function. Since we are considering intergalactic distances on the Mpc scale, and typical disk sizes are kpc size, localization in space is a good approximation. Likewise, an order one number of the CHAMPs ejected from disks occurs in a billion years or so, which is less (though not \textit{much} less) than the age of the universe.}
the solution to \eqref{eq:IGMDifEqn} is
\begin{align}
    f(r,t) &= \frac{\mathcal{A}}{(4 \pi D t)^{3/2}} \exp \left(\frac{-r^2}{4 D t}\right)
    \label{eq:singleEGSpec}
\end{align}
where $\mathcal{A} = \int f d^3 x =  d N(p)/dp = (N_\upex p_2/p^2) \theta(p - p_2)$. $N_\upex$ is the number of CHAMPs evacuated from the disk.

If we take $r = 0$ to be the position of our Milky Way, the total observed CHAMP spectrum is the superposition of all other extragalactic spectra
\begin{align}
    f &= \int \frac{\mathcal{A}(M)}{(4 \pi D t)^{3/2}} \exp \left(\frac{-r^2}{4 D t}\right) 4 \pi r^2 dr ~ dn_{\rm{gal}}(M) \label{totEGSpec1}
\end{align}
where $4 \pi r^2 dr ~ dn_{\rm{gal}}(M)$ is the differential number of galaxies with halo mass $M$ a distance $r$ away. For now we assume that $dn_{\rm{gal}}(M)$ can be treated as continuous on the scale $\sqrt{4Dt}$.  By the Press-Schechter formalism, $dn_{\rm{gal}} = M_0 dM/M^2 ~{\rm (Mpc)^{-3}}$ where $M_0 \approx 10^9 M_\odot$~\cite{2013fgu..book.....L}, so that \eqref{totEGSpec1} becomes
\begin{align}
    f &=  \int \left(\frac{M}{m} f_\upex f_{D} \frac{p_2(M)}{p^2}\right)\left(\frac{\exp \left(-r^2/4 D t \right)}{(4 \pi D t)^{3/2}} \right)\left(4 \pi r^2 dr \frac{M_0}{M^2} dM ~{\rm Mpc^{-3}}\right) \nonumber \\
        &=  \int_{10^{11} M_\odot}^{10^{12} M_\odot} \frac{f_\upex f_{D}}{m} \frac{p_2(M)}{p^2} \frac{M_0}{M} dM ~ {\rm Mpc^{-3}} \nonumber
\end{align}
Note both time and the diffusion constant drop out from the integration over $r$ when going from the first to second line. 

Integrating over the massive galaxies which dominantly eject CHAMPs relativistically, we find a steady-state, diffuse CHAMP differential momentum spectrum
\begin{align}
    f = \frac{dn}{dp}  \approx 10^{-7} ~{\rm cm^{-3}} \left(\frac{m}{\GeV}\right)^{-1} f_{\upex} f_{D}\;  \mathcal{F} \; \frac{p_2(10^{12}M_\odot)}{p^2}
        \label{eq:totalEGSpectrum}
\end{align}
where $f_{\upex} = \Omega_\upex /\Omega_{\rm{DM}}$ is the fractional abundance of CHAMPs to dark matter, $f_{D}$ is the fraction of halo CHAMPs exposed to SNs, $\mathcal{F}$ is the fraction of CHAMPs ejected from the disk as given by (\ref{eq:frem}), and
$p_2 = \gamma m v_2$, with $v_2$ given by \eqref{eq:v2Final}. The fraction $f_D$ is determined by whether or not CHAMPs collapse into the disks and the diffusion into the disks,
\begin{align}
\label{eq:fD}
f_D = \begin{cases}
1/4 & m/q^2 < 10^5 {\rm GeV}, \\
10^{-3} \times \frac{J t_0}{n_0 H_d } & 10^5  {\rm GeV} < m /q^2,
\end{cases}
\end{align}
where the diffusion current $J$ is given in Eq.~(\ref{eq:J}). Eq. \eqref{eq:totalEGSpectrum} shows the extragalactic CHAMP signal is subdominant in comparison with the galactic CHAMP signal discussed in the main text.

Finally, we investigate the validity of \eqref{eq:totalEGSpectrum} for the slowest moving CHAMPs (with speeds near $v_2$), which may not be able to diffuse intergalactic distances within the age of the universe. The intergalactic magnetic field is not well known, though flux-freezing arguments suggest magnetic fields below a nanogauss permeate the IGM, with up to hundred nanogauss fields within large galactic clusters~\cite{Adams:1997ym,Aloisio:2004jda}. In contrast to the MW disk, it is unlikely that a turbulent (nanogauss) spectrum can form over $1 ~{\rm Mpc}$ intergalactic distances within the age of the universe~\cite{Aloisio:2004jda}, so that the magnetic field lines connecting nearby galaxies are taut and the mean free path $\lambda$ of CHAMPs is just the coherence length of the field $l_c \sim 1 ~{\rm Mpc}$, \textit{independent} of magnetic field strength~\cite{Adams:1997ym}. Thus the diffusion length of a CHAMP in the IGM is approximately
\begin{align}
    R_0 = \sqrt{2 D t} = \sqrt{2 t \lambda v /3} = 2 ~ {\rm Mpc} \left(\frac{\lambda}{1 ~{\rm Mpc}}\right)^{1/2}\left(\frac{t}{10^{10} ~{\rm yr}}\right)^{1/2} \left(\frac{v}{600 ~{\rm km/s}}\right)^{1/2}. \label{eq:IGMDiffLength}
\end{align}
Consequently, if the typical intergalactic magnetic field is non-turbulent, then expelled CHAMPS easily traverse intergalactic distances and \eqref{eq:totalEGSpectrum} should be a good approximation to the diffuse CHAMP background.%
\footnote{According to the Press-Schecter halo distribution function, the large spirals we consider are typically separated slightly farther apart than $2 ~{\rm Mpc}$. However, our galactic neighbor, Andromeda, is atypically close ($\sim .8 ~{\rm Mpc}$). In fact, the slow-moving CHAMP spectrum is likely dominated by Andromeda: Inserting \eqref{eq:IGMDiffLength} into \eqref{eq:singleEGSpec}, implies the low-speed CHAMPs escaping Andromeda produce a spectrum a few times greater than the diffuse background \eqref{eq:totalEGSpectrum}.}

On the other hand, if the intergalactic magnetic field $\textit{is}$ actually turbulent, then the CHAMP mean free path is generally much less than $1 ~{\rm Mpc}$. For a turbulent, Kolmogorov spectrum, the mean free path is~\cite{Aloisio:2004jda}
\begin{align}
    \lambda_{turb} \approx .75 ~ l_c \left(\frac{r_g}{l_c}\right)^{1/3} = .01 ~{\rm Mpc} \left(\frac{m/q}{10^6 ~ \GeV}\right)^{1/3} \left(\frac{\gamma v}{10^3 ~ {\rm km/s}}\right)^{1/3}  \left(\frac{B}{1 ~{\rm nG}}\right)^{-1/3} \left(\frac{l_c}{1 {~\rm Mpc}}\right)^{2/3} 
\end{align}
and hence the slowest moving CHAMPs expelled from disks are unable to reach the Milky Way within the lifetime of the universe.

\let\oldaddcontentsline\addcontentsline
\renewcommand{\addcontentsline}[3]{}

\bibliography{CHAMPS}

\begin{thebibliography}{84}%
\makeatletter
\providecommand \@ifxundefined [1]{%
 \@ifx{#1\undefined}
}%
\providecommand \@ifnum [1]{%
 \ifnum #1\expandafter \@firstoftwo
 \else \expandafter \@secondoftwo
 \fi
}%
\providecommand \@ifx [1]{%
 \ifx #1\expandafter \@firstoftwo
 \else \expandafter \@secondoftwo
 \fi
}%
\providecommand \natexlab [1]{#1}%
\providecommand \enquote  [1]{``#1''}%
\providecommand \bibnamefont  [1]{#1}%
\providecommand \bibfnamefont [1]{#1}%
\providecommand \citenamefont [1]{#1}%
\providecommand \href@noop [0]{\@secondoftwo}%
\providecommand \href [0]{\begingroup \@sanitize@url \@href}%
\providecommand \@href[1]{\@@startlink{#1}\@@href}%
\providecommand \@@href[1]{\endgroup#1\@@endlink}%
\providecommand \@sanitize@url [0]{\catcode `\\12\catcode `\$12\catcode
  `\&12\catcode `\#12\catcode `\^12\catcode `\_12\catcode `\%12\relax}%
\providecommand \@@startlink[1]{}%
\providecommand \@@endlink[0]{}%
\providecommand \url  [0]{\begingroup\@sanitize@url \@url }%
\providecommand \@url [1]{\endgroup\@href {#1}{\urlprefix }}%
\providecommand \urlprefix  [0]{URL }%
\providecommand \Eprint [0]{\href }%
\providecommand \doibase [0]{http://dx.doi.org/}%
\providecommand \selectlanguage [0]{\@gobble}%
\providecommand \bibinfo  [0]{\@secondoftwo}%
\providecommand \bibfield  [0]{\@secondoftwo}%
\providecommand \translation [1]{[#1]}%
\providecommand \BibitemOpen [0]{}%
\providecommand \bibitemStop [0]{}%
\providecommand \bibitemNoStop [0]{.\EOS\space}%
\providecommand \EOS [0]{\spacefactor3000\relax}%
\providecommand \BibitemShut  [1]{\csname bibitem#1\endcsname}%
\let\auto@bib@innerbib\@empty
\bibitem [{\citenamefont {Nomura}\ and\ \citenamefont
  {Tweedie}(2005)}]{Nomura:2005qg}%
  \BibitemOpen
  \bibfield  {author} {\bibinfo {author} {\bibfnamefont {Y.}~\bibnamefont
  {Nomura}}\ and\ \bibinfo {author} {\bibfnamefont {B.}~\bibnamefont
  {Tweedie}},\ }\href {\doibase 10.1103/PhysRevD.72.015006} {\bibfield
  {journal} {\bibinfo  {journal} {Phys. Rev.}\ }\textbf {\bibinfo {volume}
  {D72}},\ \bibinfo {pages} {015006} (\bibinfo {year} {2005})},\ \Eprint
  {http://arxiv.org/abs/hep-ph/0504246} {arXiv:hep-ph/0504246 [hep-ph]}
  \BibitemShut {NoStop}%
\bibitem [{\citenamefont {Harigaya}\ and\ \citenamefont
  {Nomura}(2016)}]{Harigaya:2016pnu}%
  \BibitemOpen
  \bibfield  {author} {\bibinfo {author} {\bibfnamefont {K.}~\bibnamefont
  {Harigaya}}\ and\ \bibinfo {author} {\bibfnamefont {Y.}~\bibnamefont
  {Nomura}},\ }\href {\doibase 10.1007/JHEP03(2016)091} {\bibfield  {journal}
  {\bibinfo  {journal} {JHEP}\ }\textbf {\bibinfo {volume} {03}},\ \bibinfo
  {pages} {091} (\bibinfo {year} {2016})},\ \Eprint
  {http://arxiv.org/abs/1602.01092} {arXiv:1602.01092 [hep-ph]} \BibitemShut
  {NoStop}%
\bibitem [{\citenamefont {De~Luca}\ \emph {et~al.}(2018)\citenamefont
  {De~Luca}, \citenamefont {Mitridate}, \citenamefont {Redi}, \citenamefont
  {Smirnov},\ and\ \citenamefont {Strumia}}]{DeLuca:2018mzn}%
  \BibitemOpen
  \bibfield  {author} {\bibinfo {author} {\bibfnamefont {V.}~\bibnamefont
  {De~Luca}}, \bibinfo {author} {\bibfnamefont {A.}~\bibnamefont {Mitridate}},
  \bibinfo {author} {\bibfnamefont {M.}~\bibnamefont {Redi}}, \bibinfo {author}
  {\bibfnamefont {J.}~\bibnamefont {Smirnov}}, \ and\ \bibinfo {author}
  {\bibfnamefont {A.}~\bibnamefont {Strumia}},\ }\href {\doibase
  10.1103/PhysRevD.97.115024} {\bibfield  {journal} {\bibinfo  {journal} {Phys.
  Rev.}\ }\textbf {\bibinfo {volume} {D97}},\ \bibinfo {pages} {115024}
  (\bibinfo {year} {2018})},\ \Eprint {http://arxiv.org/abs/1801.01135}
  {arXiv:1801.01135 [hep-ph]} \BibitemShut {NoStop}%
\bibitem [{\citenamefont {Dunsky}\ \emph {et~al.}(2019)\citenamefont {Dunsky},
  \citenamefont {Hall},\ and\ \citenamefont {Harigaya}}]{dunsky2019higgs}%
  \BibitemOpen
  \bibfield  {author} {\bibinfo {author} {\bibfnamefont {D.}~\bibnamefont
  {Dunsky}}, \bibinfo {author} {\bibfnamefont {L.~J.}\ \bibnamefont {Hall}}, \
  and\ \bibinfo {author} {\bibfnamefont {K.}~\bibnamefont {Harigaya}},\
  }\href@noop {} {\bibfield  {journal} {\bibinfo  {journal} {arXiv preprint
  arXiv:1902.07726}\ } (\bibinfo {year} {2019})}\BibitemShut {NoStop}%
\bibitem [{\citenamefont {Holdom}(1986)}]{Holdom:1985ag}%
  \BibitemOpen
  \bibfield  {author} {\bibinfo {author} {\bibfnamefont {B.}~\bibnamefont
  {Holdom}},\ }\href {\doibase 10.1016/0370-2693(86)91377-8} {\bibfield
  {journal} {\bibinfo  {journal} {Phys. Lett.}\ }\textbf {\bibinfo {volume}
  {166B}},\ \bibinfo {pages} {196} (\bibinfo {year} {1986})}\BibitemShut
  {NoStop}%
\bibitem [{\citenamefont {Goldberg}\ and\ \citenamefont
  {Hall}(1986)}]{Goldberg:1986nk}%
  \BibitemOpen
  \bibfield  {author} {\bibinfo {author} {\bibfnamefont {H.}~\bibnamefont
  {Goldberg}}\ and\ \bibinfo {author} {\bibfnamefont {L.~J.}\ \bibnamefont
  {Hall}},\ }\href {\doibase 10.1016/0370-2693(86)90731-8} {\bibfield
  {journal} {\bibinfo  {journal} {Phys. Lett.}\ }\textbf {\bibinfo {volume}
  {B174}},\ \bibinfo {pages} {151} (\bibinfo {year} {1986})}\BibitemShut
  {NoStop}%
\bibitem [{\citenamefont {De~Rujula}\ \emph {et~al.}(1990)\citenamefont
  {De~Rujula}, \citenamefont {Glashow},\ and\ \citenamefont
  {Sarid}}]{DeRujula:1989fe}%
  \BibitemOpen
  \bibfield  {author} {\bibinfo {author} {\bibfnamefont {A.}~\bibnamefont
  {De~Rujula}}, \bibinfo {author} {\bibfnamefont {S.~L.}\ \bibnamefont
  {Glashow}}, \ and\ \bibinfo {author} {\bibfnamefont {U.}~\bibnamefont
  {Sarid}},\ }\href {\doibase 10.1016/0550-3213(90)90227-5} {\bibfield
  {journal} {\bibinfo  {journal} {Nucl. Phys.}\ }\textbf {\bibinfo {volume}
  {B333}},\ \bibinfo {pages} {173} (\bibinfo {year} {1990})}\BibitemShut
  {NoStop}%
\bibitem [{\citenamefont {Dimopoulos}\ \emph {et~al.}(1990)\citenamefont
  {Dimopoulos}, \citenamefont {Eichler}, \citenamefont {Esmailzadeh},\ and\
  \citenamefont {Starkman}}]{Dimopoulos:1989hk}%
  \BibitemOpen
  \bibfield  {author} {\bibinfo {author} {\bibfnamefont {S.}~\bibnamefont
  {Dimopoulos}}, \bibinfo {author} {\bibfnamefont {D.}~\bibnamefont {Eichler}},
  \bibinfo {author} {\bibfnamefont {R.}~\bibnamefont {Esmailzadeh}}, \ and\
  \bibinfo {author} {\bibfnamefont {G.~D.}\ \bibnamefont {Starkman}},\ }\href
  {\doibase 10.1103/PhysRevD.41.2388} {\bibfield  {journal} {\bibinfo
  {journal} {Phys. Rev.}\ }\textbf {\bibinfo {volume} {D41}},\ \bibinfo {pages}
  {2388} (\bibinfo {year} {1990})}\BibitemShut {NoStop}%
\bibitem [{\citenamefont {Dobroliubov}\ and\ \citenamefont
  {Ignatiev}(1990)}]{Dobroliubov:1989mr}%
  \BibitemOpen
  \bibfield  {author} {\bibinfo {author} {\bibfnamefont {M.~I.}\ \bibnamefont
  {Dobroliubov}}\ and\ \bibinfo {author} {\bibfnamefont {A.~{\relax Yu}.}\
  \bibnamefont {Ignatiev}},\ }\href {\doibase 10.1103/PhysRevLett.65.679}
  {\bibfield  {journal} {\bibinfo  {journal} {Phys. Rev. Lett.}\ }\textbf
  {\bibinfo {volume} {65}},\ \bibinfo {pages} {679} (\bibinfo {year}
  {1990})}\BibitemShut {NoStop}%
\bibitem [{\citenamefont {Dubovsky}\ \emph {et~al.}(2004)\citenamefont
  {Dubovsky}, \citenamefont {Gorbunov},\ and\ \citenamefont
  {Rubtsov}}]{Dubovsky:2003yn}%
  \BibitemOpen
  \bibfield  {author} {\bibinfo {author} {\bibfnamefont {S.~L.}\ \bibnamefont
  {Dubovsky}}, \bibinfo {author} {\bibfnamefont {D.~S.}\ \bibnamefont
  {Gorbunov}}, \ and\ \bibinfo {author} {\bibfnamefont {G.~I.}\ \bibnamefont
  {Rubtsov}},\ }\href {\doibase 10.1134/1.1675909} {\bibfield  {journal}
  {\bibinfo  {journal} {JETP Lett.}\ }\textbf {\bibinfo {volume} {79}},\
  \bibinfo {pages} {1} (\bibinfo {year} {2004})},\ \bibinfo {note} {[Pisma Zh.
  Eksp. Teor. Fiz.79,3(2004)]},\ \Eprint {http://arxiv.org/abs/hep-ph/0311189}
  {arXiv:hep-ph/0311189 [hep-ph]} \BibitemShut {NoStop}%
\bibitem [{\citenamefont {Burrage}\ \emph {et~al.}(2009)\citenamefont
  {Burrage}, \citenamefont {Jaeckel}, \citenamefont {Redondo},\ and\
  \citenamefont {Ringwald}}]{Burrage:2009yz}%
  \BibitemOpen
  \bibfield  {author} {\bibinfo {author} {\bibfnamefont {C.}~\bibnamefont
  {Burrage}}, \bibinfo {author} {\bibfnamefont {J.}~\bibnamefont {Jaeckel}},
  \bibinfo {author} {\bibfnamefont {J.}~\bibnamefont {Redondo}}, \ and\
  \bibinfo {author} {\bibfnamefont {A.}~\bibnamefont {Ringwald}},\ }\href
  {\doibase 10.1088/1475-7516/2009/11/002} {\bibfield  {journal} {\bibinfo
  {journal} {JCAP}\ }\textbf {\bibinfo {volume} {0911}},\ \bibinfo {pages}
  {002} (\bibinfo {year} {2009})},\ \Eprint {http://arxiv.org/abs/0909.0649}
  {arXiv:0909.0649 [astro-ph.CO]} \BibitemShut {NoStop}%
\bibitem [{\citenamefont {McDermott}\ \emph {et~al.}(2011)\citenamefont
  {McDermott}, \citenamefont {Yu},\ and\ \citenamefont
  {Zurek}}]{McDermott:2010pa}%
  \BibitemOpen
  \bibfield  {author} {\bibinfo {author} {\bibfnamefont {S.~D.}\ \bibnamefont
  {McDermott}}, \bibinfo {author} {\bibfnamefont {H.-B.}\ \bibnamefont {Yu}}, \
  and\ \bibinfo {author} {\bibfnamefont {K.~M.}\ \bibnamefont {Zurek}},\ }\href
  {\doibase 10.1103/PhysRevD.83.063509} {\bibfield  {journal} {\bibinfo
  {journal} {Phys. Rev.}\ }\textbf {\bibinfo {volume} {D83}},\ \bibinfo {pages}
  {063509} (\bibinfo {year} {2011})},\ \Eprint {http://arxiv.org/abs/1011.2907}
  {arXiv:1011.2907 [hep-ph]} \BibitemShut {NoStop}%
\bibitem [{\citenamefont {Dolgov}\ \emph {et~al.}(2013)\citenamefont {Dolgov},
  \citenamefont {Dubovsky}, \citenamefont {Rubtsov},\ and\ \citenamefont
  {Tkachev}}]{Dolgov:2013una}%
  \BibitemOpen
  \bibfield  {author} {\bibinfo {author} {\bibfnamefont {A.~D.}\ \bibnamefont
  {Dolgov}}, \bibinfo {author} {\bibfnamefont {S.~L.}\ \bibnamefont
  {Dubovsky}}, \bibinfo {author} {\bibfnamefont {G.~I.}\ \bibnamefont
  {Rubtsov}}, \ and\ \bibinfo {author} {\bibfnamefont {I.~I.}\ \bibnamefont
  {Tkachev}},\ }\href {\doibase 10.1103/PhysRevD.88.117701} {\bibfield
  {journal} {\bibinfo  {journal} {Phys. Rev.}\ }\textbf {\bibinfo {volume}
  {D88}},\ \bibinfo {pages} {117701} (\bibinfo {year} {2013})},\ \Eprint
  {http://arxiv.org/abs/1310.2376} {arXiv:1310.2376 [hep-ph]} \BibitemShut
  {NoStop}%
\bibitem [{\citenamefont {Chuzhoy}\ and\ \citenamefont
  {Kolb}(2009)}]{Chuzhoy:2008zy}%
  \BibitemOpen
  \bibfield  {author} {\bibinfo {author} {\bibfnamefont {L.}~\bibnamefont
  {Chuzhoy}}\ and\ \bibinfo {author} {\bibfnamefont {E.~W.}\ \bibnamefont
  {Kolb}},\ }\href {\doibase 10.1088/1475-7516/2009/07/014} {\bibfield
  {journal} {\bibinfo  {journal} {JCAP}\ }\textbf {\bibinfo {volume} {0907}},\
  \bibinfo {pages} {014} (\bibinfo {year} {2009})},\ \Eprint
  {http://arxiv.org/abs/0809.0436} {arXiv:0809.0436 [astro-ph]} \BibitemShut
  {NoStop}%
\bibitem [{\citenamefont {{Peacock}}(1999)}]{1999coph.book.....P}%
  \BibitemOpen
  \bibfield  {author} {\bibinfo {author} {\bibfnamefont {J.~A.}\ \bibnamefont
  {{Peacock}}},\ }\href@noop {} {\emph {\bibinfo {title} {Cosmological
  Physics}}}\ (\bibinfo {year} {1999})\ p.\ \bibinfo {pages} {704}\BibitemShut
  {NoStop}%
\bibitem [{\citenamefont {{Rees}}\ and\ \citenamefont
  {{Ostriker}}(1977)}]{1977MNRAS.179..541R}%
  \BibitemOpen
  \bibfield  {author} {\bibinfo {author} {\bibfnamefont {M.~J.}\ \bibnamefont
  {{Rees}}}\ and\ \bibinfo {author} {\bibfnamefont {J.~P.}\ \bibnamefont
  {{Ostriker}}},\ }\href {\doibase 10.1093/mnras/179.4.541} {\bibfield
  {journal} {\bibinfo  {journal} {Monthly Notices of the Royal Astronomical
  Society}\ }\textbf {\bibinfo {volume} {179}},\ \bibinfo {pages} {541}
  (\bibinfo {year} {1977})}\BibitemShut {NoStop}%
\bibitem [{\citenamefont {Loeb}(2010)}]{loeb2010}%
  \BibitemOpen
  \bibfield  {author} {\bibinfo {author} {\bibfnamefont {A.}~\bibnamefont
  {Loeb}},\ }\href {http://www.jstor.org/stable/j.ctt7rgcd} {\emph {\bibinfo
  {title} {How Did the First Stars and Galaxies Form?}}}\ (\bibinfo
  {publisher} {Princeton University Press},\ \bibinfo {year}
  {2010})\BibitemShut {NoStop}%
\bibitem [{\citenamefont {{Mo}}\ \emph {et~al.}(2010)\citenamefont {{Mo}},
  \citenamefont {{van den Bosch}},\ and\ \citenamefont
  {{White}}}]{2010gfe..book.....M}%
  \BibitemOpen
  \bibfield  {author} {\bibinfo {author} {\bibfnamefont {H.}~\bibnamefont
  {{Mo}}}, \bibinfo {author} {\bibfnamefont {F.~C.}\ \bibnamefont {{van den
  Bosch}}}, \ and\ \bibinfo {author} {\bibfnamefont {S.}~\bibnamefont
  {{White}}},\ }\href@noop {} {\emph {\bibinfo {title} {Galaxy Formation and
  Evolution, Cambridge, UK: Cambridge University Press}}}\ (\bibinfo {year}
  {2010})\BibitemShut {NoStop}%
\bibitem [{\citenamefont {{Weinberg}}\ \emph {et~al.}(1997)\citenamefont
  {{Weinberg}}, \citenamefont {{Hernquist}},\ and\ \citenamefont
  {{Katz}}}]{1997ApJ...477....8W}%
  \BibitemOpen
  \bibfield  {author} {\bibinfo {author} {\bibfnamefont {D.~H.}\ \bibnamefont
  {{Weinberg}}}, \bibinfo {author} {\bibfnamefont {L.}~\bibnamefont
  {{Hernquist}}}, \ and\ \bibinfo {author} {\bibfnamefont {N.}~\bibnamefont
  {{Katz}}},\ }\href {\doibase 10.1086/303683} {\bibfield  {journal} {\bibinfo
  {journal} {\apj}\ }\textbf {\bibinfo {volume} {477}},\ \bibinfo {pages} {8}
  (\bibinfo {year} {1997})},\ \Eprint {http://arxiv.org/abs/astro-ph/9604175}
  {arXiv:astro-ph/9604175 [astro-ph]} \BibitemShut {NoStop}%
\bibitem [{\citenamefont {Loeb}(2008)}]{Loeb:2006za}%
  \BibitemOpen
  \bibfield  {author} {\bibinfo {author} {\bibfnamefont {A.}~\bibnamefont
  {Loeb}},\ }in\ \href {\doibase 10.1007/978-3-540-74163-3_1} {\emph {\bibinfo
  {booktitle} {First Light in the Universe}}}\ (\bibinfo {year} {2008})\ pp.\
  \bibinfo {pages} {1--159},\ \Eprint {http://arxiv.org/abs/astro-ph/0603360}
  {arXiv:astro-ph/0603360 [astro-ph]} \BibitemShut {NoStop}%
\bibitem [{\citenamefont {{Draine}}(2011)}]{2011piim.book.....D}%
  \BibitemOpen
  \bibfield  {author} {\bibinfo {author} {\bibfnamefont {B.~T.}\ \bibnamefont
  {{Draine}}},\ }\href@noop {} {\emph {\bibinfo {title} {Physics of the
  Interstellar and Intergalactic Medium by Bruce T.~Draine.~Princeton
  University Press, 2011.~ISBN: 978-0-691-12214-4}}}\ (\bibinfo {year}
  {2011})\BibitemShut {NoStop}%
\bibitem [{\citenamefont {Spitzer}(1956)}]{Spitzer:1956hha}%
  \BibitemOpen
  \bibfield  {author} {\bibinfo {author} {\bibfnamefont {L.}~\bibnamefont
  {Spitzer}, \bibfnamefont {Jr.}},\ }\href@noop {} {\emph {\bibinfo {title}
  {{Physics of Fully Ionized Gases}}}}\ (\bibinfo  {publisher} {Interscience
  Publishers},\ \bibinfo {address} {New York},\ \bibinfo {year}
  {1956})\BibitemShut {NoStop}%
\bibitem [{\citenamefont {Mo}\ \emph {et~al.}(1998)\citenamefont {Mo},
  \citenamefont {Mao},\ and\ \citenamefont {White}}]{Mo:1997vb}%
  \BibitemOpen
  \bibfield  {author} {\bibinfo {author} {\bibfnamefont {H.~J.}\ \bibnamefont
  {Mo}}, \bibinfo {author} {\bibfnamefont {S.}~\bibnamefont {Mao}}, \ and\
  \bibinfo {author} {\bibfnamefont {S.~D.~M.}\ \bibnamefont {White}},\ }\href
  {\doibase 10.1046/j.1365-8711.1998.01227.x} {\bibfield  {journal} {\bibinfo
  {journal} {Mon. Not. Roy. Astron. Soc.}\ }\textbf {\bibinfo {volume} {295}},\
  \bibinfo {pages} {319} (\bibinfo {year} {1998})},\ \Eprint
  {http://arxiv.org/abs/astro-ph/9707093} {arXiv:astro-ph/9707093 [astro-ph]}
  \BibitemShut {NoStop}%
\bibitem [{\citenamefont {Navarro}\ \emph {et~al.}(1996)\citenamefont
  {Navarro}, \citenamefont {Frenk},\ and\ \citenamefont
  {White}}]{Navarro:1995iw}%
  \BibitemOpen
  \bibfield  {author} {\bibinfo {author} {\bibfnamefont {J.~F.}\ \bibnamefont
  {Navarro}}, \bibinfo {author} {\bibfnamefont {C.~S.}\ \bibnamefont {Frenk}},
  \ and\ \bibinfo {author} {\bibfnamefont {S.~D.~M.}\ \bibnamefont {White}},\
  }\href {\doibase 10.1086/177173} {\bibfield  {journal} {\bibinfo  {journal}
  {Astrophys. J.}\ }\textbf {\bibinfo {volume} {462}},\ \bibinfo {pages} {563}
  (\bibinfo {year} {1996})},\ \Eprint {http://arxiv.org/abs/astro-ph/9508025}
  {arXiv:astro-ph/9508025 [astro-ph]} \BibitemShut {NoStop}%
\bibitem [{\citenamefont {{Noh}}\ and\ \citenamefont
  {{McQuinn}}(2014)}]{2014MNRAS.444..503N}%
  \BibitemOpen
  \bibfield  {author} {\bibinfo {author} {\bibfnamefont {Y.}~\bibnamefont
  {{Noh}}}\ and\ \bibinfo {author} {\bibfnamefont {M.}~\bibnamefont
  {{McQuinn}}},\ }\href {\doibase 10.1093/mnras/stu1412} {\bibfield  {journal}
  {\bibinfo  {journal} {mnras}\ }\textbf {\bibinfo {volume} {444}},\ \bibinfo
  {pages} {503} (\bibinfo {year} {2014})},\ \Eprint
  {http://arxiv.org/abs/1401.0737} {arXiv:1401.0737} \BibitemShut {NoStop}%
\bibitem [{\citenamefont {{Altay}}\ \emph {et~al.}(2011)\citenamefont
  {{Altay}}, \citenamefont {{Theuns}}, \citenamefont {{Schaye}}, \citenamefont
  {{Crighton}},\ and\ \citenamefont {{Dalla Vecchia}}}]{2011ApJ...737L..37A}%
  \BibitemOpen
  \bibfield  {author} {\bibinfo {author} {\bibfnamefont {G.}~\bibnamefont
  {{Altay}}}, \bibinfo {author} {\bibfnamefont {T.}~\bibnamefont {{Theuns}}},
  \bibinfo {author} {\bibfnamefont {J.}~\bibnamefont {{Schaye}}}, \bibinfo
  {author} {\bibfnamefont {N.~H.~M.}\ \bibnamefont {{Crighton}}}, \ and\
  \bibinfo {author} {\bibfnamefont {C.}~\bibnamefont {{Dalla Vecchia}}},\
  }\href {\doibase 10.1088/2041-8205/737/2/L37} {\bibfield  {journal} {\bibinfo
   {journal} {apjl}\ }\textbf {\bibinfo {volume} {737}},\ \bibinfo {pages}
  {L37} (\bibinfo {year} {2011})},\ \Eprint {http://arxiv.org/abs/1012.4014}
  {arXiv:1012.4014 [astro-ph.CO]} \BibitemShut {NoStop}%
\bibitem [{\citenamefont {{Schaye}}(2001)}]{2001ApJ...559..507S}%
  \BibitemOpen
  \bibfield  {author} {\bibinfo {author} {\bibfnamefont {J.}~\bibnamefont
  {{Schaye}}},\ }\href {\doibase 10.1086/322421} {\bibfield  {journal}
  {\bibinfo  {journal} {\apj}\ }\textbf {\bibinfo {volume} {559}},\ \bibinfo
  {pages} {507} (\bibinfo {year} {2001})},\ \Eprint
  {http://arxiv.org/abs/astro-ph/0104272} {astro-ph/0104272} \BibitemShut
  {NoStop}%
\bibitem [{\citenamefont {{Mesinger}}(2016)}]{2016ASSL..423.....M}%
  \BibitemOpen
  \bibinfo {editor} {\bibfnamefont {A.}~\bibnamefont {{Mesinger}}},\ ed.,\
  \href {\doibase 10.1007/978-3-319-21957-8} {\emph {\bibinfo {title}
  {Understanding the Epoch of Cosmic Reionization: Challenges and Progress}}},\
  \bibinfo {series} {Astrophysics and Space Science Library}, Vol.\ \bibinfo
  {volume} {423}\ (\bibinfo {year} {2016})\BibitemShut {NoStop}%
\bibitem [{\citenamefont {{Stahler}}\ and\ \citenamefont
  {{Palla}}(2005)}]{2005fost.book.....S}%
  \BibitemOpen
  \bibfield  {author} {\bibinfo {author} {\bibfnamefont {S.~W.}\ \bibnamefont
  {{Stahler}}}\ and\ \bibinfo {author} {\bibfnamefont {F.}~\bibnamefont
  {{Palla}}},\ }\href@noop {} {\emph {\bibinfo {title} {The Formation of
  Stars.~Wiley-VCH}}}\ (\bibinfo {year} {2005})\BibitemShut {NoStop}%
\bibitem [{\citenamefont {McKee}\ and\ \citenamefont
  {Ostriker}(1977)}]{McKee:1977dz}%
  \BibitemOpen
  \bibfield  {author} {\bibinfo {author} {\bibfnamefont {C.~F.}\ \bibnamefont
  {McKee}}\ and\ \bibinfo {author} {\bibfnamefont {J.~P.}\ \bibnamefont
  {Ostriker}},\ }\href {\doibase 10.1086/155667} {\bibfield  {journal}
  {\bibinfo  {journal} {Astrophys. J.}\ }\textbf {\bibinfo {volume} {218}},\
  \bibinfo {pages} {148} (\bibinfo {year} {1977})}\BibitemShut {NoStop}%
\bibitem [{\citenamefont {{Martizzi}}\ \emph {et~al.}(2015)\citenamefont
  {{Martizzi}}, \citenamefont {{Faucher-Gigu{\`e}re}},\ and\ \citenamefont
  {{Quataert}}}]{2015MNRAS.450..504M}%
  \BibitemOpen
  \bibfield  {author} {\bibinfo {author} {\bibfnamefont {D.}~\bibnamefont
  {{Martizzi}}}, \bibinfo {author} {\bibfnamefont {C.-A.}\ \bibnamefont
  {{Faucher-Gigu{\`e}re}}}, \ and\ \bibinfo {author} {\bibfnamefont
  {E.}~\bibnamefont {{Quataert}}},\ }\href {\doibase 10.1093/mnras/stv562}
  {\bibfield  {journal} {\bibinfo  {journal} {mnras}\ }\textbf {\bibinfo
  {volume} {450}},\ \bibinfo {pages} {504} (\bibinfo {year} {2015})},\ \Eprint
  {http://arxiv.org/abs/1409.4425} {arXiv:1409.4425 [astro-ph.GA]} \BibitemShut
  {NoStop}%
\bibitem [{\citenamefont {{Slavin}}\ \emph {et~al.}(2015)\citenamefont
  {{Slavin}}, \citenamefont {{Dwek}},\ and\ \citenamefont
  {{Jones}}}]{2015ApJ...803....7S}%
  \BibitemOpen
  \bibfield  {author} {\bibinfo {author} {\bibfnamefont {J.~D.}\ \bibnamefont
  {{Slavin}}}, \bibinfo {author} {\bibfnamefont {E.}~\bibnamefont {{Dwek}}}, \
  and\ \bibinfo {author} {\bibfnamefont {A.~P.}\ \bibnamefont {{Jones}}},\
  }\href {\doibase 10.1088/0004-637X/803/1/7} {\bibfield  {journal} {\bibinfo
  {journal} {apj}\ }\textbf {\bibinfo {volume} {803}},\ \bibinfo {eid} {7}
  (\bibinfo {year} {2015})},\ \Eprint {http://arxiv.org/abs/1502.00929}
  {arXiv:1502.00929 [astro-ph.GA]} \BibitemShut {NoStop}%
\bibitem [{\citenamefont {{Yamazaki}}\ \emph {et~al.}(2006)\citenamefont
  {{Yamazaki}}, \citenamefont {{Kohri}}, \citenamefont {{Bamba}}, \citenamefont
  {{Yoshida}}, \citenamefont {{Tsuribe}},\ and\ \citenamefont
  {{Takahara}}}]{2006MNRAS.371.1975Y}%
  \BibitemOpen
  \bibfield  {author} {\bibinfo {author} {\bibfnamefont {R.}~\bibnamefont
  {{Yamazaki}}}, \bibinfo {author} {\bibfnamefont {K.}~\bibnamefont {{Kohri}}},
  \bibinfo {author} {\bibfnamefont {A.}~\bibnamefont {{Bamba}}}, \bibinfo
  {author} {\bibfnamefont {T.}~\bibnamefont {{Yoshida}}}, \bibinfo {author}
  {\bibfnamefont {T.}~\bibnamefont {{Tsuribe}}}, \ and\ \bibinfo {author}
  {\bibfnamefont {F.}~\bibnamefont {{Takahara}}},\ }\href {\doibase
  10.1111/j.1365-2966.2006.10832.x} {\ \textbf {\bibinfo {volume} {371}},\
  \bibinfo {pages} {1975} (\bibinfo {year} {2006})},\ \Eprint
  {http://arxiv.org/abs/astro-ph/0601704} {astro-ph/0601704} \BibitemShut
  {NoStop}%
\bibitem [{\citenamefont {{Bell}}(2013)}]{2013APh....43...56B}%
  \BibitemOpen
  \bibfield  {author} {\bibinfo {author} {\bibfnamefont {A.~R.}\ \bibnamefont
  {{Bell}}},\ }\href {\doibase 10.1016/j.astropartphys.2012.05.022} {\bibfield
  {journal} {\bibinfo  {journal} {Astroparticle Physics}\ }\textbf {\bibinfo
  {volume} {43}},\ \bibinfo {pages} {56} (\bibinfo {year} {2013})}\BibitemShut
  {NoStop}%
\bibitem [{\citenamefont {{Thompson}}\ \emph {et~al.}(2009)\citenamefont
  {{Thompson}}, \citenamefont {{Quataert}},\ and\ \citenamefont
  {{Murray}}}]{2009MNRAS.397.1410T}%
  \BibitemOpen
  \bibfield  {author} {\bibinfo {author} {\bibfnamefont {T.~A.}\ \bibnamefont
  {{Thompson}}}, \bibinfo {author} {\bibfnamefont {E.}~\bibnamefont
  {{Quataert}}}, \ and\ \bibinfo {author} {\bibfnamefont {N.}~\bibnamefont
  {{Murray}}},\ }\href {\doibase 10.1111/j.1365-2966.2009.14889.x} {\bibfield
  {journal} {\bibinfo  {journal} {mnras}\ }\textbf {\bibinfo {volume} {397}},\
  \bibinfo {pages} {1410} (\bibinfo {year} {2009})},\ \Eprint
  {http://arxiv.org/abs/0902.1755} {arXiv:0902.1755 [astro-ph.HE]} \BibitemShut
  {NoStop}%
\bibitem [{\citenamefont {Longair}(1981)}]{Longair:1981jc}%
  \BibitemOpen
  \bibfield  {author} {\bibinfo {author} {\bibfnamefont {M.~S.}\ \bibnamefont
  {Longair}},\ }\href@noop {} {\emph {\bibinfo {title} {High energy
  astrophysics}}}\ (\bibinfo  {publisher} {Cambridge [England] ; New York :
  Cambridge University Press, 1981},\ \bibinfo {year} {1981})\BibitemShut
  {NoStop}%
\bibitem [{\citenamefont {Jones}\ \emph {et~al.}(2001)\citenamefont {Jones},
  \citenamefont {Lukasiak}, \citenamefont {Ptuskin},\ and\ \citenamefont
  {Webber}}]{Jones:2000qd}%
  \BibitemOpen
  \bibfield  {author} {\bibinfo {author} {\bibfnamefont {F.~C.}\ \bibnamefont
  {Jones}}, \bibinfo {author} {\bibfnamefont {A.}~\bibnamefont {Lukasiak}},
  \bibinfo {author} {\bibfnamefont {V.}~\bibnamefont {Ptuskin}}, \ and\
  \bibinfo {author} {\bibfnamefont {W.}~\bibnamefont {Webber}},\ }\href
  {\doibase 10.1086/318358} {\bibfield  {journal} {\bibinfo  {journal}
  {Astrophys. J.}\ }\textbf {\bibinfo {volume} {547}},\ \bibinfo {pages} {264}
  (\bibinfo {year} {2001})},\ \Eprint {http://arxiv.org/abs/astro-ph/0007293}
  {arXiv:astro-ph/0007293 [astro-ph]} \BibitemShut {NoStop}%
\bibitem [{\citenamefont {Strong}\ \emph {et~al.}(2007)\citenamefont {Strong},
  \citenamefont {Moskalenko},\ and\ \citenamefont {Ptuskin}}]{Strong:2007nh}%
  \BibitemOpen
  \bibfield  {author} {\bibinfo {author} {\bibfnamefont {A.~W.}\ \bibnamefont
  {Strong}}, \bibinfo {author} {\bibfnamefont {I.~V.}\ \bibnamefont
  {Moskalenko}}, \ and\ \bibinfo {author} {\bibfnamefont {V.~S.}\ \bibnamefont
  {Ptuskin}},\ }\href {\doibase 10.1146/annurev.nucl.57.090506.123011}
  {\bibfield  {journal} {\bibinfo  {journal} {Ann. Rev. Nucl. Part. Sci.}\
  }\textbf {\bibinfo {volume} {57}},\ \bibinfo {pages} {285} (\bibinfo {year}
  {2007})},\ \Eprint {http://arxiv.org/abs/astro-ph/0701517}
  {arXiv:astro-ph/0701517 [astro-ph]} \BibitemShut {NoStop}%
\bibitem [{\citenamefont {{Kulsrud}}(2005)}]{2005ppfa.book.....K}%
  \BibitemOpen
  \bibfield  {author} {\bibinfo {author} {\bibfnamefont {R.~M.}\ \bibnamefont
  {{Kulsrud}}},\ }\href@noop {} {\emph {\bibinfo {title} {Plasma physics for
  astrophysics}}}\ (\bibinfo {year} {2005})\BibitemShut {NoStop}%
\bibitem [{\citenamefont {Drury}(1983)}]{drury1983introduction}%
  \BibitemOpen
  \bibfield  {author} {\bibinfo {author} {\bibfnamefont {L.~O.}\ \bibnamefont
  {Drury}},\ }\href@noop {} {\bibfield  {journal} {\bibinfo  {journal} {Reports
  on Progress in Physics}\ }\textbf {\bibinfo {volume} {46}},\ \bibinfo {pages}
  {973} (\bibinfo {year} {1983})}\BibitemShut {NoStop}%
\bibitem [{\citenamefont {Bell}(1978)}]{bell1978acceleration2}%
  \BibitemOpen
  \bibfield  {author} {\bibinfo {author} {\bibfnamefont {A.}~\bibnamefont
  {Bell}},\ }\href@noop {} {\bibfield  {journal} {\bibinfo  {journal} {Monthly
  Notices of the Royal Astronomical Society}\ }\textbf {\bibinfo {volume}
  {182}},\ \bibinfo {pages} {443} (\bibinfo {year} {1978})}\BibitemShut
  {NoStop}%
\bibitem [{\citenamefont {{Hu}}\ \emph {et~al.}(2017)\citenamefont {{Hu}},
  \citenamefont {{Kusenko}},\ and\ \citenamefont
  {{Takhistov}}}]{2017PhLB..768...18H}%
  \BibitemOpen
  \bibfield  {author} {\bibinfo {author} {\bibfnamefont {P.-K.}\ \bibnamefont
  {{Hu}}}, \bibinfo {author} {\bibfnamefont {A.}~\bibnamefont {{Kusenko}}}, \
  and\ \bibinfo {author} {\bibfnamefont {V.}~\bibnamefont {{Takhistov}}},\
  }\href {\doibase 10.1016/j.physletb.2017.02.035} {\bibfield  {journal}
  {\bibinfo  {journal} {Physics Letters B}\ }\textbf {\bibinfo {volume}
  {768}},\ \bibinfo {pages} {18} (\bibinfo {year} {2017})},\ \Eprint
  {http://arxiv.org/abs/1611.04599} {arXiv:1611.04599 [hep-ph]} \BibitemShut
  {NoStop}%
\bibitem [{\citenamefont {{Wu}}\ \emph {et~al.}(1997)\citenamefont {{Wu}},
  \citenamefont {{Yoon}},\ and\ \citenamefont {{Chao}}}]{1997PhPl....4..856W}%
  \BibitemOpen
  \bibfield  {author} {\bibinfo {author} {\bibfnamefont {C.~S.}\ \bibnamefont
  {{Wu}}}, \bibinfo {author} {\bibfnamefont {P.~H.}\ \bibnamefont {{Yoon}}}, \
  and\ \bibinfo {author} {\bibfnamefont {J.~K.}\ \bibnamefont {{Chao}}},\
  }\href {\doibase 10.1063/1.872176} {\bibfield  {journal} {\bibinfo  {journal}
  {Physics of Plasmas}\ }\textbf {\bibinfo {volume} {4}},\ \bibinfo {pages}
  {856} (\bibinfo {year} {1997})}\BibitemShut {NoStop}%
\bibitem [{\citenamefont {Gaensler}\ \emph {et~al.}(2008)\citenamefont
  {Gaensler}, \citenamefont {Madsen}, \citenamefont {Chatterjee},\ and\
  \citenamefont {Mao}}]{Gaensler:2008ec}%
  \BibitemOpen
  \bibfield  {author} {\bibinfo {author} {\bibfnamefont {B.~M.}\ \bibnamefont
  {Gaensler}}, \bibinfo {author} {\bibfnamefont {G.~J.}\ \bibnamefont
  {Madsen}}, \bibinfo {author} {\bibfnamefont {S.}~\bibnamefont {Chatterjee}},
  \ and\ \bibinfo {author} {\bibfnamefont {S.~A.}\ \bibnamefont {Mao}},\ }\href
  {\doibase 10.1071/AS08004} {\bibfield  {journal} {\bibinfo  {journal} {Publ.
  Astron. Soc. Austral.}\ }\textbf {\bibinfo {volume} {25}},\ \bibinfo {pages}
  {184} (\bibinfo {year} {2008})},\ \Eprint {http://arxiv.org/abs/0808.2550}
  {arXiv:0808.2550 [astro-ph]} \BibitemShut {NoStop}%
\bibitem [{\citenamefont {Longair}(2011)}]{ucb.b1836189520110101}%
  \BibitemOpen
  \bibfield  {author} {\bibinfo {author} {\bibfnamefont {M.~S.}\ \bibnamefont
  {Longair}},\ }\href
  {https://libproxy.berkeley.edu/login?qurl=http%3a%2f%2fsearch.ebscohost.com%2flogin.aspx%3fdirect%3dtrue%26db%3dcat04202a%26AN%3ducb.b18361895%26site%3deds-live}
  {\emph {\bibinfo {title} {High energy astrophysics}}}\ (\bibinfo  {publisher}
  {Cambridge, UK ; New York : Cambridge University Press, 2011.},\ \bibinfo
  {year} {2011})\BibitemShut {NoStop}%
\bibitem [{\citenamefont {Meyer}(1969)}]{Meyer:1969we}%
  \BibitemOpen
  \bibfield  {author} {\bibinfo {author} {\bibfnamefont {P.}~\bibnamefont
  {Meyer}},\ }\href {\doibase 10.1146/annurev.aa.07.090169.000245} {\bibfield
  {journal} {\bibinfo  {journal} {Ann. Rev. Astron. Astrophys.}\ }\textbf
  {\bibinfo {volume} {7}},\ \bibinfo {pages} {1} (\bibinfo {year}
  {1969})}\BibitemShut {NoStop}%
\bibitem [{\citenamefont {{Jokipii}}(1971)}]{1971RvGSP...9...27J}%
  \BibitemOpen
  \bibfield  {author} {\bibinfo {author} {\bibfnamefont {J.~R.}\ \bibnamefont
  {{Jokipii}}},\ }\href {\doibase 10.1029/RG009i001p00027} {\bibfield
  {journal} {\bibinfo  {journal} {Reviews of Geophysics and Space Physics}\
  }\textbf {\bibinfo {volume} {9}},\ \bibinfo {pages} {27} (\bibinfo {year}
  {1971})}\BibitemShut {NoStop}%
\bibitem [{\citenamefont {Ziegler}(1988)}]{ZIEGLER19883}%
  \BibitemOpen
  \bibfield  {author} {\bibinfo {author} {\bibfnamefont {J.}~\bibnamefont
  {Ziegler}},\ }in\ \href {\doibase
  https://doi.org/10.1016/B978-0-12-780621-1.50005-8} {\emph {\bibinfo
  {booktitle} {Ion Implantation Science and Technology (Second Edition)}}},\
  \bibinfo {editor} {edited by\ \bibinfo {editor} {\bibfnamefont
  {J.}~\bibnamefont {Ziegler}}}\ (\bibinfo  {publisher} {Academic Press},\
  \bibinfo {year} {1988})\ \bibinfo {edition} {second edition}\ ed.,\ pp.\
  \bibinfo {pages} {3 -- 61}\BibitemShut {NoStop}%
\bibitem [{\citenamefont {Berger}\ \emph {et~al.}(2017)\citenamefont {Berger},
  \citenamefont {Coursey}, \citenamefont {Zucker},\ and\ \citenamefont
  {Chang}}]{Berger:399381}%
  \BibitemOpen
  \bibfield  {author} {\bibinfo {author} {\bibfnamefont {M.}~\bibnamefont
  {Berger}}, \bibinfo {author} {\bibfnamefont {J.~S.}\ \bibnamefont {Coursey}},
  \bibinfo {author} {\bibfnamefont {M.~A.}\ \bibnamefont {Zucker}}, \ and\
  \bibinfo {author} {\bibfnamefont {J.}~\bibnamefont {Chang}},\ }\href
  {\doibase 10.18434/T4NC7P} {\emph {\bibinfo {title} {{Stopping-Power and
  Range Tables for Electrons, Protons, and Helium Ions}}}}\ (\bibinfo
  {publisher} {NIST},\ \bibinfo {year} {2017})\BibitemShut {NoStop}%
\bibitem [{\citenamefont {Helm}(1956)}]{Helm:1956zz}%
  \BibitemOpen
  \bibfield  {author} {\bibinfo {author} {\bibfnamefont {R.~H.}\ \bibnamefont
  {Helm}},\ }\href {\doibase 10.1103/PhysRev.104.1466} {\bibfield  {journal}
  {\bibinfo  {journal} {Phys. Rev.}\ }\textbf {\bibinfo {volume} {104}},\
  \bibinfo {pages} {1466} (\bibinfo {year} {1956})}\BibitemShut {NoStop}%
\bibitem [{\citenamefont {Lewin}\ and\ \citenamefont
  {Smith}(1996)}]{Lewin:1995rx}%
  \BibitemOpen
  \bibfield  {author} {\bibinfo {author} {\bibfnamefont {J.~D.}\ \bibnamefont
  {Lewin}}\ and\ \bibinfo {author} {\bibfnamefont {P.~F.}\ \bibnamefont
  {Smith}},\ }\href {\doibase 10.1016/S0927-6505(96)00047-3} {\bibfield
  {journal} {\bibinfo  {journal} {Astropart. Phys.}\ }\textbf {\bibinfo
  {volume} {6}},\ \bibinfo {pages} {87} (\bibinfo {year} {1996})}\BibitemShut
  {NoStop}%
\bibitem [{\citenamefont {Aprile}\ \emph {et~al.}(2018)\citenamefont {Aprile}
  \emph {et~al.}}]{Aprile:2018dbl}%
  \BibitemOpen
  \bibfield  {author} {\bibinfo {author} {\bibfnamefont {E.}~\bibnamefont
  {Aprile}} \emph {et~al.} (\bibinfo {collaboration} {XENON}),\ }\href
  {\doibase 10.1103/PhysRevLett.121.111302} {\bibfield  {journal} {\bibinfo
  {journal} {Phys. Rev. Lett.}\ }\textbf {\bibinfo {volume} {121}},\ \bibinfo
  {pages} {111302} (\bibinfo {year} {2018})},\ \Eprint
  {http://arxiv.org/abs/1805.12562} {arXiv:1805.12562 [astro-ph.CO]}
  \BibitemShut {NoStop}%
\bibitem [{\citenamefont {Davidson}\ \emph {et~al.}(1991)\citenamefont
  {Davidson}, \citenamefont {Campbell},\ and\ \citenamefont
  {Bailey}}]{Davidson:1991si}%
  \BibitemOpen
  \bibfield  {author} {\bibinfo {author} {\bibfnamefont {S.}~\bibnamefont
  {Davidson}}, \bibinfo {author} {\bibfnamefont {B.}~\bibnamefont {Campbell}},
  \ and\ \bibinfo {author} {\bibfnamefont {D.~C.}\ \bibnamefont {Bailey}},\
  }\href {\doibase 10.1103/PhysRevD.43.2314} {\bibfield  {journal} {\bibinfo
  {journal} {Phys. Rev.}\ }\textbf {\bibinfo {volume} {D43}},\ \bibinfo {pages}
  {2314} (\bibinfo {year} {1991})}\BibitemShut {NoStop}%
\bibitem [{\citenamefont {Prinz}\ \emph {et~al.}(1998)\citenamefont {Prinz}
  \emph {et~al.}}]{Prinz:1998ua}%
  \BibitemOpen
  \bibfield  {author} {\bibinfo {author} {\bibfnamefont {A.~A.}\ \bibnamefont
  {Prinz}} \emph {et~al.},\ }\href {\doibase 10.1103/PhysRevLett.81.1175}
  {\bibfield  {journal} {\bibinfo  {journal} {Phys. Rev. Lett.}\ }\textbf
  {\bibinfo {volume} {81}},\ \bibinfo {pages} {1175} (\bibinfo {year}
  {1998})},\ \Eprint {http://arxiv.org/abs/hep-ex/9804008}
  {arXiv:hep-ex/9804008 [hep-ex]} \BibitemShut {NoStop}%
\bibitem [{\citenamefont {Davidson}\ \emph {et~al.}(2000)\citenamefont
  {Davidson}, \citenamefont {Hannestad},\ and\ \citenamefont
  {Raffelt}}]{Davidson:2000hf}%
  \BibitemOpen
  \bibfield  {author} {\bibinfo {author} {\bibfnamefont {S.}~\bibnamefont
  {Davidson}}, \bibinfo {author} {\bibfnamefont {S.}~\bibnamefont {Hannestad}},
  \ and\ \bibinfo {author} {\bibfnamefont {G.}~\bibnamefont {Raffelt}},\ }\href
  {\doibase 10.1088/1126-6708/2000/05/003} {\bibfield  {journal} {\bibinfo
  {journal} {JHEP}\ }\textbf {\bibinfo {volume} {05}},\ \bibinfo {pages} {003}
  (\bibinfo {year} {2000})},\ \Eprint {http://arxiv.org/abs/hep-ph/0001179}
  {arXiv:hep-ph/0001179 [hep-ph]} \BibitemShut {NoStop}%
\bibitem [{\citenamefont {Chatrchyan}\ \emph {et~al.}(2013)\citenamefont
  {Chatrchyan} \emph {et~al.}}]{CMS:2012xi}%
  \BibitemOpen
  \bibfield  {author} {\bibinfo {author} {\bibfnamefont {S.}~\bibnamefont
  {Chatrchyan}} \emph {et~al.} (\bibinfo {collaboration} {CMS}),\ }\href
  {\doibase 10.1103/PhysRevD.87.092008} {\bibfield  {journal} {\bibinfo
  {journal} {Phys. Rev.}\ }\textbf {\bibinfo {volume} {D87}},\ \bibinfo {pages}
  {092008} (\bibinfo {year} {2013})},\ \Eprint {http://arxiv.org/abs/1210.2311}
  {arXiv:1210.2311 [hep-ex]} \BibitemShut {NoStop}%
\bibitem [{\citenamefont {Magill}\ \emph {et~al.}(2018)\citenamefont {Magill},
  \citenamefont {Plestid}, \citenamefont {Pospelov},\ and\ \citenamefont
  {Tsai}}]{Magill:2018tbb}%
  \BibitemOpen
  \bibfield  {author} {\bibinfo {author} {\bibfnamefont {G.}~\bibnamefont
  {Magill}}, \bibinfo {author} {\bibfnamefont {R.}~\bibnamefont {Plestid}},
  \bibinfo {author} {\bibfnamefont {M.}~\bibnamefont {Pospelov}}, \ and\
  \bibinfo {author} {\bibfnamefont {Y.-D.}\ \bibnamefont {Tsai}},\ }\href@noop
  {} {\  (\bibinfo {year} {2018})},\ \Eprint {http://arxiv.org/abs/1806.03310}
  {arXiv:1806.03310 [hep-ph]} \BibitemShut {NoStop}%
\bibitem [{\citenamefont {Chang}\ \emph {et~al.}(2018)\citenamefont {Chang},
  \citenamefont {Essig},\ and\ \citenamefont {McDermott}}]{Chang:2018rso}%
  \BibitemOpen
  \bibfield  {author} {\bibinfo {author} {\bibfnamefont {J.~H.}\ \bibnamefont
  {Chang}}, \bibinfo {author} {\bibfnamefont {R.}~\bibnamefont {Essig}}, \ and\
  \bibinfo {author} {\bibfnamefont {S.~D.}\ \bibnamefont {McDermott}},\ }\href
  {\doibase 10.1007/JHEP09(2018)051} {\bibfield  {journal} {\bibinfo  {journal}
  {JHEP}\ }\textbf {\bibinfo {volume} {09}},\ \bibinfo {pages} {051} (\bibinfo
  {year} {2018})},\ \Eprint {http://arxiv.org/abs/1803.00993} {arXiv:1803.00993
  [hep-ph]} \BibitemShut {NoStop}%
\bibitem [{\citenamefont {Vogel}\ and\ \citenamefont
  {Redondo}(2014)}]{Vogel:2013raa}%
  \BibitemOpen
  \bibfield  {author} {\bibinfo {author} {\bibfnamefont {H.}~\bibnamefont
  {Vogel}}\ and\ \bibinfo {author} {\bibfnamefont {J.}~\bibnamefont
  {Redondo}},\ }\href {\doibase 10.1088/1475-7516/2014/02/029} {\bibfield
  {journal} {\bibinfo  {journal} {JCAP}\ }\textbf {\bibinfo {volume} {1402}},\
  \bibinfo {pages} {029} (\bibinfo {year} {2014})},\ \Eprint
  {http://arxiv.org/abs/1311.2600} {arXiv:1311.2600 [hep-ph]} \BibitemShut
  {NoStop}%
\bibitem [{\citenamefont {Ahmed}\ \emph {et~al.}(2010)\citenamefont {Ahmed}
  \emph {et~al.}}]{Ahmed:2009zw}%
  \BibitemOpen
  \bibfield  {author} {\bibinfo {author} {\bibfnamefont {Z.}~\bibnamefont
  {Ahmed}} \emph {et~al.} (\bibinfo {collaboration} {CDMS-II}),\ }\href
  {\doibase 10.1126/science.1186112} {\bibfield  {journal} {\bibinfo  {journal}
  {Science}\ }\textbf {\bibinfo {volume} {327}},\ \bibinfo {pages} {1619}
  (\bibinfo {year} {2010})},\ \Eprint {http://arxiv.org/abs/0912.3592}
  {arXiv:0912.3592 [astro-ph.CO]} \BibitemShut {NoStop}%
\bibitem [{\citenamefont {Essig}\ \emph {et~al.}(2017)\citenamefont {Essig},
  \citenamefont {Volansky},\ and\ \citenamefont {Yu}}]{Essig:2017kqs}%
  \BibitemOpen
  \bibfield  {author} {\bibinfo {author} {\bibfnamefont {R.}~\bibnamefont
  {Essig}}, \bibinfo {author} {\bibfnamefont {T.}~\bibnamefont {Volansky}}, \
  and\ \bibinfo {author} {\bibfnamefont {T.-T.}\ \bibnamefont {Yu}},\ }\href
  {\doibase 10.1103/PhysRevD.96.043017} {\bibfield  {journal} {\bibinfo
  {journal} {Phys. Rev.}\ }\textbf {\bibinfo {volume} {D96}},\ \bibinfo {pages}
  {043017} (\bibinfo {year} {2017})},\ \Eprint
  {http://arxiv.org/abs/1703.00910} {arXiv:1703.00910 [hep-ph]} \BibitemShut
  {NoStop}%
\bibitem [{\citenamefont {Kachulis}\ \emph {et~al.}(2018)\citenamefont
  {Kachulis} \emph {et~al.}}]{Kachulis:2017nci}%
  \BibitemOpen
  \bibfield  {author} {\bibinfo {author} {\bibfnamefont {C.}~\bibnamefont
  {Kachulis}} \emph {et~al.} (\bibinfo {collaboration} {Super-Kamiokande}),\
  }\href {\doibase 10.1103/PhysRevLett.120.221301} {\bibfield  {journal}
  {\bibinfo  {journal} {Phys. Rev. Lett.}\ }\textbf {\bibinfo {volume} {120}},\
  \bibinfo {pages} {221301} (\bibinfo {year} {2018})},\ \Eprint
  {http://arxiv.org/abs/1711.05278} {arXiv:1711.05278 [hep-ex]} \BibitemShut
  {NoStop}%
\bibitem [{\citenamefont {{Abbasi}}\ \emph {et~al.}(2011)\citenamefont
  {{Abbasi}}, \citenamefont {{Abdou}}, \citenamefont {{Abu-Zayyad}},
  \citenamefont {{Ackermann}}, \citenamefont {{Adams}}, \citenamefont
  {{Aguilar}}, \citenamefont {{Ahlers}}, \citenamefont {{Allen}}, \citenamefont
  {{Altmann}}, \citenamefont {{Andeen}},\ and\ \citenamefont
  {et~al.}}]{2011A&A...535A.109A}%
  \BibitemOpen
  \bibfield  {author} {\bibinfo {author} {\bibfnamefont {R.}~\bibnamefont
  {{Abbasi}}}, \bibinfo {author} {\bibfnamefont {Y.}~\bibnamefont {{Abdou}}},
  \bibinfo {author} {\bibfnamefont {T.}~\bibnamefont {{Abu-Zayyad}}}, \bibinfo
  {author} {\bibfnamefont {M.}~\bibnamefont {{Ackermann}}}, \bibinfo {author}
  {\bibfnamefont {J.}~\bibnamefont {{Adams}}}, \bibinfo {author} {\bibfnamefont
  {J.~A.}\ \bibnamefont {{Aguilar}}}, \bibinfo {author} {\bibfnamefont
  {M.}~\bibnamefont {{Ahlers}}}, \bibinfo {author} {\bibfnamefont {M.~M.}\
  \bibnamefont {{Allen}}}, \bibinfo {author} {\bibfnamefont {D.}~\bibnamefont
  {{Altmann}}}, \bibinfo {author} {\bibfnamefont {K.}~\bibnamefont {{Andeen}}},
  \ and\ \bibinfo {author} {\bibnamefont {et~al.}},\ }\href {\doibase
  10.1051/0004-6361/201117810} {\bibfield  {journal} {\bibinfo  {journal}
  {aap}\ }\textbf {\bibinfo {volume} {535}},\ \bibinfo {eid} {A109} (\bibinfo
  {year} {2011})},\ \Eprint {http://arxiv.org/abs/1108.0171} {arXiv:1108.0171
  [astro-ph.HE]} \BibitemShut {NoStop}%
\bibitem [{\citenamefont {Patrignani}(2018)}]{PhysRevD.98.030001}%
  \BibitemOpen
  \bibfield  {author} {\bibinfo {author} {\bibfnamefont {C.}~\bibnamefont
  {Patrignani}} (\bibinfo {collaboration} {Particle Data Group}),\ }\href
  {\doibase 10.1103/PhysRevD.98.030001} {\bibfield  {journal} {\bibinfo
  {journal} {Phys. Rev. D}\ }\textbf {\bibinfo {volume} {98}},\ \bibinfo
  {pages} {030001} (\bibinfo {year} {2018})}\BibitemShut {NoStop}%
\bibitem [{\citenamefont {Jackson}(1998)}]{Jackson:1998nia}%
  \BibitemOpen
  \bibfield  {author} {\bibinfo {author} {\bibfnamefont {J.~D.}\ \bibnamefont
  {Jackson}},\ }\href@noop {} {\emph {\bibinfo {title} {{Classical
  Electrodynamics}}}}\ (\bibinfo  {publisher} {Wiley},\ \bibinfo {year}
  {1998})\BibitemShut {NoStop}%
\bibitem [{\citenamefont {Alvis}\ \emph {et~al.}(2018)\citenamefont {Alvis}
  \emph {et~al.}}]{Alvis:2018yte}%
  \BibitemOpen
  \bibfield  {author} {\bibinfo {author} {\bibfnamefont {S.~I.}\ \bibnamefont
  {Alvis}} \emph {et~al.} (\bibinfo {collaboration} {Majorana}),\ }\href
  {\doibase 10.1103/PhysRevLett.120.211804} {\bibfield  {journal} {\bibinfo
  {journal} {Phys. Rev. Lett.}\ }\textbf {\bibinfo {volume} {120}},\ \bibinfo
  {pages} {211804} (\bibinfo {year} {2018})},\ \Eprint
  {http://arxiv.org/abs/1801.10145} {arXiv:1801.10145 [hep-ex]} \BibitemShut
  {NoStop}%
\bibitem [{\citenamefont {Ambrosio}\ \emph {et~al.}(2004)\citenamefont
  {Ambrosio} \emph {et~al.}}]{Ambrosio:2004ub}%
  \BibitemOpen
  \bibfield  {author} {\bibinfo {author} {\bibfnamefont {M.}~\bibnamefont
  {Ambrosio}} \emph {et~al.} (\bibinfo {collaboration} {MACRO}),\ }\href@noop
  {} {\  (\bibinfo {year} {2004})},\ \Eprint
  {http://arxiv.org/abs/hep-ex/0402006} {arXiv:hep-ex/0402006 [hep-ex]}
  \BibitemShut {NoStop}%
\bibitem [{\citenamefont {Kajino}\ \emph {et~al.}(1984)\citenamefont {Kajino},
  \citenamefont {Matsuno}, \citenamefont {Kitamura}, \citenamefont {Aoki},
  \citenamefont {Yuan}, \citenamefont {Mitsui}, \citenamefont {Ohashi},\ and\
  \citenamefont {Okada}}]{Kajino:1984ug}%
  \BibitemOpen
  \bibfield  {author} {\bibinfo {author} {\bibfnamefont {F.}~\bibnamefont
  {Kajino}}, \bibinfo {author} {\bibfnamefont {S.}~\bibnamefont {Matsuno}},
  \bibinfo {author} {\bibfnamefont {T.}~\bibnamefont {Kitamura}}, \bibinfo
  {author} {\bibfnamefont {T.}~\bibnamefont {Aoki}}, \bibinfo {author}
  {\bibfnamefont {Y.~K.}\ \bibnamefont {Yuan}}, \bibinfo {author}
  {\bibfnamefont {K.}~\bibnamefont {Mitsui}}, \bibinfo {author} {\bibfnamefont
  {Y.}~\bibnamefont {Ohashi}}, \ and\ \bibinfo {author} {\bibfnamefont
  {A.}~\bibnamefont {Okada}},\ }\href {\doibase 10.1088/0305-4616/10/4/007}
  {\bibfield  {journal} {\bibinfo  {journal} {J. Phys.}\ }\textbf {\bibinfo
  {volume} {G10}},\ \bibinfo {pages} {447} (\bibinfo {year}
  {1984})}\BibitemShut {NoStop}%
\bibitem [{\citenamefont {{Barish}}\ \emph {et~al.}(1987)\citenamefont
  {{Barish}}, \citenamefont {{Liu}},\ and\ \citenamefont
  {{Lane}}}]{1987PhRvD..36.2641B}%
  \BibitemOpen
  \bibfield  {author} {\bibinfo {author} {\bibfnamefont {B.}~\bibnamefont
  {{Barish}}}, \bibinfo {author} {\bibfnamefont {G.}~\bibnamefont {{Liu}}}, \
  and\ \bibinfo {author} {\bibfnamefont {C.}~\bibnamefont {{Lane}}},\ }\href
  {\doibase 10.1103/PhysRevD.36.2641} {\bibfield  {journal} {\bibinfo
  {journal} {\prd}\ }\textbf {\bibinfo {volume} {36}},\ \bibinfo {pages} {2641}
  (\bibinfo {year} {1987})}\BibitemShut {NoStop}%
\bibitem [{\citenamefont {{Alexeyev}}\ \emph {et~al.}(1983)\citenamefont
  {{Alexeyev}}, \citenamefont {{Boliev}}, \citenamefont {{Chudakov}},
  \citenamefont {{Mikheyev}},\ and\ \citenamefont
  {{Shkvorets}}}]{1983ICRC....5...52A}%
  \BibitemOpen
  \bibfield  {author} {\bibinfo {author} {\bibfnamefont {E.~N.}\ \bibnamefont
  {{Alexeyev}}}, \bibinfo {author} {\bibfnamefont {M.~M.}\ \bibnamefont
  {{Boliev}}}, \bibinfo {author} {\bibfnamefont {A.~E.}\ \bibnamefont
  {{Chudakov}}}, \bibinfo {author} {\bibfnamefont {S.~P.}\ \bibnamefont
  {{Mikheyev}}}, \ and\ \bibinfo {author} {\bibfnamefont {O.~I.}\ \bibnamefont
  {{Shkvorets}}},\ }\href@noop {} {\bibfield  {journal} {\bibinfo  {journal}
  {International Cosmic Ray Conference}\ }\textbf {\bibinfo {volume} {5}},\
  \bibinfo {pages} {52} (\bibinfo {year} {1983})}\BibitemShut {NoStop}%
\bibitem [{\citenamefont {{Alexeyev}}\ \emph {et~al.}(1985)\citenamefont
  {{Alexeyev}}, \citenamefont {{Boliev}}, \citenamefont {{Chudakov}},\ and\
  \citenamefont {{Mikheyev}}}]{1985ICRC....8..250A}%
  \BibitemOpen
  \bibfield  {author} {\bibinfo {author} {\bibfnamefont {E.~N.}\ \bibnamefont
  {{Alexeyev}}}, \bibinfo {author} {\bibfnamefont {M.~M.}\ \bibnamefont
  {{Boliev}}}, \bibinfo {author} {\bibfnamefont {A.~E.}\ \bibnamefont
  {{Chudakov}}}, \ and\ \bibinfo {author} {\bibfnamefont {S.~P.}\ \bibnamefont
  {{Mikheyev}}},\ }\href@noop {} {\bibfield  {journal} {\bibinfo  {journal}
  {International Cosmic Ray Conference}\ }\textbf {\bibinfo {volume} {8}}
  (\bibinfo {year} {1985})}\BibitemShut {NoStop}%
\bibitem [{\citenamefont {Mount}\ \emph {et~al.}(2017)\citenamefont {Mount}
  \emph {et~al.}}]{Mount:2017qzi}%
  \BibitemOpen
  \bibfield  {author} {\bibinfo {author} {\bibfnamefont {B.~J.}\ \bibnamefont
  {Mount}} \emph {et~al.},\ }\href@noop {} {\  (\bibinfo {year} {2017})},\
  \Eprint {http://arxiv.org/abs/1703.09144} {arXiv:1703.09144
  [physics.ins-det]} \BibitemShut {NoStop}%
\bibitem [{\citenamefont {Chu}\ \emph {et~al.}(2012)\citenamefont {Chu},
  \citenamefont {Hambye},\ and\ \citenamefont {Tytgat}}]{Chu:2011be}%
  \BibitemOpen
  \bibfield  {author} {\bibinfo {author} {\bibfnamefont {X.}~\bibnamefont
  {Chu}}, \bibinfo {author} {\bibfnamefont {T.}~\bibnamefont {Hambye}}, \ and\
  \bibinfo {author} {\bibfnamefont {M.~H.~G.}\ \bibnamefont {Tytgat}},\ }\href
  {\doibase 10.1088/1475-7516/2012/05/034} {\bibfield  {journal} {\bibinfo
  {journal} {JCAP}\ }\textbf {\bibinfo {volume} {1205}},\ \bibinfo {pages}
  {034} (\bibinfo {year} {2012})},\ \Eprint {http://arxiv.org/abs/1112.0493}
  {arXiv:1112.0493 [hep-ph]} \BibitemShut {NoStop}%
\bibitem [{\citenamefont {Aghanim}\ \emph {et~al.}(2018)\citenamefont {Aghanim}
  \emph {et~al.}}]{Aghanim:2018eyx}%
  \BibitemOpen
  \bibfield  {author} {\bibinfo {author} {\bibfnamefont {N.}~\bibnamefont
  {Aghanim}} \emph {et~al.} (\bibinfo {collaboration} {Planck}),\ }\href@noop
  {} {\  (\bibinfo {year} {2018})},\ \Eprint {http://arxiv.org/abs/1807.06209}
  {arXiv:1807.06209 [astro-ph.CO]} \BibitemShut {NoStop}%
\bibitem [{\citenamefont {Bowman}\ \emph {et~al.}(2018)\citenamefont {Bowman},
  \citenamefont {Rogers}, \citenamefont {Monsalve}, \citenamefont {Mozdzen},\
  and\ \citenamefont {Mahesh}}]{bowman2018absorption}%
  \BibitemOpen
  \bibfield  {author} {\bibinfo {author} {\bibfnamefont {J.~D.}\ \bibnamefont
  {Bowman}}, \bibinfo {author} {\bibfnamefont {A.~E.}\ \bibnamefont {Rogers}},
  \bibinfo {author} {\bibfnamefont {R.~A.}\ \bibnamefont {Monsalve}}, \bibinfo
  {author} {\bibfnamefont {T.~J.}\ \bibnamefont {Mozdzen}}, \ and\ \bibinfo
  {author} {\bibfnamefont {N.}~\bibnamefont {Mahesh}},\ }\href@noop {}
  {\bibfield  {journal} {\bibinfo  {journal} {Nature}\ }\textbf {\bibinfo
  {volume} {555}},\ \bibinfo {pages} {67} (\bibinfo {year} {2018})}\BibitemShut
  {NoStop}%
\bibitem [{\citenamefont {Berlin}\ \emph {et~al.}(2018)\citenamefont {Berlin},
  \citenamefont {Hooper}, \citenamefont {Krnjaic},\ and\ \citenamefont
  {McDermott}}]{Berlin:2018sjs}%
  \BibitemOpen
  \bibfield  {author} {\bibinfo {author} {\bibfnamefont {A.}~\bibnamefont
  {Berlin}}, \bibinfo {author} {\bibfnamefont {D.}~\bibnamefont {Hooper}},
  \bibinfo {author} {\bibfnamefont {G.}~\bibnamefont {Krnjaic}}, \ and\
  \bibinfo {author} {\bibfnamefont {S.~D.}\ \bibnamefont {McDermott}},\ }\href
  {\doibase 10.1103/PhysRevLett.121.011102} {\bibfield  {journal} {\bibinfo
  {journal} {Phys. Rev. Lett.}\ }\textbf {\bibinfo {volume} {121}},\ \bibinfo
  {pages} {011102} (\bibinfo {year} {2018})},\ \Eprint
  {http://arxiv.org/abs/1803.02804} {arXiv:1803.02804 [hep-ph]} \BibitemShut
  {NoStop}%
\bibitem [{\citenamefont {{Bell}}(1978)}]{1978MNRAS.182..147B}%
  \BibitemOpen
  \bibfield  {author} {\bibinfo {author} {\bibfnamefont {A.~R.}\ \bibnamefont
  {{Bell}}},\ }\href {\doibase 10.1093/mnras/182.2.147} {\bibfield  {journal}
  {\bibinfo  {journal} {Monthly Notices of the Royal Astronomical Society}\
  }\textbf {\bibinfo {volume} {182}},\ \bibinfo {pages} {147} (\bibinfo {year}
  {1978})}\BibitemShut {NoStop}%
\bibitem [{\citenamefont {{Loeb}}\ and\ \citenamefont
  {{Furlanetto}}(2013)}]{2013fgu..book.....L}%
  \BibitemOpen
  \bibfield  {author} {\bibinfo {author} {\bibfnamefont {A.}~\bibnamefont
  {{Loeb}}}\ and\ \bibinfo {author} {\bibfnamefont {S.~R.}\ \bibnamefont
  {{Furlanetto}}},\ }\href@noop {} {\emph {\bibinfo {title} {The First Galaxies
  in the Universe}}}\ (\bibinfo {year} {2013})\BibitemShut {NoStop}%
\bibitem [{\citenamefont {{Toomre}}(1964)}]{1964ApJ...139.1217T}%
  \BibitemOpen
  \bibfield  {author} {\bibinfo {author} {\bibfnamefont {A.}~\bibnamefont
  {{Toomre}}},\ }\href {\doibase 10.1086/147861} {\bibfield  {journal}
  {\bibinfo  {journal} {\apj}\ }\textbf {\bibinfo {volume} {139}},\ \bibinfo
  {pages} {1217} (\bibinfo {year} {1964})}\BibitemShut {NoStop}%
\bibitem [{\citenamefont {{Hopkins}}\ \emph {et~al.}(2012)\citenamefont
  {{Hopkins}}, \citenamefont {{Quataert}},\ and\ \citenamefont
  {{Murray}}}]{2012MNRAS.421.3488H}%
  \BibitemOpen
  \bibfield  {author} {\bibinfo {author} {\bibfnamefont {P.~F.}\ \bibnamefont
  {{Hopkins}}}, \bibinfo {author} {\bibfnamefont {E.}~\bibnamefont
  {{Quataert}}}, \ and\ \bibinfo {author} {\bibfnamefont {N.}~\bibnamefont
  {{Murray}}},\ }\href {\doibase 10.1111/j.1365-2966.2012.20578.x} {\bibfield
  {journal} {\bibinfo  {journal} {Monthly Notices of the Royal Astronomical
  Society}\ }\textbf {\bibinfo {volume} {421}},\ \bibinfo {pages} {3488}
  (\bibinfo {year} {2012})},\ \Eprint {http://arxiv.org/abs/1110.4636}
  {arXiv:1110.4636 [astro-ph.CO]} \BibitemShut {NoStop}%
\bibitem [{\citenamefont {{Stinson}}\ \emph {et~al.}(2015)\citenamefont
  {{Stinson}}, \citenamefont {{Dutton}}, \citenamefont {{Wang}}, \citenamefont
  {{Macci{\`o}}}, \citenamefont {{Herpich}}, \citenamefont {{Bradford}},
  \citenamefont {{Quinn}}, \citenamefont {{Wadsley}},\ and\ \citenamefont
  {{Keller}}}]{2015MNRAS.454.1105S}%
  \BibitemOpen
  \bibfield  {author} {\bibinfo {author} {\bibfnamefont {G.~S.}\ \bibnamefont
  {{Stinson}}}, \bibinfo {author} {\bibfnamefont {A.~A.}\ \bibnamefont
  {{Dutton}}}, \bibinfo {author} {\bibfnamefont {L.}~\bibnamefont {{Wang}}},
  \bibinfo {author} {\bibfnamefont {A.~V.}\ \bibnamefont {{Macci{\`o}}}},
  \bibinfo {author} {\bibfnamefont {J.}~\bibnamefont {{Herpich}}}, \bibinfo
  {author} {\bibfnamefont {J.~D.}\ \bibnamefont {{Bradford}}}, \bibinfo
  {author} {\bibfnamefont {T.~R.}\ \bibnamefont {{Quinn}}}, \bibinfo {author}
  {\bibfnamefont {J.}~\bibnamefont {{Wadsley}}}, \ and\ \bibinfo {author}
  {\bibfnamefont {B.}~\bibnamefont {{Keller}}},\ }\href {\doibase
  10.1093/mnras/stv1985} {\bibfield  {journal} {\bibinfo  {journal} {Monthly
  Notices of the Royal Astronomical Society}\ }\textbf {\bibinfo {volume}
  {454}},\ \bibinfo {pages} {1105} (\bibinfo {year} {2015})},\ \Eprint
  {http://arxiv.org/abs/1506.08785} {arXiv:1506.08785 [astro-ph.GA]}
  \BibitemShut {NoStop}%
\bibitem [{\citenamefont {Dekel}\ and\ \citenamefont
  {Silk}(1986)}]{dekel1986origin}%
  \BibitemOpen
  \bibfield  {author} {\bibinfo {author} {\bibfnamefont {A.}~\bibnamefont
  {Dekel}}\ and\ \bibinfo {author} {\bibfnamefont {J.}~\bibnamefont {Silk}},\
  }\href@noop {} {\bibfield  {journal} {\bibinfo  {journal} {The Astrophysical
  Journal}\ }\textbf {\bibinfo {volume} {303}},\ \bibinfo {pages} {39}
  (\bibinfo {year} {1986})}\BibitemShut {NoStop}%
\bibitem [{\citenamefont {Adams}\ \emph {et~al.}(1997)\citenamefont {Adams},
  \citenamefont {Freese}, \citenamefont {Laughlin}, \citenamefont {Tarle},\
  and\ \citenamefont {Schwadron}}]{Adams:1997ym}%
  \BibitemOpen
  \bibfield  {author} {\bibinfo {author} {\bibfnamefont {F.~C.}\ \bibnamefont
  {Adams}}, \bibinfo {author} {\bibfnamefont {K.}~\bibnamefont {Freese}},
  \bibinfo {author} {\bibfnamefont {G.}~\bibnamefont {Laughlin}}, \bibinfo
  {author} {\bibfnamefont {G.}~\bibnamefont {Tarle}}, \ and\ \bibinfo {author}
  {\bibfnamefont {N.}~\bibnamefont {Schwadron}},\ }\href {\doibase
  10.1086/304962} {\bibfield  {journal} {\bibinfo  {journal} {Astrophys. J.}\
  }\textbf {\bibinfo {volume} {491}},\ \bibinfo {pages} {6} (\bibinfo {year}
  {1997})},\ \Eprint {http://arxiv.org/abs/astro-ph/9710113}
  {arXiv:astro-ph/9710113 [astro-ph]} \BibitemShut {NoStop}%
\bibitem [{\citenamefont {Aloisio}\ and\ \citenamefont
  {Berezinsky}(2004)}]{Aloisio:2004jda}%
  \BibitemOpen
  \bibfield  {author} {\bibinfo {author} {\bibfnamefont {R.}~\bibnamefont
  {Aloisio}}\ and\ \bibinfo {author} {\bibfnamefont {V.}~\bibnamefont
  {Berezinsky}},\ }\href {\doibase 10.1086/421869} {\bibfield  {journal}
  {\bibinfo  {journal} {Astrophys. J.}\ }\textbf {\bibinfo {volume} {612}},\
  \bibinfo {pages} {900} (\bibinfo {year} {2004})},\ \Eprint
  {http://arxiv.org/abs/astro-ph/0403095} {arXiv:astro-ph/0403095 [astro-ph]}
  \BibitemShut {NoStop}%
\end{thebibliography}%

\let\addcontentsline\oldaddcontentsline
  
\end{document}